\documentclass[12pt,preprint]{aastex}
\usepackage{apjfonts}
\usepackage{natbib}
\usepackage{amssymb, amsmath, amsbsy, epsfig, epsf, slashbox, rotating}
\usepackage[dvips]{color}
\usepackage{fancyvrb}
\usepackage{enumerate}
\usepackage{afterpage}
\usepackage{lscape}

\newcommand{\msun}{\ensuremath{M_{\odot}}}
\newcommand{\lsun}{\ensuremath{L_{\odot}}}
\newcommand{\kms}{\ensuremath{\mathrm{km~s^{-1}}}}

\newcommand{\chisq}{\ensuremath{\chi^2}}

\newcommand{\ha}{H\ensuremath{\alpha}}
\newcommand{\hb}{H\ensuremath{\beta}}

\newcommand{\oiii}{[O\,{\footnotesize III}]}

\newcommand{\feii}{{\rm Fe\,{\footnotesize II}}}

\newcommand{\mgii}{Mg\,{\footnotesize II}}

\newcommand{\civ}{C\,{\footnotesize IV}}

\newcommand{\hei}{He\,{\footnotesize I}*}

\newcommand{\colnh}{\ensuremath{N_\mathrm{H}}}

\newcommand{\etal}{et~al.}

\slugcomment{}
\shorttitle{\hei\,$\lambda3889$ BAL Quasars}
\shortauthors{W.-J.~Liu \etal}

\begin{document}

\title{A Comprehensive Study of Broad Absorption Line Quasars: I. Prevalence of \hei\ Absorption Line Multiplets 
in Low-Ionization Objects}

\author{ Wen-Juan~Liu\altaffilmark{1,2}, Hongyan~Zhou\altaffilmark{2,1}, Tuo~Ji\altaffilmark{2,1},
Weimin~Yuan\altaffilmark{3}, Ting-Gui~Wang\altaffilmark{1}, Ge~Jian\altaffilmark{4}, Xiheng~Shi\altaffilmark{2}, 
Shaohua~Zhang\altaffilmark{2}, Peng~Jiang\altaffilmark{1,2}, Xinwen~Shu\altaffilmark{5}, 
Huiyuan~Wang\altaffilmark{1}, Shu-Fen~Wang\altaffilmark{1,2}, Luming~Sun\altaffilmark{1,2}, 
Chenwei~Yang\altaffilmark{1}, Bo~Liu\altaffilmark{1,2}, and Wen~Zhao\altaffilmark{1} }

\altaffiltext{1}{Key Laboratory for Research in Galaxies and Cosmology, Department of Astronomy, University 
  of Sciences and Technology of China, Chinese Academy of Sciences,
Hefei, Anhui 230026, China, zoey@mail.ustc.edu.cn}
\altaffiltext{2}{Polar Research Institute of China, 451 Jinqiao Rd, Shanghai 200136, China, zhouhongyan@pric.org.cn, 
jituo1982@gmail.com}
\altaffiltext{3}{National Astronomical Observatories, Chinese Academy of Sciences, 20A Datun Road, Beijing 100012, 
China}
\altaffiltext{4}{Astronomy Department, University of Florida, 211 Bryant Space Science Center, P.O. Box 112055, 
Gainesville, FL 32611, USA}
\altaffiltext{5}{CEA Saclay, DSM/Irfu/Service d'Astrophysique,
Orme des Merisiers, 91191 Gif-sur-Yvette Cedex, France}
\begin{abstract}
  Neutral Helium multiplets, \hei\,$\lambda\lambda3189,3889,10830$ are very useful diagnostics to the
  geometry and physical conditions of the absorbing gas in quasars. 
  So far only a handful of \hei\ detections have been reported.
  Using a newly developed method, we detected \hei$\lambda3889$ absorption line in 101 sources of a well-defined 
  sample of 285 \mgii\ BAL quasars selected from SDSS DR5.
  This has increased the number of \hei\ BAL quasars by more than one order of magnitude.
  We further detected \hei$\lambda3189$ in 50\% (52/101) quasars in the sample. 
  The detection fraction of \hei\ BALs in \mgii\ BAL quasars is $\sim$35\% as a whole, and
  increases dramatically with increasing spectral signal-to-noise ratios, from $\sim$18\% at S/N $\leq$~10 
  to $\sim$93\% at S/N $\geq$~35. This suggests that \hei\ BALs could be detected in most \mgii\ LoBAL
  quasars, provided spectra S/N is high enough. 
  Such a surprisingly high \hei\ BAL fraction is actually predicted from photo-ionization calculations based on a 
  simple BAL model.
  The result indicates that \hei\ absorption lines can be used to search for BAL quasars at low-$z$, which cannot 
  be identified by ground-based optical spectroscopic survey with commonly seen UV absorption lines. 
  Using \hei\,$\lambda$3889, we discovered 19 BAL quasars at $z<0.3$ from available SDSS spectral database.
  The fraction of \hei\ BAL quasars is similar to that of LoBAL objects. 
\end{abstract}

\keywords{galaxies: active --- quasars: absorption lines  --- quasars:  general}

\setcounter{footnote}{0}
\setcounter{section}{0}

\section{Introduction\label{sec:intro}}

   Broad absorption line quasars are a small yet important population of active galactic nuclei (AGNs) that 
   have a continuous broad absorption trough spanning a large range of velocities up to several times 
   10$^4$~km~s$^{-1}$ \citep{1991ApJ....373....23W,2006ApJS..165.....1T,2009ApJ....692..758G}.
   The broad absorption lines are generally blueshifted with respect to the systematic redshift of their emission 
   counterparts by up to 0.1-0.2 of light speed 
   \citep[e.g.][]{1991ApJ....373....23W,1992ApJ...401..529K,1993ApJS...88..357K,2013MNRAS.435..133H}.
   Traditionally, BAL quasars are classified into three subcategories depending on which absorption features are 
   seen.
   Hi-ionization BAL quasars (HiBALs) show absorption in \ion{N}{5}\,$\lambda\lambda1238,1242$, 
   \ion{Si}{4}\,$\lambda\lambda1393,1402$ and \civ\,$\lambda\lambda1548,1550$, and comprise about 85\% of the BAL 
   quasars.
   Low-ionization quasars (LoBALs) show, besides all the HiBAL features, absorption troughs in low-ionization species 
   like \mgii\,$\lambda\lambda2796,2803$, \ion{Al}{3}\,$\lambda\lambda1854,1862$ 
   (hereafter N\,V, Si\,IV, \civ, \mgii, Al\,III), and comprise about 15\% of the whole BAL population 
   \citep[e.g.][]{2002ApJ....578L..31T,2003AJ....126.2594R,2003AJ....125.1784H}.
   Additionally, a rare class of LoBAL termed FeLoBAL quasars, show absorption features arising from excited-state 
   of \feii.

   BALs are generally believed to be associated with AGN outflows from an accretion disk
   \citep[e.g.][]{1995ApJ...451..498M}.
   Outflows may carry away huge amounts of material, energy, and angular momentum, and are believed to be one of the 
   most important feedback processes that connects AGNs and their host galaxies 
   \citep[e.g.][]{2005Natur.433..604D,2006MmSAI..77..573E,2010ApJ...709..611D,2012ApJ...745..178F}.
   Regarding the origin of BALs in quasars, there are two theoretical scenarios.
   The first is that BAL quasars are essentially normal quasars viewed along a line of sight that penetrates the 
   outflow gas \citep{2000ApJ...545...63E}.
   The second is an evolution scenario, in which BALs are associated with youthful quasars enshrouded 
   heavily with gas and dust \citep[e.g.][]{1999MNRAS.310..913W}.
   Observationally, multi-wavelength comparisons find basically no intrinsic difference between BALs and non-BALs, 
   except that BAL quasars, particularly the LoBAL ones, typically have redder ultraviolet (UV) continua 
   than non-BAL quasars \citep{1991ApJ....373....23W, 2003AJ....126.2594R,2003ApJ...596L..35L, 2007ApJ...665..157G}.
   This fact is consistent with both the orientation and evolutionary scenarios.
   Recently, evidence has mounted that LoBAL quasars, and FeLoBAL ones in particular, are highly reddened objects 
   with high IR luminosities, associated with ultraluminous infrared galaxies (ULIRGs) in some way 
   \citep{2007ApJ...662L..59F,2009ApJ...700..395F,2012ApJ...757...51G}. 
   This leads to the interpretation that LoBALs may be just at a 
   transition phase of the evolutionary sequence from major mergers of galaxies to star-bursting ULIRGs, 
   dust-enshrouded BAL quasars, and finally to unobscured luminous quasars 
   \citep[e.g.][]{1988ApJ...325...74S,2002ApJ...564..133G,2012ApJ...748..131S}.
   Therefore, the study of BALs has important implications to understanding both the structure and 
   emission/absorption physics of AGNs and the co-evolution of AGNs and their host galaxies.

   Absorption lines provide abundant information about the outflows of quasars, such as velocity, 
   column density, ionization state and density, and furthermore distances from the central black holes and even 
 kinetic energy.
   Column densities could be determined by simply integrating over the apparent optical depth profile for unsaturated
   absorption lines \citep[e.g.][]{1991ApJ...379..245S}.
   Two or more lines from the same lower level are very helpful to jointly determine the column density and covering 
   factor of the outflow \citep[e.g.][]{1997ApJS..109..279H,2005ApJ...620..665A}.
   However, the commonly seen absorption lines, such as \civ\ and \mgii\, are easily saturated when the ion column 
   densities are high.
   Furthermore the wavelength separations of the \civ\ and \mgii\
   doublets are less than 10 \AA, so blending is a serious problem for BALs.
   By contrast, the \hei\ absorption lines offer a number of advantages in determining the physical conditions of 
   outflow gas.
   The metastable 2$s$ state in the Helium triplet, \hei, which is populated via recombination of He$^+$ ions, are  
   photo-ionized by photons with energies of $h\nu \ge 24.56$ eV. 
   \hei\ has multiple upward transitions in a wide wavelength span from UV to NIR, which are easy to
   observe.
   The strongest three transitions are 10830, 3889 and 3189 \AA\ from metastable state to 2p, 3p, 4p state 
   respectively. 
   Since these transitions are widely separated, there is no blending problem.
   Compared to C\,IV and \mgii\ ions, the \hei\ atoms have much smaller abundance, and 
   the optical depth, $\tau \varpropto \lambda~f_{ik}N_{ion}$, is much smaller than those of \civ\ and \mgii\ ions. 
   Therefore the \hei\ may remain optically thin even when \civ\ and \mgii\ lines are 
   saturated \citep{2011ApJ....728....94L}.
   The \hei\ absorption lines are also sensitive to the ionization state of the outflow gas, and thus they can,
   combined with other lines, set a tight constraint on the outflow 
   \citep[e.g.][]{2001ApJ....546..140A,2014arXiv1412.2881J}.
   Because the wavelength coverage of the \hei\ lines are between NUV and 
   NIR, it is convenient to search more low redshift BAL AGNs via the \hei\ absorption lines, 
   which can be potentially used to study the host galaxy properties of BALs quasars.

   However, the valuable \hei\ absorption lines have received little attention for a long time. Known \hei\
   BAL quasars are very rare to date, with only eleven being reports in the literature:
   Mrk~231 \citep[e.g.,][]{1985ApJ....288..531R,2014ApJ...788..123L},
   FBQS~J1151$+$3822 \citep{2011ApJ....728....94L,2014ApJ....783....58L},
   NGC~4151 \citep[e.g.,][]{1974ApJ....189..195A,2009MNRAS.394.1148S},
   NVSS~J2359$-$1241 \citep{2001ApJ...546..134B,2001ApJ....546..140A,2008ApJ...681..954A,2008ApJ...688..108K,2010ApJ...713...25B}, 
   AKARI J1757$+$5907 \citep{2011PASJ....63S.457A},
   LBQS~1206$+$1052 \citep{2012RAA.....12..369J},
   IRAS~14026$+$4341 \citep{2013AJ.....145..157J},
   SDSS~J080248.18$+$551328.9 \citep{2014arXiv1412.2881J},
   SDSS~J030000.56$+$004828.0 \citep{2002ApJS..141..267H,2003ApJ....593..189H},
     SDSS~J151249.29$+$111929.36 \citep{2012ApJ...758...69B,2013ApJ...762...49B},
    SDSS~J110645.05$+$193929.1 \citep{2013ApJ...762...49B}), which are summarized in Table \ref{tab:heiknown}.
   Having investigating these well studied objects, we noticed with surprise that all of these show
   the \mgii\ absorption troughs at the corresponding velocities of \hei. 
   This suggests that the \hei\ and \mgii\ absorption lines may have some physical connections.
   Given the aforementioned merits of \hei\ absorption, it is important to find more \hei\ BAL quasars,
   yet we even don't know whether \hei\ BAL quasars are intrinsically rare or not.
   In light of these considerations, as a first step, we carry out a systematic search campaign for the 
   \hei$\lambda3889$ absorption line among \mgii\ LoBAL quasars.

   This paper is arranged as follows.
   In \S\ref{sec:heisample}, a parent sample of \mgii\ LoBAL quasars is constructed and the pair-matching method 
   is introduced.
   We then adopt this method to compile a \hei\ $\lambda3889$ sample.
   In \S\ref{sec:fraction}, we present discussion on the fraction of the \hei\ absorption line in \mgii\ BALs,
   which is found to be strongly dependent on the S/N of the spectra.
   In \S\ref{sec:implication}, we calculate a series of photoionization models to investigate the physical
   conditions of the \hei\ absorption gas and give a physical explanation to the high fraction of the \hei\ in 
   \mgii\ BALs. We show one example of determining the physical condition of the outflow gas by jointly using 
   \hei$\lambda\lambda10830,3889,3189$ lines.
   We also compile a sample of 19 low-$z$ ($z < 0.3$) BAL candidates selected by detecting the \hei$\lambda3889$ 
   absorption line.
   The summary and future works are given in \S\ref{sec:summary}.
   The details of supportive NIR observations and other results are described in Appendix.
   Throughout this work we use a cosmology with $H_{0} = 70$~km~s$^{-1}$~Mpc$^{-1}$, $\Omega_M = 0.3$, and
  $\Omega_{\Lambda} = 0.7$.

\section{Sample of \hei$\lambda3889$ BAL Quasars \label{sec:heisample}}

\subsection{Existing samples of \mgii\ LoBAL quasars \label{subsec:parentsample}}

   We start with existing samples in the literature of \mgii\ BAL quasars at $0.4 \leq z \leq 1.35$.
   The redshift cutoffs are chosen such that both \hei$\lambda3889$ and \mgii\
   fall in the wavelength coverage of
   the SDSS spectrograph (3800--9200\AA), which enables us
   to use \mgii\ BALs as a reference in this first systematic search for \hei$\lambda3889$ BALs.

   There were more studies on the C\,IV BAL than on \mgii\ BAL in previous studies.
   Among the existing BAL samples, \citet[hereafter T06]{2006ApJS..165.....1T},
   \citet[hereafter G09]{2009ApJ....692..758G} and \citet[hereafter Z10]{2010ApJ....714..367Z}
   have conducted systematic searches for \mgii\ BALs in  the SDSS data set.
   T06 identified 4784 BAL quasars in SDSS DR3
   quasar catalog, and G09 identified 5039 from the SDSS DR5, in \civ\ and/or \mgii.
   Specifically aimed at \mgii\ BALs,
   Z10 compiled 68 \mgii\ BAL quasars in the SDSS DR5
   at $0.4 \leq z \leq 0.8$.
   Those samples were compiled by using different spectral fitting procedures and somehow different
   BAL definitions (see Appendix \ref{app:parentsample} for the detail).
   We find considerable discrepancies among the three samples:
   32\% objects of T06, 19.5\% of G09, and 11.4\% of Z10, respectively, are not included in the other two samples,
   reflecting to what a extent of the incompleteness of the samples could be caused by their different selection 
   procedures and criteria.
   We combined these samples into a merged \mgii\ BAL quasar sample, 
   from which we build our parent sample of this work. 
   Selection biases in the individual samples, which are partially complementary to one another, 
   are also reduced to some extent in the combined sample.

   The main uncertainty caused by the selection procedure lies in determining the unabsorbed spectrum
   (the intrinsic quasar continuum and/or emission lines). There are basically two schemes in the literature.
   One scheme is to use some kinds of quasar composite spectra to model the unabsorbed spectrum 
   \citep[e.g.,][T06]{2003AJ.....125.1711R} in light of the striking global similarity of ultraviolet/optical spectra
   of most quasars \citep[e.g.,][]{2001AJ.....121.2308R,2001AJ.....122..549V}.
   With the advantages of simplicity and fast speed, the disadvantage of this scheme is also obvious:
   on the wavelength scale as far as the BAL features
   are concerned, there are considerable object-to-object variations in the quasar continuum shape, broad emission
   line profile, iron emission multiplets (pseudo-continuum) and Balmer continuum shape, etc.,
   which cannot be incorporated in a single composite spectrum.
   The other scheme
   is to recover the unabsorbed continuum and emission lines via $\chisq-$minimization fitting with (analytic)
   models \citep[e.g.,][G09, Z10]{2002ApJ....578L..31T}.
   In this case, the unabsorbed UV/optical continuum is modeled with a single or a broken power law or polynomial,
   emission lines with Gaussian(s) or Lorentzian(s), and iron emission with tabulated or analytic templates
   \citep[e.g.][]{1992ApJS...80..109B,2011ApJ...736...86D}.
   The advantages are the flexibility of fitting individual spectral components, as well as its quick speed.
   Yet there are still several disadvantages.
   For example, some spectral components (such as Balmer continua) are hard to constrain by fitting 
   \citep[cf.][]{2009ApJ....707.1334W}; particularly, the Fe\,II multiplet emissions are considerably diverse 
   among quasars and are hard to be fitted well in many cases using the existing Fe\,II templates, 
   because those templates are all based on the famous NLS1, I~Zw~1
   \citep[see a recent discussion in][]{2011ApJ...736...86D}.
   To sum up, though the two traditional schemes have an advantage of being fast in computation,
   the best-fit model of the unabsorbed spectrum is not necessarily the real one,
   and---even worse---this {\it systematic error} is not accounted for.
   
   A better method is required, and actually some attempts have been made.
   \citet{2011ApJ....728....94L} used about a dozen quasar NIR spectra as templates to measure the 
   \hei$\lambda$10830\AA\ BAL of the quasar FBQS~J1151+3822.
   They deemed the best matched template spectrum as the intrinsic spectrum underlying the \hei\ BAL
   and the variation in the best-fits with those different templates as the {\em systematic uncertainty}.
   We have developed a pair-matching method to select \civ\ BALs \citep{2014ApJ...786...42Z}, which is similar 
   to those used in the studies of extinction curves in the literature \citep[e.g.][]{2012ApJ...760...42W}.
   The postulate of the pair-matching method is the similarity of continua and emission line profiles between BAL 
   and non-BAL quasars except  the (possible) dust reddening effect 
   \citep[see, e.g.,][]{1991ApJ....373....23W,2007ApJ....665..990G}.
   Therefore, we can always find one or more non-BAL quasars whose spectra resemble the 
   spectral features surrounding the BAL of a given BAL quasar,
   provided the library of the non-BAL quasar spectra is large enough. 
   In order to investigate the \hei\,$\lambda3889$ BALs, which are much weaker than the usually studied BALs in 
   \civ\ and \mgii, we improve the pair-matching method on two aspects. 
   First, we refine carefully the matching procedure to suit the case of \hei$\lambda3889$ absorption; 
   second, we implement its merit of being able to quantify the systematic error.
   
   Combining all together the \mgii\ BAL quasars detected by T06, G09 and Z10, regardless of their own selection 
   criteria, we obtain a large sample of 351 distinct \mgii\ BAL quasars.%
   \footnote{The published \mgii\ BALs in Z10 are only at $0.4 \leq z \leq 0.8$, so we use the same pipeline of Z10
     to enlarge this sample to 175 \mgii\ BALs at $0.4 \leq z \leq 1.35$ in the SDSS DR5.}
   In the following subsection, we will first apply the pair-matching method to the \mgii\ BAL quasars, and obtain
   measurements of the BAL properties uniformly by this method; meanwhile the method will be described in detail.
   Then, based on our own measurements, we will build the parent \mgii\ BAL sample which comprises 285 objects 
   (see below).

   \subsection{\mgii\ BAL measurements by the pair-matching method \label{subsec:mg2pairmethod}}

   First of all, we apply the pair-matching method to the  351 objects compiled above to measure the \mgii\ BALs.
   Our purpose is two-fold:
   (1) to test our pair-matching method (including the BAL selection procedure and criteria)
   with \mgii\ BALs, which is much stronger and easier to measure than \hei$\lambda3889$;
   (2) to measure \mgii\ BAL parameters uniformly and in the same way as we will do with the \hei$\lambda3889$
   BAL in \S~\ref{subsec:heisamplebuild}.
   The SDSS spectra%
   \footnote{This parent sample is compiled from the SDSS DR5, yet here we use the spectral data
   reduced by the improved SDSS pipeline, version {\it rerun 26}, as released since the SDSS DR7 (downloadable at
   http://das.sdss.org/spectro/ss\_tar\_26/\,), instead of the data reductions archived in the SDSS DR5
   (version {\it rerun 23}).}
   are corrected for the Galactic extinction using the extinction map of \citet{1998ApJ....500..525S}
   and the reddening curve of \citet{1999PASP..111...63F},
   and transformed into the rest frame using the redshifts provided by \citet{2010AJ.....139.2360S}.
   Our automated procedure consists of the following steps.

   \begin{description}
     \item[1. Build up the library  of unabsorbed quasar spectra as templates. ]
          The unabsorbed quasar spectra for the template library are all selected from the SDSS DR7 quasar catalog 
	  \citep{2010AJ.....139.2360S}. Only quasars at 0.4~$\le$~z~$\le$~2.2 are considered to ensure 
	  \mgii\ lies in the spectral coverage. We use only spectra with median 
	  S/N $> 25$ pixel$^{-1}$ in the 2400-3000 \AA\ region, and exclude any possible BAL quasars in previous 
	  BAL catalogs. Finally, a library of 1343 quasar spectra are selected as templates, which are visually 
	  examined to have no \mgii\ absorption features.

    \item[2. Loop over the templates for target BAL quasars.]

      For each of the target BAL quasars, we fit the continuum and emission lines with each template in a looping
      way. We picked up acceptable fits with reduced $\chisq < 1.5$.
      To fit the spectrum of a target quasar ($f_{\rm obj}(\lambda)$),
      the template spectrum ($f_{\rm t}(\lambda)$) is multiplied by a second-order polynomial,
      accounting for possible reddening and flux calibration problem, and finally yields a best-matched spectrum 
      $f_{\rm t'}(\lambda)$.
      This can be expressed analytically as follows,
      \begin{equation}
         f_{\rm obj} (\lambda)~=~ f_{\rm t'} (\lambda)~=~ f_{\rm t}(\lambda)\,\cdot\,(a + b\lambda + c\lambda^{2})~~,
      \end{equation}
      where $a$, $b$ and $c$ are free variables in the fitting.

      In the case of searching for \mgii\ BALs, only the spectral region in the range of 2400--3200~\AA\ is used.
      Bad pixels identified by the SDSS reduction pipeline and absorption troughs are masked out.
      Initially, the wavelength range of 2613--2809~\AA\ is masked out as a possible absorption-affected region;
      A refined absorption-masking region, as described below, will be obtained according to the best-fit 
      unabsorbed spectrum, and the fitting is then re-done in iteration.
      Generally for a object spectrum there are more than 20 acceptable fittings with reduced $\chisq \leq 1.5$.
      However, in some spectra with iron absorption and/or peculiar \feii\ emission surrounding \mgii, 
      the number of acceptable fittings is less than 20. 
      For such objects, we carefully mask out those features until we get about 20 acceptable fittings.

      Then we make a mean model spectrum ($f_{model}(\lambda)$\,) of all the acceptable fits 
      ($f_{\rm t'}(\lambda)$\,), and normalize the object spectrum with it:
      \begin{equation}
        I(\lambda) = \frac{f_{\rm obj}(\lambda)}{f_{\rm model}(\lambda)} ~~.
      \end{equation}
      In order to determine the region of possible absorption trough, we smooth $I(\lambda)$ to reduce noise and 
      unresolved absorption features.
      The smoothing is performed with a 5-point wide Savitsky-Golay filter of degree 2.
      The absorption trough region, as defined by the minimum velocity ($v_{\rm min}$) and maximum velocity 
      ($v_{\rm max}$),
      \footnote{Zero velocity is defined by using the improved redshifts for SDSS quasars provided by 
      \citet{2010MNRAS.405.2302H}, with 2798.75~\AA\ for \mgii\ and 3889.74~\AA\ for \hei\ (vacuum wavelength).}
      is determined as pixels with flux densities lower than unity by twice the root-mean-square (rms) fluctuation;
      i.e., $I(\lambda) < 1 - 2$\,rms, where the rms is calculated in the 2400--2600 and 2900--3000~\AA\ regions 
      of $I(\lambda)$.
      The absorption width, $W_{\rm abs}$, is defined as $|v_{\rm max} - v_{\rm min} |$. 
      Likewise, the absorption depth ($d_{\rm abs}$), defined as the maximum depth in the trough,%
      \footnote{Note that here for convenience we define $d_{\rm abs}$ to be the deepest point as measured
       from the smoothed spectra and use it throughout the paper.
       We would caution that this $d_{\rm abs}$ is easily affected by smoothing, noisy pixels and spectral 
       resolution. 
       In this work, as the noisy pixels are masked and smoothed out, and the absorption trough is broad 
       enough to be insensitive to the smoothing and the spectral resolution, this $d_{\rm abs}$ is robust;
       otherwise, the absorption-averaged depth of the trough is a better definition for the depth of trough.} 
       is calculated.

      As described in the above, the thus calculated absorption region serves as part of the masking regions 
      to feed the fitting routine for a better iterated fit.
      The convergence criterion is set to be that both $W_{\rm abs}$ and $d_{\rm abs}$ change less than 10\% 
      between iterations.
      Normally, two iterations are enough to get convergent results.

      We note that in previous studies on \civ\ and \mgii\ BALs the BAL region is usually set to be the continual 
      pixels deeper than 10\% of the normalized continuum \citep[e.g.][]{1991ApJ....373....23W}.
      However, it is {\em not proper for weak absorption troughs} based on our following tests 
      (e.g., in the case of \hei$\lambda3889$ generally).
      So we adopt the above criterion that is set by trial and error, and by which the contrast (significance) of the
      absorption is {\em with respect to the spectral quality instead of to the continuum strength}.
      Anyway, the previous treatment (10\% of the continuum strength) is just for the ease to handle, yet not 
      physically or statistically meaningful.

    \item[3. Identify and measure BALs.]
      For each of the targets, there are $\gtrsim 20$ best-fit models of the unabsorbed spectrum.
      In practice, we deem all those models equally possible 
      (in terms of their reduced $\chisq$ being $\leq~1.5$).%
      \footnote{According to our experiment, the fit of the unabsorbed spectrum
	predicted from the unabsorbed regions with minimum reduced \chisq\
	is not necessarily the real one underlying the absorption region.
	This is because there is essentially no ideal, well-defined relationship of quasar spectra
	between different wavelength regions to a very precise degree;
	this is even true for the local AGN continuum usually modeled with a power-law.
	According to our trial and error (cf. Appendix \ref{app:Tests}), 
	the best-fit models with reduced \chisq\ $\lesssim~1.5$ are basically indistinguishable.}

	For each accepted model $f_{\rm t',i}(\lambda)$, the absorber rest-frame EW of the possibly 
	existing BAL can be calculated as follows,
      \begin{equation}
        EW_{i} = \int^{\lambda_{u}}_{\lambda_{l}}[1~-~I_{i}(\lambda)]d\lambda~~,
      \end{equation}
      where $\lambda_{\rm u}$ and $\lambda_{\rm l}$ are the wavelength corresponding to the 
      $v_{\rm max}$ and $v_{\rm min}$ as described in the above in rest wavelength frame, and 
      $I_{\rm i}(\lambda)$~=~$f_{\rm obj}(\lambda)$/$f_{t',i}(\lambda)$.

      Then the mean of all the EW$_{\rm i}$ values, $\overline{\rm EW}$,
      is a good estimate of the true value (EW), and the standard
      deviation indicates the systematic error of this pair-matching method ($\sigma_{\rm sys}$),
      as discussed in \S~\ref{subsec:parentsample} (see also \citet{2011ApJ....728....94L}).
      On the other hand, we can estimate the purely statistical error
      ($\sigma_{\rm n,i}$; i.e., caused by random noises) for every EW$_{\rm i}$.
      We follow the same way of T06 to calculate $\sigma_{\rm n, i}$,
      accounting for both the error in the fit of the
      unabsorbed spectrum and the measurement error in every pixel comprising the absorption trough.
      Then we take the mean of the $\sigma_{\rm n,i}$ values as the final random error ($\sigma_{\rm n}$).
      The total measurement error of the BAL EW is thus estimated as
      \begin{equation}
        \sigma_{\rm total} = \sqrt{\sigma_{\rm sys}^2 + \sigma_{\rm n}^2}~~,
      \end{equation}
      Based on the above calculated quantities, we set the criteria of {\it bona fide} BALs as follows:
      \begin{enumerate}[(i)]
        \item EW $\geq 2\, \sigma_{\rm total}   $ ~~~~~~~~~~~~~~~~~~~~~~~~~~~~ {\rm (the intensity criterion) },
        \item $|v_{\rm max}~-~v_{\rm min}| \geq 1600 \kms\ $  ~~~ {\rm (the width criterion) } .
        \end{enumerate}
   \end{description}

   The whole fitting and identification procedures are summarized in Figure~\ref{fig:procedures}.
   Six \mgii\ BALs are chosen as examples to show the fitted continua in \mgii\ and \hei\ region in
   Figure~\ref{fig:fitshow}.
   The acceptable fits for each source are shown by green dotted lines, and the composite spectrum built 
   from these fits is shown by red solid line. 
   The standard deviation of the acceptable fits is thus the systematic error of this method.
   Among the 351 sources in the combined sample, 53 sources are rejected according to our
   criteria,  285 \mgii\ BAL quasars are culled as our parent sample,
   and the left 13 sources cannot be categorized because their intrinsic spectra are unusual and 
   difficult to be fitted well by the template quasars.
   Those 13 objects, some of which have been studied by \citet{2002ApJS..141..267H}, 
    are listed in Table~\ref{tab:unusual} (see Appendix~\ref{app:unusual})
   and not included in our parent \mgii\ BAL sample. %

   Figure~\ref{fig:mg2abscompare} is the direct comparison of the measurements of EWs (AIs), $v_{\rm max}$,
   $W_{\rm abs}$ and $d_{\rm abs}$ among T06, G09, Z10 and our sample.
   Basically our parameter measurements are consistent with those of T06, G09 and Z10, especially for the BAL quasars
   that detected by all the three samples (red dots).
   Apparent discrepancies are mostly shown in maximum velocities and widths of FeLoBALs (blue dots), for which the
   boundary of the \mgii\ absorption lines are easy to be contaminated by \feii\ absorption lines.
   Four FeLoBAL quasars are shown in Figure~\ref{fig:fitshow} (J074554.74+18187.0; J080248.18+551328.8;
   J084044.41+363327.8; J104459.60+365605.1), which are also marked in Figure~\ref{fig:mg2abscompare} by green
   pentagrams.

   Before applying the pair-matching method to the detection of \hei$\lambda3889$, we perform a series of
   tests to evaluate the pair-matching method (see Appendix \ref{app:Tests}) both for \mgii\ and \hei.
   The tests show that spectral S/N and absorption depth ($d_{\rm abs}$) are the primary factors
   that affect absorption detection.
   Figure~\ref{fig:sndepth} shows the contour plots of the relative error of the absorption EW and 
   recovered detected fraction in the S/N and $d_{\rm abs}$ space.
   From this figure, with increasing S/N and $d_{\rm abs}$, the relative error decreases and the detection 
   fraction increases.
   This figure shows that the relative error decreases and the detection fraction increases with increasing 
   S/N and $d_{\rm abs}$.
   An important implication is that we can estimate the accuracy (relative error of the absorption EW) and 
   efficiency (detection fraction of the BAL) of the pair-matching method as long as we know the spectral S/N 
   and $d_{\rm abs}$ of a given absorption line.
   The detection by the pair-matching method is almost complete for absorption troughs with high S/N ($\geq$35)
   or large absorption depth ($d_{\rm abs} \geq 0.5$).
   For absorption troughs with either a low S/N ($\sim$10) or a small depth ($\sim$0.15), which is common for 
   \hei\,$\lambda3889$, the detection rate can be still as high as $\sim$60\%.

\subsection{The \hei$\lambda3889$ BAL sample \label{subsec:heisamplebuild}}

  Now we use the pair-matching method to search for \hei$\lambda3889$ absorption troughs in the parent 
  sample of 285 \mgii\ BAL quasars.
  The procedure to identify \hei$\lambda3889$ absorption is essentially the same
  as described in Section \ref{subsec:mg2pairmethod} for \mgii\ BALs.
  As \hei$\lambda3889$ absorption is much weaker than \mgii, we make a few minor adjustments
  suitable to detect weak absorption troughs (e.g., the templates of unabsorbed quasars,
  the masking regions, and a loosening of the width criterion).
  The details are as follows.

\begin{figure}[tbp]
    \centering
    \includegraphics[width=6.1in]{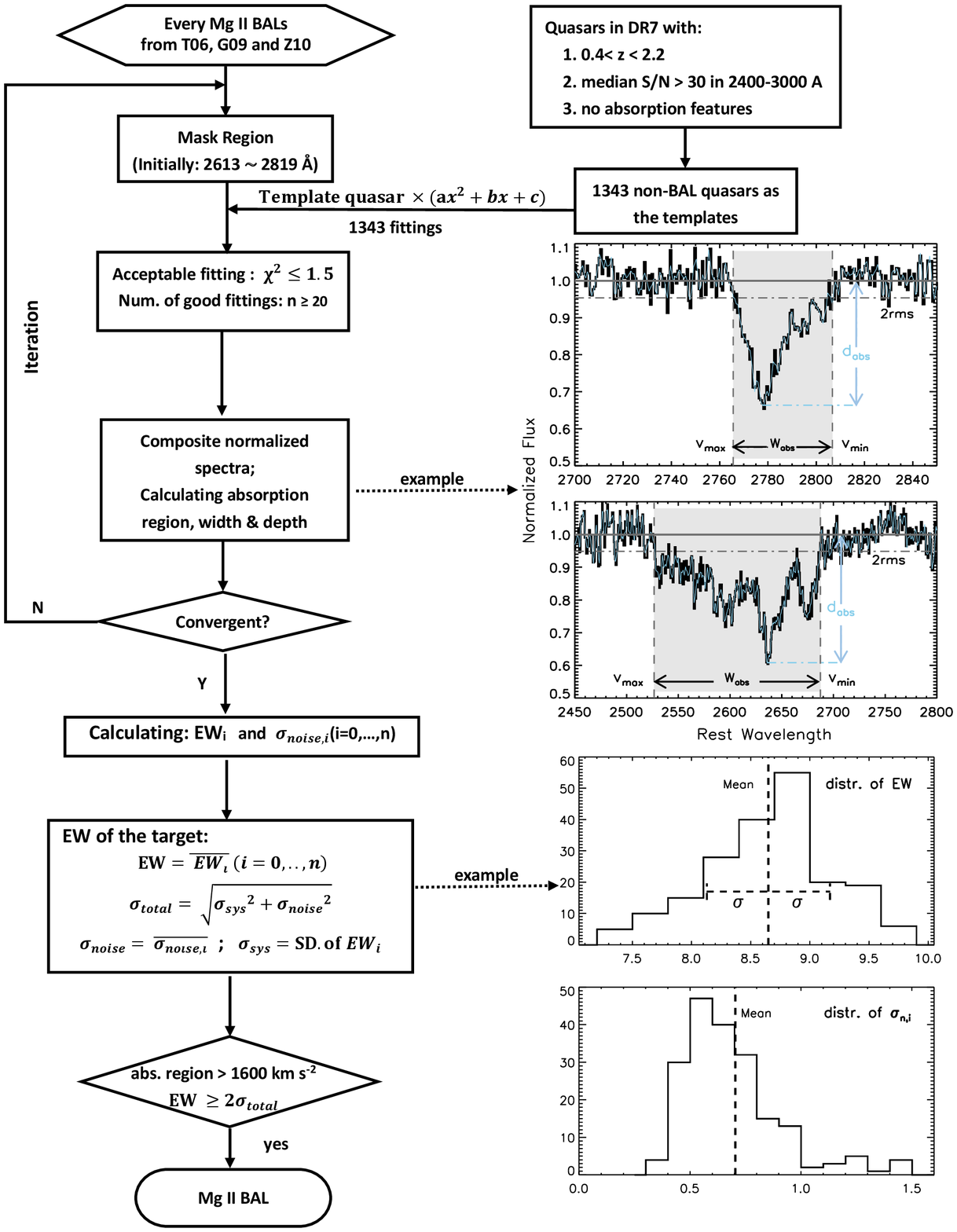}
    \caption{\footnotesize Flowchart of the procedure to select broad absorption lines using the pair-matching method
      (see \S \ref{subsec:mg2pairmethod} for details). The Blue solid line is the absorption line smoothed 
      with a 5-point wide Savitsky-Golay filter of degree 2, and absorption depth ($d_{\rm abs}$) are determined 
      as the maximum depth of this.\label{fig:procedures}}
  \end{figure}

\begin{figure}[tbp]
    \centering
    \includegraphics[width=6.1in]{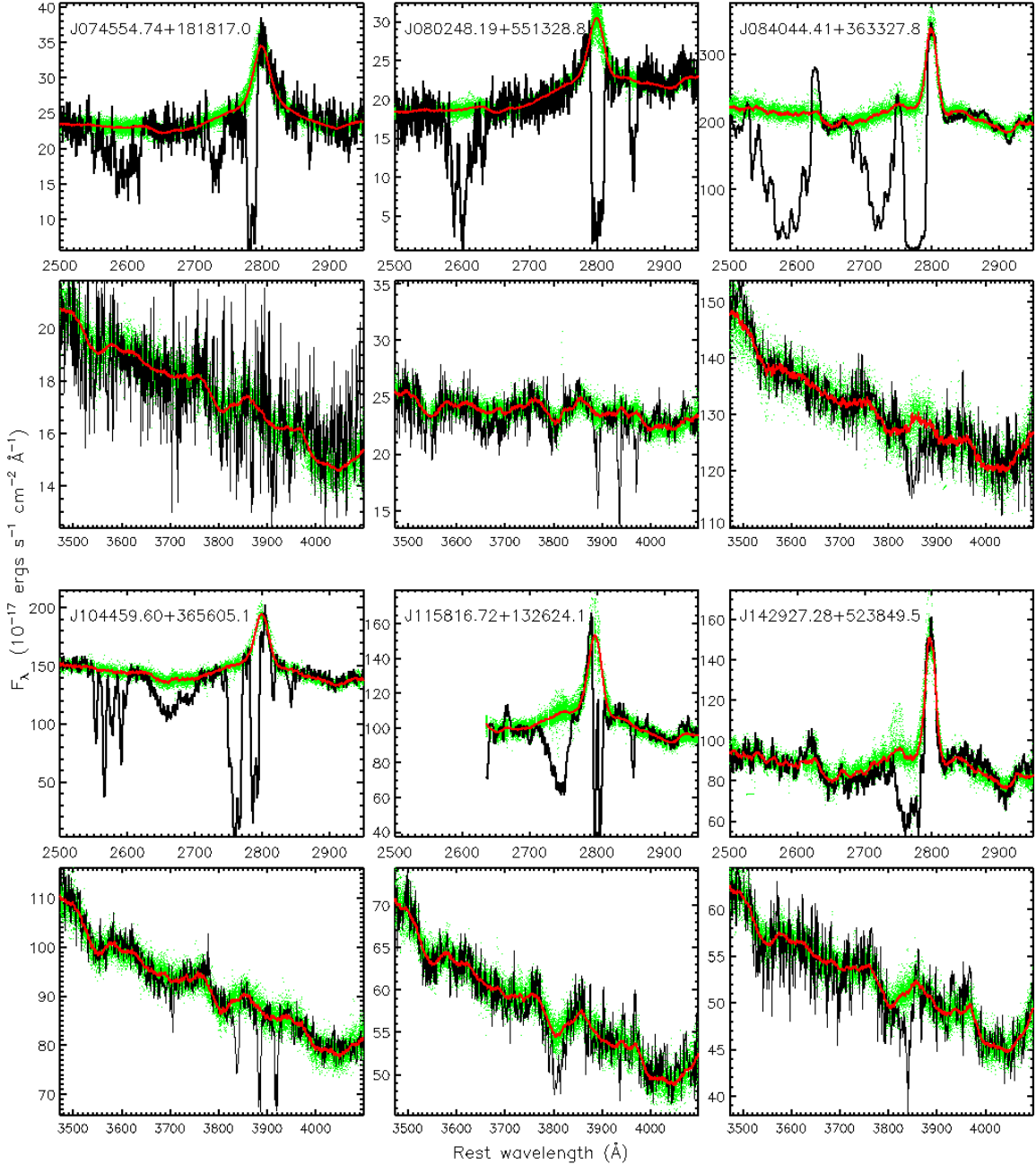}
        \caption{\footnotesize Example SDSS spectra of six He\,I$\lambda3889$ BAL quasars. 
        For every objects, the fittings in the Mg\,II region (upper panel) and in the 
        He\,I*$\lambda3889$ region (bottom panel) are shown.
        The acceptable fits of the unabsorbed spectrum in every panels are denoted as green dotted lines,
        the mean of these acceptable fits are denoted as the red solid line.
	  Note that J0745$+$1818, J0802$+$5513, J0840$+$3633 
	  and J1044$+$3656 are FeLoBAL quasars.\label{fig:fitshow}}
  \end{figure}

\begin{figure}[htbp]
        \centering
    \includegraphics[width=6.5in]{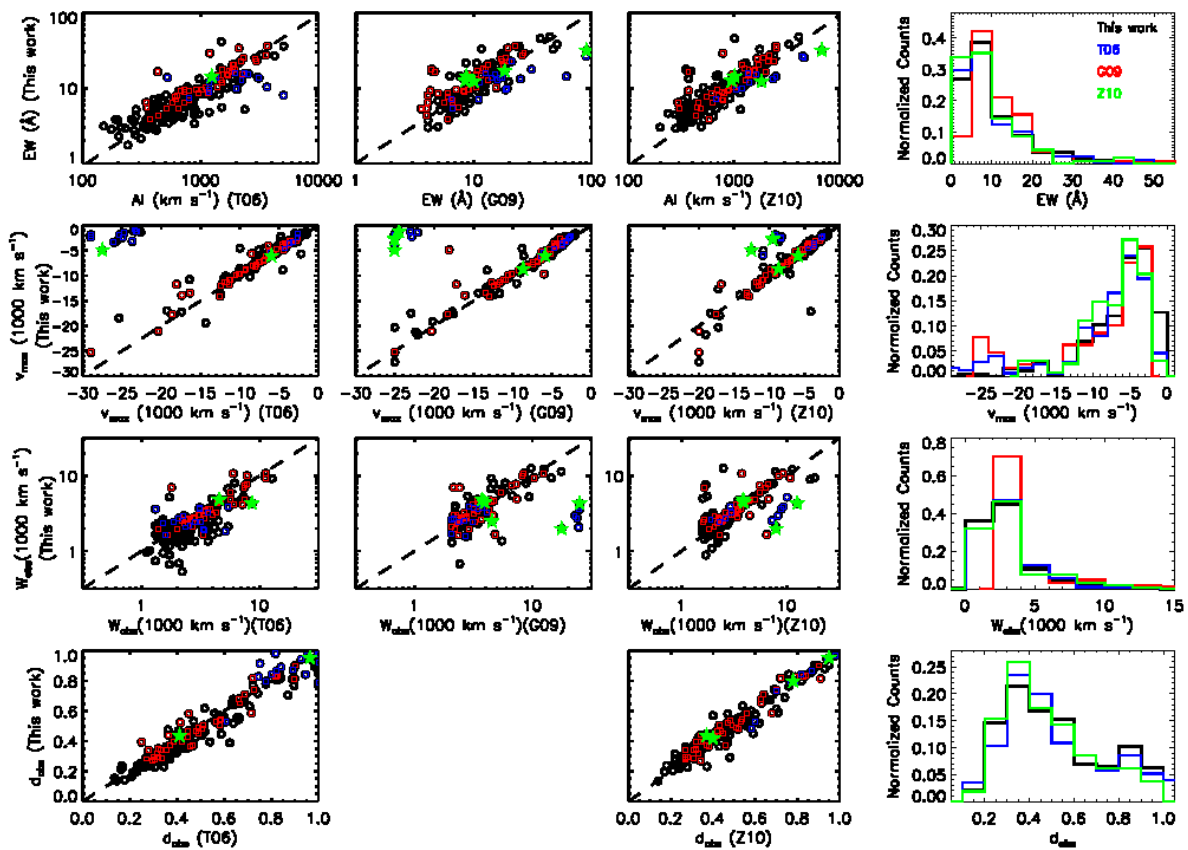}
    \caption{\footnotesize {\it Left}: Comparison of Mg\,II BAL parameters between our measurements and 
    those by T06, G09, and Z10.
     Red dots are sources common in the three samples; green pentagrams are the six sources shown in Figure~\ref{fig:fitshow};
     blue circles are FeLoBAL quasars of which the \mgii\ absorption region is 
	  difficult to determine and thus the measurements are the most divergent in the plots.
    {\it Right}: Histograms of the BAL parameters measured by these different groups. 	  
	  \label{fig:mg2abscompare}}
\end{figure}

  A template library of 316 unabsorbed quasar spectra at 0.4~$\leq~z~\leq$~1.35
  is compiled from the SDSS DR7 quasar catalog, which are selected to have
  median S/N >25 pixel$^{-1}$ in 3500-4000 \AA\ region and have no any absorption features.%
  \footnote{Some template spectra have strong [O\,II]$\lambda3727$ and [Ne\,III]$\lambda3869$ emission lines,
  we apply models of one polynomial plus one or two Gaussian profiles to fit the spectra locally,
  and subtract these two emission lines from the template spectra.}
  Each of the Mg\,II BAL quasars spectra is fitted with the 316 templates.
  Only the line-free continuum region in the spectral range of 3500--4000~\AA\ is used.
  Bad pixels are masked out according to the tags by the SDSS pipeline.
  We carefully mask the regions contaminated by
  the [O\,II]$\lambda3727$ and [Ne\,III]$\lambda3869$ emission lines
  to make sure that the placement of AGN continuum around \hei$\lambda3889$
  will not be affected by those nearby narrow emission lines.
  Initially the possible \hei\ absorption region is masked out
  according to the measured \mgii\ absorption region,
  and later on in the iterated fittings it is masked out according to
  the calculated $v_{\rm min}$ and $v_{\rm max}$.
  Other absorption lines such as Ca\,II$\lambda\lambda3949,3969$ and high order Balmer absorption lines,
  if present, are also masked.
  Following the procedure as described in Section \ref{subsec:mg2pairmethod} (see also Figure~\ref{fig:procedures}),
  fittings with reduced $\chi^2$ smaller than 1.5 for each candidate are picked up as acceptable,
  and the $v_{\rm min}$, $v_{\rm max}$, $d_{\rm abs}$, EW, and $\sigma_{\rm total}$
  are calculated for the (possible) \hei$\lambda3889$ absorption.
  Then the intensity criterion is applied to select {\it bona fide} \hei\ BALs.
  The width criterion is not applied to \hei$\lambda3889$, because under the typical ionization parameter ($U$)
  and hydrogen column density ($N_{\rm H}$) conditions
  the optical depth of \hei$\lambda3889$ is much smaller than
  that of \mgii\ \citep[cf.,][their Figure~15]{2011ApJ....728....94L}.
  Although the observed widths of \hei$\lambda3889$ absorption troughs are narrower than the conventional 
  definition of BAL, we still refer to them as \hei\ BALs in this paper, given most likely that they are
  physically associated with the BAL outflows as shown in this work.

  As a sanity check, we visually inspect the SDSS spectra of the selected \hei\ BALs.
  The presence of \hei$\lambda3189$ and/or even higher order \hei\ absorption lines in many objects
  is a strong support for the genuineness  of our measured \hei$\lambda3889$ absorption trough.
  And a few spurious objects are excluded as they are mimicked by
  absorption features from H9, H10 or H11 Balmer absorption lines.
  Finally we obtain 101 \hei$\lambda3889$ BAL quasars, which are listed in Table~\ref{tab:dr5he1table}.

  We further apply the pair-matching method to detect \hei$\lambda3189$ absorption line,
  and 52 objects out of the \hei$\lambda3889$ BAL sample pass the intensity criterion.
  The measurements of the 52 \hei$\lambda3189$ BAL subsample are summarized in Table~\ref{tab:dr5he13189table}.

  On the other hand, the optical depth ($\tau \varpropto \lambda f_{\rm ik}N_{\rm ion}$) of 
  \hei$\lambda10830$ is 23.5 times that of \hei$\lambda3889$ (according to Table \ref{tab:lineinfo}), so 
  absorption trough of \hei$\lambda10830$ should be much stronger than that of \hei$\lambda3889$.
  Thus, if we detect BALs in \hei$\lambda10830$, the detection probability should be higher.
  As a check, we have made NIR spectroscopic observations for
  five sources in this present \hei$\lambda3889$ BAL sample.
  It turns out that all of them show evident \hei$\lambda10830$ BALs.
  The details of the NIR observations and data are presented in \ref{app:nirspectra}.

  Based on our measurements of the \mgii\ and \hei$\lambda3889$ absorption troughs using the same technique,
  we present a direct comparison of the two in Figure~\ref{fig:absmg2he1measure}.
  The parameters such as the EW, $d_{\rm abs}$, $v_{\rm max}$ and $W_{\rm abs}$ of \hei$\lambda3889$ 
  are smaller than those of \mgii\ consistently.
  Moreover, the absorption-weighted average velocity ($v_{avg}$, namely the velocity centroid)
  for \mgii\ and \hei$\lambda3889$ agrees fairly well, with a very small scatter.
  These facts indicate a dynamical connection between the gas producing the two absorption lines.

\section{Fraction of \hei$\lambda3889$ Absorption in \mgii\ BAL Quasars \label{sec:fraction}}

\subsection{The measured fraction of \hei\ $\lambda3889$ BALs}

  There are 101 (35.4\%) \hei$\lambda 3889$ BAL quasars in the parent 285 \mgii\ BAL quasar sample.
  Such a high detection rate of \hei\ BALs in \mgii\ BAL quasars is unexpected, particularly
  considering the scarceness of \hei\ BALs reported so far.

  According to the aforementioned tests (see Appendix \ref{app:Tests}) using simulated spectra,
  the detection probability of absorption lines depends mainly on two factors: the spectral S/N and 
  $d_{\rm abs}$ (Figure \ref{fig:testresults} and Figure \ref{fig:sndepth}).
  Now we carry out a quick investigation of the effect of the spectral S/N
  to the measured fraction of \hei\ BALs in the \mgii\ BAL sample, $f({\rm He\,I*|Mg\,II})$.
  We divide the \mgii\ BAL sample into four S/N bins, with roughly equal number of objects in every bins
  (S/N~<~10, 10~<~S/N~<~20, 20~<~S/N~<~30, and 30~<~S/N~<~50), and then
  calculate the $f({\rm He\,I*|Mg\,II})$ of every bins.
  Here the S/N refers to the median spectral S/N per pixel around \hei\ line (3700--4000~\AA),
  and we take the mean S/N of every bins as the fiducial one.
  The measurement uncertainty of the faction values is estimated using a bootstrap technique.
  For every an aforementioned bin, we extract randomly a considerable number of objects from it
  and calculate the \hei\ BAL fraction in these extracted objects, $f_{\rm i}$.
  This process is repeated for 100 times, and the standard deviation of
  the $f_{\rm i}$ values is regarded as the 1-$\sigma$ error to the \hei\ BAL fraction of this bin.
  The observed $f({\rm He\,I*|Mg\,II})$, as a function of spectral S/N, is shown in
  Figure~\ref{fig:fraction}.
  We find that $f({\rm He\,I*|Mg\,II})$ increases monotonically from 18.1\% at S/N $\sim$ 6.5 to 92.9\% at
  S/N $\sim$ 35.5.

  As a comparison, the fraction $f({\rm He\,I*|Mg\,II})$ with respect to the spectral S/N has been predicted in
  our tests with simulated spectra (see the Appendix \ref{app:Tests} and Figure~\ref{fig:modelfrac}, left panel).
  The steeply rising trend of the observed one (Figure~\ref{fig:fraction}) is well consistent with that of
  the tests using both assumed Gaussian absorption profiles and actual absorption profiles.
  It is worth noting that the fraction in low S/N bins predicted by the tests is much higher than
  the what is observed.
  This can be understood in the following way.
  Right panel of Figure~\ref{fig:modelfrac} shows the detection probability of the pair-matching method
  for absorption troughs with different $d_{\rm abs}$ under different S/N bins, which is marked by different colors.
  For absorption troughs with $d_{\rm abs}$ larger than 0.5, the detections are almost complete in all the S/N bins
  using the pair-matching method. While according to measurements for real sample, the distribution of
  \hei$\lambda3889$ absorption troughs peaks at $\sim$~0.2, and few of them reach $d_{\rm abs}~\sim~0.5$
  (Figure~\ref{fig:absmg2he1measure}).
  This is also proved by the photoionization models in \S \ref{sec:implication}.
  Under normal ionization condition (log$U$~$\sim$~-1.5), the $d_{\rm abs}$ of \hei$\lambda3889$ trough
  is much weaker than that of \mgii.
  We infer that essential number of \hei$\lambda3889$ absorption line with large $d_{\rm abs}$ should be 
  smaller than the one with small $d_{\rm abs}$.
  Compared with the uniform distribution of $d_{\rm abs}$ in the simulations, it is expected that the predicted 
  fraction is much higher than the real sample in low S/N bins. 
  It is also remarkable that in the bins of S/N $\ge~30$, the fractions of both real sample and simulation are 
  higher than 90\%.
  As shown in the right panel of Figure~\ref{fig:modelfrac}, the recovered percentage is very high even for
  $d_{\rm abs}$~$\sim~0.1$ in the bin of S/N~$\ge~25$.
  Therefore the distribution of $d_{\rm abs}$ has little effect on the fraction of \hei$\lambda3889$ in 
  high S/N bins.
  That is to say, in high S/N bins, the measured fraction of \hei$\lambda3889$ absorption line
  is very close to its intrinsic value.
  Our results imply that \hei$\lambda3889$ absorption in fact has a fairly high incidence in \mgii\ LoBAL quasars.

\subsection{Unveiling the \hei$\lambda3889$ absorption in low-S/N spectra}

  As shown above, the detection of the \hei$\lambda3889$ absorption troughs in the SDSS spectra of
  the \mgii\ BAL quasars is seriously hindered by the low spectral S/N (e.g., $\lesssim 20$).
  In order to test our hypothesis that most, if not all, \mgii\ BAL quasars have \hei\ BAL features,
  in this subsection we examine possible \hei$\lambda3889$ absorption troughs
  that are below our detection threshold in the low-S/N SDSS spectra.
  For this purpose we take two approaches: one is by combining the multiple observations in the SDSS legacy survey
  and/or the SDSS-III/BOSS; the second is by stacking spectra of low S/N.

  In the parent \mgii\ BAL sample, 61 sources have repeated spectroscopic observations
  in the SDSS Legacy programme and/or in the SDSS DR10 \citep[SDSS-III/BOSS;][]{2014ApJS..211...17A}
  which are summarized in Table~\ref{tab:multiobserve}.
  Some interesting sources with repeated observations are commented in Appendix \ref{app:multiobs},
  including several variable sources (e.g., J14264704.7$+$401250.8).
  Combining the multiple spectra of one source could enhance the spectral S/N.   
  Moreover, the BOSS spectra have longer exposure time than the SDSS ones, and therefore higher S/N generally; 
  their S/N is typically $\sqrt 2$ times that of the SDSS DR7 spectra.
  According to the analysis of their SDSS DR7 spectra (see Section \ref{sec:heisample}),
  34 of the 61 sources do not have detected \hei$\lambda3889$ absorption troughs.
  Twenty-one of the 34 objects have combined spectra with S/N $> 15$, which is sufficient to detect 
  \hei$\lambda3889$ absorption line according to our simulation.
  We apply the same procedure of the pair-matched method to the 21 combined spectra.
  It turns out that 4 more sources pass the \hei$\lambda3889$ BAL criteria,
  which are shown in Figure~\ref{fig:dr10example}.
  Among the rest 17 sources that we still do not detect \hei\ absorption,
  5 have strong [Ne\,III]$\lambda3869$ emission line, which hinders the detection of \hei$\lambda3869$ absorption;
  4 sources have a flat and shallow \mgii\ absorption trough, so the potential \hei$\lambda3889$ would be too 
  shallow to be detected (cf. Figure~\ref{fig:absmg2he1measure});
  For the rest 8 sources, we failed to detect \hei$\lambda3889$ absorption line even in the combined spectra.

  Secondly, we stack respectively the sources with and without detection of \hei$\lambda3889$ absorption
  in every S/N bins. As there are only 2 non-detection sources in the highest S/N bin (30~<~S/N~<~40),
  we coalesce that bin and the 20~<~S/N~<~30 one.
  The two kinds of stacked spectra in every bins are shown in Figure~\ref{fig:superposition}.
  We can see that now the absorption troughs emerges apparently in the stacked spectrum of non-detections in each 
  bin.
  The $d_{\rm abs}$ of those emerging \hei$\lambda3889$ troughs is around 0.05, which is too weak to detect 
  in spectra of moderate S/N level (S/N $\lesssim 15$; Figure~\ref{fig:sndepth}).
  Note that the $d_{\rm abs}$ decreases in the bins with increasing S/N;
  this is because the higher S/N, the higher completeness of detection in individual sources,
  and thus the weaker absorption troughs in the non-detections.
  These results, therefore, strongly support our conclusion that the
  \hei\ absorption line has a high incidence in \mgii\ LoBAL quasars.

  \subsection{Spectral S/N or luminosity?}

  We would like to mention in passing that a similar trend that the BAL fraction increases with increasing S/N 
  and luminosity has also been found for \civ\ and \mgii\ BAL quasars in the literature (e.g., G09; Z10).
  \citet{2007ApJ....665..990G} also suggested that the most luminous quasars are more likely to show BALs. 
  As to our sample, there is also a dependence of the \hei\ BAL fraction on luminosity, which is shown in the 
  upper-left panel of Figure~\ref{fig:snorlumi}. The \hei\ BAL fraction is growing up slowly between
  44.5 < Log$L_{\lambda}$(3000 \AA) < 46 erg s$^{-1}$, and increases sharply in the highest $L_{\lambda}$(3000 \AA) 
  bin.
  As a check, the upper-right panel of Figure~\ref{fig:snorlumi} shows the relation between spectral S/N 
  and luminosity.
  A clear edge appears in the spectral S/N-luminosity plane, which indicates that each spectral S/N corresponds 
  to a minimum luminosity. 
  While for the sources with highest luminosities (Log$L_{\lambda}$(3000 \AA) > 46 erg s$^{-1}$),
  their spectral S/N are all larger than 25. 
  Because the spectral S/N and luminosities are somewhat degenerate, it is important to know which is the major 
  factor. 
  In order to test this, we select two subsamples from the parent \mgii\ BAL sample. 
  The two subsamples are selected in a narrow bin of luminosity/spectral S/N, so as to check the dependence of 
  fraction on the other parameter.
  The first subsample is selected as sources in luminosity bin of 45.4 < Log$L_{\lambda}$(3000 \AA) < 46 erg s$^{-1}$
  (see the dotted lines), in which the spread of spectral S/N is equivalent to the parent sample. 
  The sample size is 184, which is large enough to do the statistical analysis. 
  For this subsample, luminosity is taken as the controlled factor, and we can see the \hei\ BAL fraction has no 
  significant dependence on luminosity. But the \hei\ BAL fraction still has a strong 
  dependence on spectral S/N, which is similar to that of parent sample (see Figure~\ref{fig:fraction}).
  The second subsample is selected as sources in median spectral S/N bin of 5 < median S/N < 11 
  (see the dashed lines), in which the spread of luminosity covers from 10$^{44}$ to 10$^{46}$ erg s$^{-1}$. 
  As the lower-right panel of Figure~\ref{fig:snorlumi} shows, though the \hei\ BAL fraction is growing slowly with 
  increasing of spectral S/N in this subsample, we can approximatively think spectra S/N has a limited influence on 
  the fraction for this subsample.
  We do not find any significant dependence of \hei\ BAL fraction on luminosity in this subsample.
  This test strongly suggests that the spectral S/N is the major factor in determining the \hei\ BAL fraction.
  \footnote{It is worth noting that the problem of the \hei\ BAL fraction in \mgii\ BAL is very different with that 
    of C\,IV or \mgii\ BAL fraction in all quasars. There is a very strong physical connection between \mgii\ and 
    \hei\ absorption lines, which is strongly indicated by both the observation and the photoionization models (see 
    \S \ref{sec:implication}). In other words, the probability f(HeI*|MgII) is so high that finding \hei\ 
    absorption line in \mgii\ BAL may be only a problem of detection. 
    The C\,IV or \mgii\ BAL fraction in all quasars is a much more complex problem. Whether a quasar is classified 
    a BAL quasar or not depends on many factors, which involves the physical condition for a quasar to have
    high-speed outflows to produce absorption lines, the chance of the outflow gas in our line of sight
    and if the absorption lines are strong enough for us to detect. }
  
  In order to further test our hypothesis, we perform direct comparisons
  of the absorption-line, emission-line and continuum properties between our \hei\ BAL sample and
  those without detection of \hei\ absorption in the parent \mgii\ BAL sample (hereinafter, dubbed as
  `non-\hei\ sample').
  The fittings of optical and UV continuum and emission lines are performed with
  the routines as described by \citet{2008MNRAS.383..581D} and \citet{2009ApJ....707.1334W};
  the results are presented in Appendix~\ref{app:emissionline}.
  As shown in Figure~\ref{fig:distrofmg2abs},
  the \mgii\ BAL properties (EW and $d_{\rm abs}$, $v_{\rm max}$, and $W_{\rm abs}$)
  between the two samples display no significant difference,
  which is confirmed by the Kolmogorov-Smirnov (K-S) test with all the chance probabilities
  $P_{\rm null}> 0.01$ or even close to 1.
  The chance probabilities are also denoted in each panel of this figure.
  We also compare the continuum and emission-line properties between the two samples.
  Again, there are no any differences in the UV and optical continuum luminosities,
  widths of broad \mgii\ and \hb\ emission lines, optical and UV \feii\ strength, \oiii$\lambda5007$ strength,
  and continuum slope, which are apparent in Figure~\ref{fig:distruvopt} and confirmed by the K-S test.

  As a reference, we also show the distributions for non-BAL quasars in Figure~\ref{fig:distruvopt}.
  We select the non-BAL quasar sample from SDSS DR7 to match the \mgii\ BAL sample in redshift and spectra S/N.
  Therefore the distribution of luminosities of non-BALs agrees well with that of BAL sample.
  Besides luminosity, non-BAL quasar have different continuum and emission-line properties with the BAL sample,
  which is confirmed by the K-S test.
  The most significant difference is the distributions of $\beta_{[3k,4k]}$, that LoBAL quasars have
  redder UV continue than non-BAL quasars, which have been proved by many previous studies.
  Our results are in accordance with theirs.

\section{Interpretation of the Observed \hei\ Absorption \label{sec:implication}}
  \subsection{Theoretical modeling  \label{subsec:models}}

    We generate a series of over-simplified photo-ionization models using 
    CLOUDY \citep[c13.03; c.f.][]{1998PASP..110..761F}
    to investigate the physical relationships between \hei$\lambda3889$ and \mgii\ absorption lines.
    We start by considering a gas slab, illuminated by a quasar, with a density of $n_{\rm e}$ and total column
    density of $N_{\rm H}$.
    Sources in our sample have various absorption intensities for \hei$\lambda3889$ and \mgii\, and some of
    them also show Fe\,II, H Balmer, Ca\,II and even Na\,I absorption lines, and these imply the detailed physical 
    condition of outflow gases are very different.
    According to previous studies on the outflows of known \hei\ BALs (see Table~\ref{tab:heiknown}),
    the $n_{\rm e}$ spans the range of 10$^{3.75}$~$\sim$~10$^8$ cm$^{-3}$ and the log$U$ is varying from
    -2.42~$\sim$~-0.5.
    To fully covered the parameter space of our sample, we calculated a grid of models with log$n_{\rm e}$
    varying from 3 to 9 with step of 1 and log$U$ varying from -2.5 to -0.1.
    We adopt optically thick models to generate a fully developed ionization front, and the stop column densities 
    are setted as as $N_{\rm H}$ = 10$^{24}$ cm$^{-2}$.
    All these models are assuming solar abundance, which can satisfactorily reproduce the observed $N_{\rm ion}$ 
    in previous studies on individual \hei\ absorbers \citep[e.g.][]{2001ApJ....546..140A,2014arXiv1412.2881J}.
      The SED incident on the outflowing gas has important consequences for the ionization and thermal 
      structures within the outflow. 
      The commonly used AGN SED is the one constructed by \citet{1987ApJ...323..456M} (hereafter MF87), 
      which is given as Table AGN in CLOUDY package. 
      Subsequent UV and X-ray observations thenceforth have indicated that the FUV slopes of radio quiet 
      quasars are generally softer than MF87 (see detailed discussion in section 4.2 of 
      \citealt{2010ApJ...709..611D}.
      Here we calculated a grid of models of ionized clouds using a realistic, UV-soft SED, which 
      is a superposition of a blackbody ``big bump'' and power laws.
      This UV-soft SED is set to be the default parameter given in the {\tt Hazy} document of CLOUDY as 
      follow: $T=150,000$~K, $\alpha_{\rm ox} = -1.4$, $\alpha_{\rm uv} = -0.5$, $\alpha_{x} = -1$, and the 
      UV bump of which peaks at around 1~Ryd, softer than the MF87 one.
      For comparison, we also calculated CLOUDY models using MF87 SED.
      In the case of the MF87 SED,  HeI~2$^{3}$S grows steeper before the ionization front of hydrogen, 
      and Mg$^{+}$ also grows steeper around ionization front than those in the case of the UV-soft SED.
      However, the differences between the two sets of models is small when we consider only the properties of 
      \hei\ absorption, especially in optically thick clouds as considered in this work.
      In this paper, we use only the results of CLOUDY modelling with the UV-soft SED.
      We would like to mention in passing that ionizing SED would be more important in optically thin clouds or 
      for jointly considering other absorption lines \citep{2001ApJ....546..140A}, and this need 
      a series of work on case studies.

    Figure~\ref{fig:modeloverview} shows the overview of the models.
    The \mgii\ and \hei\ show different behaviours nearby the hydrogen ionization front.
    $N_{\rm Mg\,II}$ increases sharply around the ionization front of hydrogen, while
    $N_{\rm He\,I*}$ grows in the front of the ionization front of hydrogen and stops growing behind of that.
    This is consistent with the result of \citet[see their Figure 8]{2001ApJ....546..140A} and 
    \citet[see their Figure 7]{2014arXiv1412.2881J}. 
    Therefore, as absorption gas grows thicker,
    the \hei\ and \mgii\ absorption lines will appear in order. It is also worth noting that $N_{\rm ion}$ of
    different models set apart from each other according to ionization parameter ($U$), while models with the same 
    $U$ but different $n_{\rm e}$ show no large divergences. This indicates that \hei\ and \mgii\ are more
    sensitive to ionization state, which agrees well with \citet{2014arXiv1412.2881J}.

    Our goal is to compare the measurements for observed sample with the results of calculated models. 
    To realize this,
    we are going to transform the relationship between $N_{\rm Mg\,II}$ and $N_{\rm He\,I }$ of models to the
    relationship between absorption depths.
    To simplify the discussion, we assume that the absorption gas fully covers the incident continua,
    so that the apparent column densities ($N_{ion}$) can be derived from an apparent optical depth profile
    $\tau$ = -ln($I_{r}$) directly, where $I_{r}$ is the normalized residual intensity of the absorption trough.
    That is
    \begin{equation}
        N_{ion} = \frac{m_{e} c}{\pi e^2 f \lambda}\int\tau(\upsilon)d\upsilon =
          \frac{3.7679\times10^{14}}{f\lambda}\int\tau(\upsilon)d\upsilon \ \ (cm^{-2})
    \end{equation}
    where $\lambda$ is the transition's wavelength and f is the oscillator strength, and where the velocity is
    measured in km s$^{-1}$. As a reference, we provide a table for first 5 \hei\ lines, Mg\,II and C\,IV doublets,
    including their wavelengths, oscillator strengths (see Table \ref{tab:lineinfo}).
    
    According to the measurements for \hei\ sample, $v_{avg}$ of the \hei$\lambda3889$ and \mgii\ 
    absorption lines have strong correlation, which indicates the two lines are associated dynamically 
    (Figure~\ref{fig:absmg2he1measure}).
    In the sample, measured \hei\ is much narrower than \mgii\ for most sources. This is because, on one hand,
    \mgii\ absorption lines we measured are a blend of \mgii\ doublets, and on the other hand, 
    the shape of absorption feature of low ionization lines is usually different from that of high ionization 
    lines for BALs.
    Here we roughly build up the relation of \mgii\ and \hei\ absorption profile from the sample.
    Four composites for \mgii\ absorption profile were constructed in four width ($W_{\rm abs}$) 
    ranges: 1600< width < 2500, 2500 < width < 3500, 3500 < width < 4500 and width > 5000. Simultaneously,
    composites for \hei$\lambda3889$ absorption line are also built up.
    The followings are the procedure for the simulation.
    A series of \mgii\ absorption profiles with different $d_{\rm abs}$ are generated based on these composites.
    In each bin of ($n_{\rm H}$, $U$, $d_{\rm abs,MgII}$),
    we calculated $N_{\rm Mg\,II}$ by integrating the \mgii\ absorption profile, and $N_{\rm He\,I*}$ was also 
    obtained according to relationship of $N_{\rm Mg\,II}$ and $N_{\rm He\,I*}$ predicated by models. 
    Then the corresponding \hei$\lambda3889$ absorption profile was generated, and we measured its $d_{\rm abs}$.
    Figure~\ref{fig:samplemodeldepths} shows the simulated procedures for a model with parameters of BAL
    outflow gas: log$n_{\rm e}$= 7.0, log$U$=-1.5.
    Figure~\ref{fig:modeldepth} shows the relationship between \mgii\ and \hei$\lambda3889$ absorption depths of 
    the simulation.
    In the left panel, we can see that results of simulation well coincide with the measurements of our sample
    (gray dots).
    In the right panel, $d_{\rm abs}$ of \hei$\lambda3889$ produced by outflows with different $n_{\rm e}$ only show
    slight differences. 
    While in the middle panel, $d_{\rm abs}$ of \hei$\lambda3889$ absorption line are layered according to ionization 
    parameter ($U$).
    These again indicate that the $d_{\rm abs}$ of \hei$\lambda3889$ absorption line is sensitive to the 
    ionization parameter but insensitive to $n_{\rm e}$.
    Outflows with large $U$ will produce \hei$\lambda3889$ absorption trough with depths equal to or even larger 
    than that of \mgii.

    Figure~\ref{fig:modelsequence} shows the ionization structure in a cloud slab of C\,IV, \mgii\ and \hei 
    predicated by the photoionization models we described before. Here we adopt models with 
    log$n_{\rm H}$(cm$^{-3}$) = 7 as the example.
    The distance from the illuminated surface of the cloud ($r$, thickness of the cloud) is represented by
    Hydrogen column density ($N_{\rm H}$) that is the total column density of Hydrogen integrated from the 
    illuminated surface to $r$. The column densities of other species are calculated in the same way.
    We can see that the column density of respective species (thus absorption strength) depends on ionization 
    parameter ($U$) and the cloud/outflow thickness (namely the cloud's column density, $N_{\rm H}$).
    When the $N_{\rm H}$ of an outflow is too small to develop a Hydrogen ionization front (the case of 
    optically thin clouds), $N_{\rm MgII}$ decrease with increasing $U$; when the cloud has a sufficiently 
    large $N_{\rm H}$ (optically thick), both high and low ionization absorption lines would be detected.
    Based on the simulation for the \hei$\lambda3889$ absorption depths, we can investigate the possible
    $N_{\rm Mg\,II}$ and $N_{\rm He\,I}$ range that can be observed. Using the composites for \mgii\ and 
    \hei$\lambda3889$ we constructed for the simulation, we calculated the ionic column densities corresponding to 
    absorption trough with $d_{\rm abs} \sim 0.05$, which is the minimum $d_{\rm abs}$ we can detect using 
    pair-matching method. 
    Therefore, these ion column densities are the boundary of detection for each absorption line.
    Red, orange and black shades in this figure represent the detection boundary for \hei$\lambda3889$, 
    \hei$\lambda10830$ and \mgii\ respectively.
    According to the figure, as an outflow is growing thick, \hei$\lambda10830$ absorption line will be 
    detected first. In the outflow with low ionization parameter (e.g., log$U$=-2), \mgii\ absorption line
    will be detected before \hei$\lambda3889$, while in the outflow with intermediate (e.g., log$U$=-1.2) and high
    (e.g., log$U$=-0.5) ionization parameter, \hei$\lambda3889$ absorption line will be detected before \mgii.
    This sequence explain the why the \hei$\lambda3889$ absorption lines so prevalent in \mgii\ BAL quasars.
    We may also expect that if we observe \mgii\ BALs in its near infrared band, the fraction of \hei\ BALs 
    appear in \mgii\ LoBALs will be much higher.
    According to this figure and Figure~\ref{fig:modeldepth}, there should be a kind of BAL with obvious 
    \hei$\lambda3889$ or strong \hei$\lambda10830$ absorption line but without \mgii\ absorption lines. 
    In fact we did have found three such cases SDSS J0352-0711, J1413+4400 and J0936+5331, which are shown in
    Appendix \ref{app:nirspectra}.
    SDSS J0352-0711 is a BAL quasar rejected by T06, G09 and Z10 due to the weakness of \mgii\ absorption.
    However, as Figure~\ref{fig:J0352J1413} shows, they have significant \civ\ absorption line and
    \hei$\lambda\lambda3889,10830$ absorption lines.
    In spectra of low-$z$ BALs J1413+4400 and J0936+5331, we can hardly find \mgii\ absorption lines,
    but \hei$\lambda\lambda3889,10830$ are significant, which are associated with \civ\ absorption trough.
    In addition, the published \hei\ BAL SDSS~J1512+1119 \citep{2012ApJ...758...69B,2013ApJ...762...49B} is 
    also in this case. The strength of \hei$\lambda2946$ absorption line is roughly same as \mgii$\lambda2796$ 
    absorption line \citep[Figure 3]{2012ApJ...758...69B},
    so \hei$\lambda3889$ or \hei$\lambda10830$ absorption lines should be much stronger.

     \subsection{Case study: deriving physical conditions of the outflow gas using \hei \label{sec:application}}

     The \hei\ absorption lines are important diagnostics of the physical conditions of the AGN outflows.
     The metastable state \hei\ has multiple upward transitions in a large wavelength span from UV to NIR 
     band, and these absorption troughs are well separated.
     Particularly, the very small abundance of the metastable state (with a number abundance of about
     $6\times10^{-7}$) makes \hei\ lines hard to saturate, valuable to probe outflows of high column density.
     The different oscillator strengths of the \hei\ lines allow us to determine the real optical depth
     and covering factor of the \hei\ absorbers \citep[e.g.,][]{2011ApJ....728....94L},
     as well as the ionization state of the outflow gas.

     Here we present an individual FeLoBAL quasar FBQS~0840$+$3633 in our \hei\ BAL sample
     to exemplify the power of the \hei\ absorption lines in probing the physical condition of the outflows.
     Its UV spectrum, taken by Keck/HIRES, has been studied  detailedly by \citet{2002ApJ....570..514D}.
     According to their measurements, the outflow covers a range of velocities from $-$700 to
     $-$3500~km~s$^{-1}$ (negative values denoting blueshift),
     with two main components centered at $-$900 and
     $-$2800 km~s$^{-1}$, respectively.
     The physical conditions of the two components are found to be significantly different,
     the low-velocity gas being of lower density ($n_{e} < 500$ cm$^{-3}$) and farther away from
     the active nucleus ($\sim$ 230 pc) than the high-velocity one.
     $N_{\rm He\,I*}$ is growing fast before the ionization front, therefore both the detection and
     non-detection of \hei\ absorption lines can place a strong constrain on column density of the gas.
     In light of  the absence of \hei$\lambda2830$ absorption line in Keck/HIRES spectrum,
     they set an upper limit to the hydrogen column density, $N_{\rm H} \lesssim 2\times10^{21}$~cm$^{-2}$.

     We detect significant \hei$\lambda3889$ and \hei$\lambda3189$ absorption lines in its SDSS spectrum.
     Then we obtain a NIR spectrum by the Triple-Spec spectrograph on the P200 telescope
     on 2014 January 17 to measure the expected broad \hei$\lambda10830$ absorption line (see Appendix
     \ref{app:nirspectra}).
     We apply the  pair-matching method to measure \mgii\ and \hei$\lambda\lambda3189,3889,10830$ absorption lines.
     To fit \hei$\lambda10830$, a NIR spectral library of 76 high-S/N unabsorbed quasars from 
     \citet{2006ApJ....640..579G}, \citet{2006A&A....457....61R} and \citet{2008ApJS..174..282L} is built up.
     Our measurements show that the \hei$\lambda10830$ absorption covers from $-$1300 to $-$4400 \kms,
     roughly corresponding to the high-velocity gas in \citet{2002ApJ....570..514D}, with which the \hei$\lambda3889$ 
     and \hei$\lambda3189$ troughs agree well.
     Figure~\ref{fig:J0840abs} shows the fitting results for \mgii\ and \hei$\lambda\lambda3189,3889,10830$ 
     absorption troughs.
     The left panels show the best-fit local continuum (blue) for each line and right panels show the zoomed-in 
     normalized absorption troughs.

     We use the \hei\ absorption lines to explore the partial coverage situation for the outflow gas.
     The optical depth ($\varpropto \lambda f_{\rm ik} N_{\rm ion}$) of \hei$\lambda10830$ is over 23 times that of 
     \hei$\lambda3889$, therefore
     \hei$\lambda10830$ should be heavily saturated provided the observed depth of \hei$\lambda3889$ trough.
     The non-zero flux in the 10830 trough infers a partial coverage situation for the \hei\ absorber 
     (see \citealt{2011ApJ....728....94L} for the arguments in detail).
     Besides, since the contribution of residual flux of 10830 trough to the spectrum is lager than the 
     contribution of broad emission lines, the absorber should only cover a fraction of the accretion disk.
     Therefore, we should subtract the emission lines and normalize the involved part of spectrum with respect to
     the AGN continuum only.
     The observed, normalized spectrum can be expressed as follows,
     \begin{equation}
          R = (1 - C_{f}(v)) + C_{f}(v)e^{-\tau(v)} .
     \end{equation}
     Here $C_{f}(v)$ is the cover fraction, and the $\tau$ ratios ($\varpropto \lambda~f_{\rm ik} N_{\rm ion}$) 
     of \hei$\lambda\lambda10830,3889,3189$ are 23.5:1:0.33.
     Following the methodology of \citet{2011ApJ....728....94L}, we derive the covering fraction, the optical depth
     and column density of \hei, and their 1-$\sigma$ uncertainties, as a function of velocity
     (see Figure~\ref{fig:J0840partialcover}).
     The \hei\ column density and its uncertainty is calculated by integrating, which is
     log$N_{\rm He\,I*}$ = 14.9$\pm0.07$~cm$^{-2}$.
     The velocity-averaged covering fraction is $\sim 50$\%.
     According to relation between the $N_{\rm He\,I*}$ and ionization parameter obtained by 
     \citet[see their Figure~8]{2014arXiv1412.2881J}, we estimate the ionization parameter 
     log$U$ is between $-$1.7 and $-$1.5.
     Assuming the $n_{\rm He\,I 2^{3}S}/n_{\rm He\,II} \sim 6\times10^{-6}$\citep{2002ApJ....570..514D},
     we estimate the total hydrogen column density $N_{\rm H}$ to be $\approx1.36\times10^{21}$~cm$^{-2}$.

     A second approach is taken to probe the physical conditions of the outflow gas,
     in the cases of the presence of BALs of low-ionization species such as Fe\,II.
     Besides the covering fraction,  \hei\ and total hydrogen column densities and ionization parameter 
     derived by the first approach, this second approach can further constrain the electron density of the gas.
     The strategy, called ``synthetic-spectra fitting'', is as follows 
     (see \citealt{2014ApJ....578L..31T} submitted to APJ, for the detail).
     Assuming that the absorption of lowly-ionized metal ions shares the same profile as \hei,
     we run a grid of photoionization simulations with Cloudy 13.03
     to produce synthetic spectra that cover the full space of the above-mentioned parameters,
     and match them to the observed spectrum
     (including the continuum, emission lines, and particularly all the absorption lines of interest);
     Then from the best-matched synthetic spectrum we derived the physical parameters of the absorbing gas.
     We apply this approach to FBQS~0840$+$3633,
     and derive Hydrogen column density log$N_{\rm H}$ =22~cm$^{-2}$ and ionization parameter
     log$U$=-1.7, that are
     consistent with those obtained by the first approach, and the electron density log$n_{\rm e}$=7.5~cm$^{-3}$.
     Figure~\ref{fig:J0840cloudymodel} displays the best-matched synthetic spectra, in the NUV region (left panel)
     and NIR region (right panel), respectively.

     In our \hei$\lambda3889$ BAL sample, more than half sources have \hei$\lambda3189$ absorption lines,
     and a few of them also have higher order \hei\ lines.
     We can also obtain \hei$\lambda10830$ with followup NIR spectroscopic observations.
     Thus we can derive important physical parameters of the outflow gas such as
       the covering factor ($C_{\rm f}$), \hei\ column density ($N_{\rm He\,I*}$) using \hei\ absorption 
       lines directly. 
     Since \hei\ absorption lines are sensitive to ionization state of clouds, ionization parameter ($U$)
     can be determined by \hei\ lines alone provide optically thick and in a moderate range of $n_{\rm e}$ 
     \citep{2014arXiv1412.2881J}, 
     or by joint use of \hei\ and lowly ionized absorption lines such as \mgii.
     The electron density ($n_{\rm e}$) of the outflow gas can be determined by absorption lines sensitive 
     to $n_{\rm e}$, such as absorption lines from excited Fe\,II state.
     With the above parameters, we can further locate the outflows from the AGN central engine,
     which is very important to study the connection between SMBH growth and host-galaxy buildup.

\subsection{Searching for low-$z$ BAL AGNs via \hei$\lambda3889$}

       So far low-$z$ BAL AGNs are still very rare due to the difficulty of carrying out UV spectroscopic 
       observations for the absorption lines such as \civ\ and \mgii.
       Yet low-$z$ BAL AGNs (particularly the high-luminosity version, quasars) are of great importance.
       First of all, their proximity enables us to investigate not only the properties of their host galaxies
       but also the spatially resolved outflows per se on the host galaxy scale and the interplay between the 
       outflows and the host-galaxy ISM.
       Additionally, a number of important spectral diagnostics can be obtained easily through optical 
       spectroscopic observations, e.g., narrow emission lines to calculate the ISM temperature,
       and broad emission lines to estimate the SMBH mass.

       Here we demonstrate the power of using \hei$\lambda3889$ absorption troughs to select low-$z$ BAL AGNs by 
       the pair-matching method.
       We simply set the redshift cutoff to be $z<0.3$, and start from all the quasar catalogs of the SDSS-I, II, 
       and III \citep{2010AJ.....139.2360S,2014A&A...563A..54P}.
       The selection procedure follows the method as described in Section~\ref{sec:heisample}.
       Note that it is more difficult to identify \hei$\lambda3889$ absorption lines
       without the reference of \mgii\ absorption.
       As the result of our above \hei\ BAL sample shows, the measured widths of \hei$\lambda3889$ absorption troughs
       are narrower than those of the corresponding \mgii\ BALs.
       So we set the width criteria to be |$V_{\rm max} - V_{\rm min}$| $>800$~km~s$^{-1}$ for low-$z$ 
       \hei$\lambda3889$ BALs.
       Then, we carefully examine the candidate spectra that pass the \hei\ BAL selection criteria.
       As for the \hei$\lambda3889$ absorption of those low-$z$ objects, the high-order Balmer absorption lines 
       from host-galaxy starlight are the principal contamination.
       Once we find Balmer absorption lines in a spectrum, we omit the object from the sample.
       Besides, some other absorption lines located at the same velocities as the candidate \hei$\lambda3889$ 
       absorption, such as \hei$\lambda3189$, \hei$\lambda2946$, and even Na\,D absorption in some spectra,
       are useful for our identification.
       Finally, 19 low-$z$ \hei\ BALs are identified, which are summarized in Table~\ref{tab:lowzbal}.
       Note that we only included the most secure detection of broad \hei$\lambda3889$ line.
       The real number of \hei\ BAL quasars might be much larger than this.
       For the typical data quality of SDSS spectra, the detection fraction of \hei\ BAL is about 35\%.
       This suggests that there might be at least 54 low-$z$ \hei\ BAL quasars.
       Thus we can roughly estimate the fraction of \hei\ BALs. There are 2539 quasars with $z < 0.3$ in 
       \citet{2010AJ.....139.2360S} for SDSS DR7 and \citet{2014A&A...563A..54P} for BOSS DR10 in total.
       The conservative fraction of \hei$\lambda3889$ BAL quasars is about 2.13\%. 
       This value is similar to \mgii\ LoBAL quasar fraction.
       \footnote{The observed C\,IV BAL fraction in quasars has been calculated as $\sim 15\%$ 
	 (e.g., \citealt{2003AJ....125.1784H,2003AJ....126.2594R}; G09), and LoBALs comprise $\sim 15\%$ 
	 of C\,IV BAL 
	 quasars \citep[e.g.,][]{1991ApJ....373....23W}}

       The combination of \hei$\lambda\lambda3889,10830$ is very useful to determine the physical properties of 
       absorption gas.
       We carried out NIR spectroscopic observations for 5 low-$z$ \hei$\lambda3889$ BAL AGNs, 
       namely J0752$+$1935, J0936$+$5331, J1535$+$564406.5, J1634$+$2049 
       and J2220$+$0109, using the Triple-Spec spectrograph on the Palomar 200-inch Hale telescope.
       All the 5 objects show evident \hei$\lambda10830$ absorption features, which confirms our detection of 
       \hei$\lambda3889$. Nearly full coverage is found in four of the five absorbers. 
       The rest one, J0752+1935, is about 50\%.
       
       The details of the NIR observation and data are described in Appendix \ref{app:nirspectra}.
       In addition, J0936+5331 and J1305+1819 have archival FUV and NUV spectra, taken by the 
       COS and STIS spectrographs aboard {\it Hubble Space Telescope} (HST). J0936+5331 shows C\,IV BAL, 
       while the \mgii\ absorption feature is absent in its NUV spectra; it is thus another example of HiBAL quasar 
       discussed in Section~\ref{sec:implication}. J1305+1819 show both C\,IV and \mgii\ BALs at the same 
       velocities of our identified \hei$\lambda3889$.

  \section{Summary and Future Works \label{sec:summary}}

   The primary findings of this work are as follows.

   1. We have carried out the first systematic search for \hei\ broad absorption line (BAL) quasars,
   yielding a sample of 101 quasars with \hei$\lambda3889$ absorption troughs culled from the SDSS DR5 data set.
   This increases the number of the quasars with any \hei\ absorption
   lines---all discovered serendipitously in the literature and including those found
   via \hei$\lambda10830$ line---by more than an order of magnitude.
   Besides, 52 objects of this \hei$\lambda3889$ BAL sample have even \hei$\lambda3189$ absorption
   detected from their SDSS spectra.

   2. We have developed an effective pair-matching method and selection procedure,
   aiming at to select uniformly weak/shallow BALs.
   Careful treatments are given to the definition of the absorption region
   (namely the maximum and minimum velocities of the absorbing outflows)
   and to the BAL criteria
   by invoking several statistical measures, particularly taking advantage of the
   statistical merit of being able to estimate
   the {\em systematic error} of the pair-matching method.
   This methodology, with minor adaption, can be applied to detect
   other weak spectral features that are highly blended
   with their complicated surrounding components.

   3. This search for \hei$\lambda3889$ BALs is based on a large parent sample of
   \mgii\ BAL quasars compiled from the literature,
   with the information of the \mgii absorption used as
   auxiliary reference to guarantee the genuineness of \hei$\lambda3889$.
   We find that the observed fraction of He\,I*$\lambda3889$ BAL quasars in the parent \mgii\ sample
   is 35.4\%.
   When only the spectra with S/N $> 30$ are considered, the fraction is 93\%, which is surprisingly high.
   According to our simulations, we conclude that the observed overall fraction is mainly affected
   by spectral S/N, and we suggest that (almost) all LoBAL quasars have associated \hei\ BAL lines.

   4. We demonstrate the power of the \hei\ absorption line as a probe of the physical parameters of 
   AGN outflows by performing a case study on FeLoBAL FBQS~0840+3633.
   Through detailed analysing the \hei$\lambda\lambda3189,3889,10830$ absorption lines, we estimate 
   the covering factor ($\sim 50\%$), Hydrogen column density ($N_{\rm H} = 1.36 \times 10^{21}$ cm$^{-2}$)
   and ionization parameter (log$U$=-1.7$\sim$-1.5) of the outflow gas.
   We also run a grid of photoionization models using Cloudy 13.03 to produce synthetic spectra, and match 
   them to the observed spectrum. According to the best-matched synthetic spectrum, we derive the physical
   parameter log$N_{\rm H} = 22$ cm$^{-2}$, log$U$=-1.7, which are consistent with the estimation above.
   Specially, the electron density is determined as log$n_{\rm e} = 7.5$ cm$^{-3}$ in this way.

   5. As another extended application of the \hei\ absorption lines,
   we have conducted a pilot search for low-$z$ BAL AGNs via \hei$\lambda3889$
   in all the SDSS spectroscopic archives.
   Those AGNs have to be free of host-galaxy starlight
   (e.g., the stellar high-order Balmer absorption lines) in their SDSS spectra.
   Finally we find 19 AGNs at $z<0.3$ with \hei$\lambda3889$ absorption troughs.
   These low-$z$ BAL AGNs are valuable in studying the outflow launching mechanism,
   the interplay between the AGN outflow and the host-galaxy ISM,
   the properties of the host galaxies, and the AGN feedback.

   There are several lines of fruitful work for the future.
   First of all,
   follow-up NIR observation for the \hei$\lambda10830$ absorption lines would be instructive
   to confirm their BAL nature. Moreover, the joint use of
   \hei$\lambda\lambda10830,3889$ absorption lines will enable us
   to determine the covering factor, column density and the ionization parameter of the outflow.
   In the optical band,
   by exploiting the huge volume data set of spectra with sufficient S/N and spectral resolution
   (e.g., the SDSS), we can further develop elaborate procedures to carefully separate stellar absorption 
   features and identify \hei$\lambda3889$ BALs. This would be very useful for identifying BALs in AGNs at
   low redshifts whose spectra are contaminated by starlight. 

   We thank the anonymous referee for helpful report that significantly improve this paper, 
   thank Xiaobo Dong for reading the manuscript and correcting the English writing. 
   H.Z. thanks Xuebing Wu and Stefanie Komossa for the helpful discussion.
   This work is supported by the SOC program (CHINARE2012-02-03), 
   Natural Science Foundation of China grants (NSFC 11473025, 11033007, 11421303),
   National Basic Research Program of China (the 973 Program 2013CB834905).
   T.J. acknowledges supports from Fundamental Research Funds for the Central Universities (WK 2030220010). 
   S.Z. acknowledges supports from the Natural Science Foundation of China with grants NSFC 11203021;
   P.J. acknowledges supports from the Natural Science Foundation of China with grants NSFC 11233002, NSFC 11203022.
   W.Z. acknowledges supports from the Natural Science Foundation of China with grants NSFC 11173021, NSFC 11322324.
   This work has made use of the data products of the SDSS and data obtained through the Telescope Access 
   Program (TAP) in 2012B (PI: Xinwen~Shu), 2013A (PI: Tuo~Ji) and 2014A (PI: Wenjuan~Liu).
   TAP is funded by the Strategic Priority Research Program ``The Emergence of Cosmological Structures'' 
   (Grant No. XDB09000000), National Astronomical Observatories, Chinese Academy of Sciences, 
   and the Special Fund for Astronomy from the Ministry of Finance.
   Observations obtained with the Hale Telescope at Palomar Observatory were obtained as part of an agreement
   between the National Astronomical Observatories, Chinese Academy of Sciences, and the California Institute of
   Technology.

\clearpage

 
\begin{figure}[htbp]
        \centering
        \includegraphics[width=3.4in]{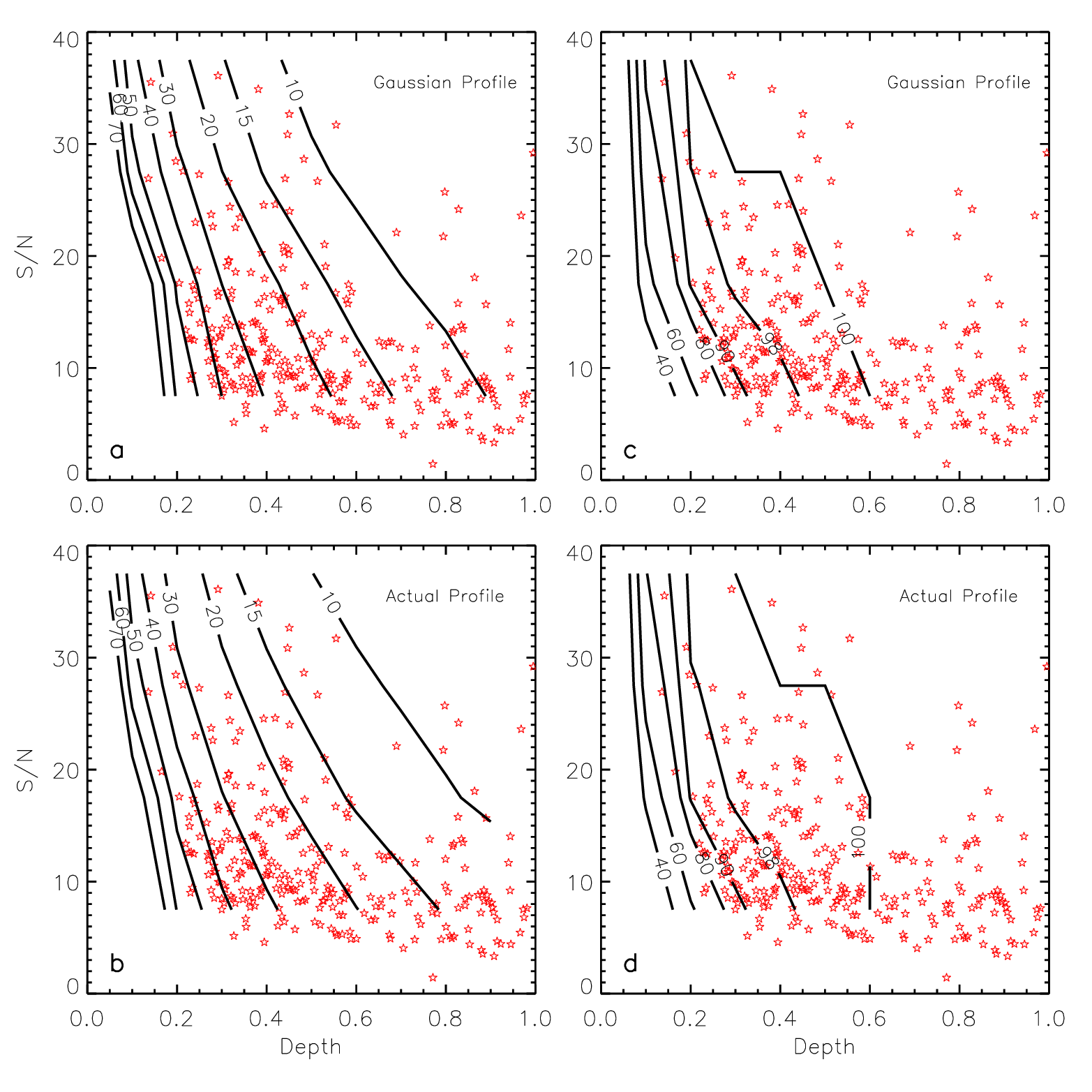}\includegraphics[width=3.4in]{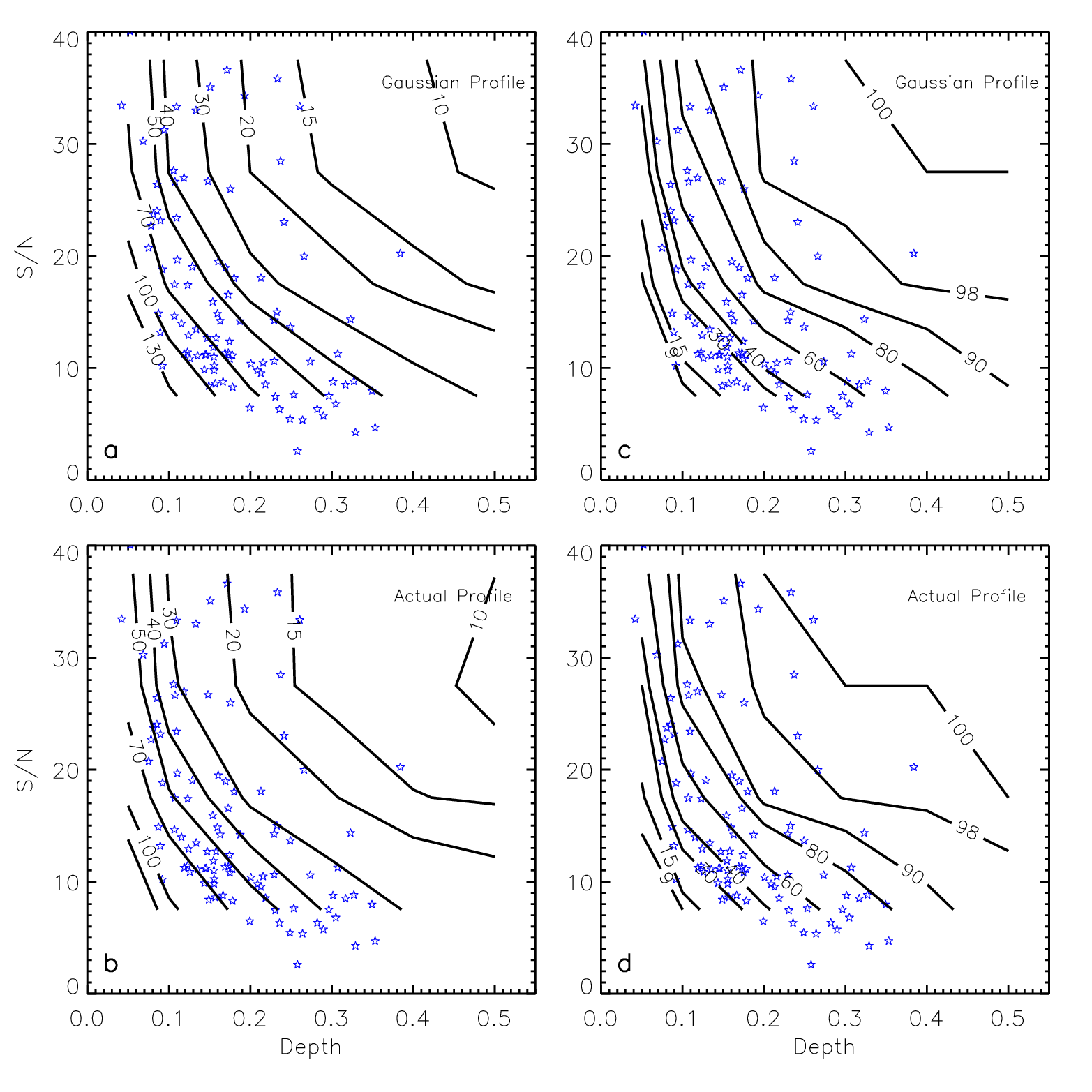}
        \caption{\footnotesize Contours of the relative total error (panels a and b)  and recovered detected 
	  fraction (panels c and d) in the spectral S/N and absorption depth plane based on our simulations 
	  (see Appendix~\ref{app:Tests}), for \mgii\ (left) and  \hei\,$\lambda3889$ (right) BALs.
          The results based on simulated BAL spectra with Gaussian profiles are shown in top panels;
          those with actual BAL profiles, in bottom panels. 
          Also denoted are the observed parent sample of \mgii\ BAL quasars (red stars) and the 
          \hei\,$\lambda3889$ BAL quasars (blue stars). \label{fig:sndepth}}
\end{figure}

\begin{figure}[htbp] 
        \centering
        \includegraphics[width=5.8in]{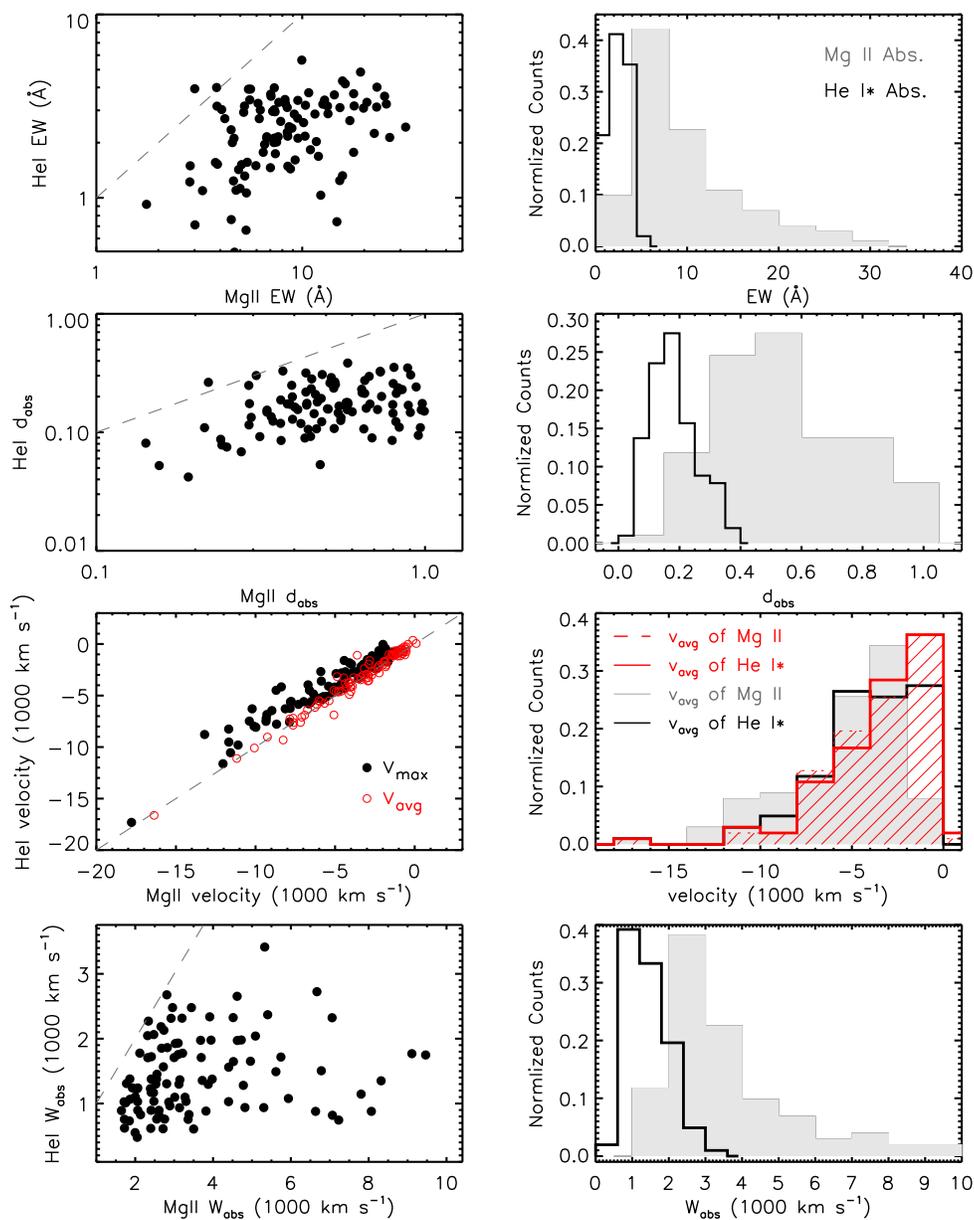}
    \caption{\footnotesize {\it Left}: Comparison of BAL parameters such as EW, depth ($d_{\rm abs}$), 
    maximum and average velocities ($v_{\rm max}$ and $v_{\rm avg}$),  
    and width ($W_{\rm abs}$, namely |$v_{max} - v_{min}$|)
    between \mgii\ and \hei\,$\lambda3889$ in the \hei\ BAL sample.
    {\it Right}: The normalized histograms of these parameters.
    \label{fig:absmg2he1measure}}
\end{figure}

\begin{figure}[htbp]
        \centering
    \includegraphics[width=5in]{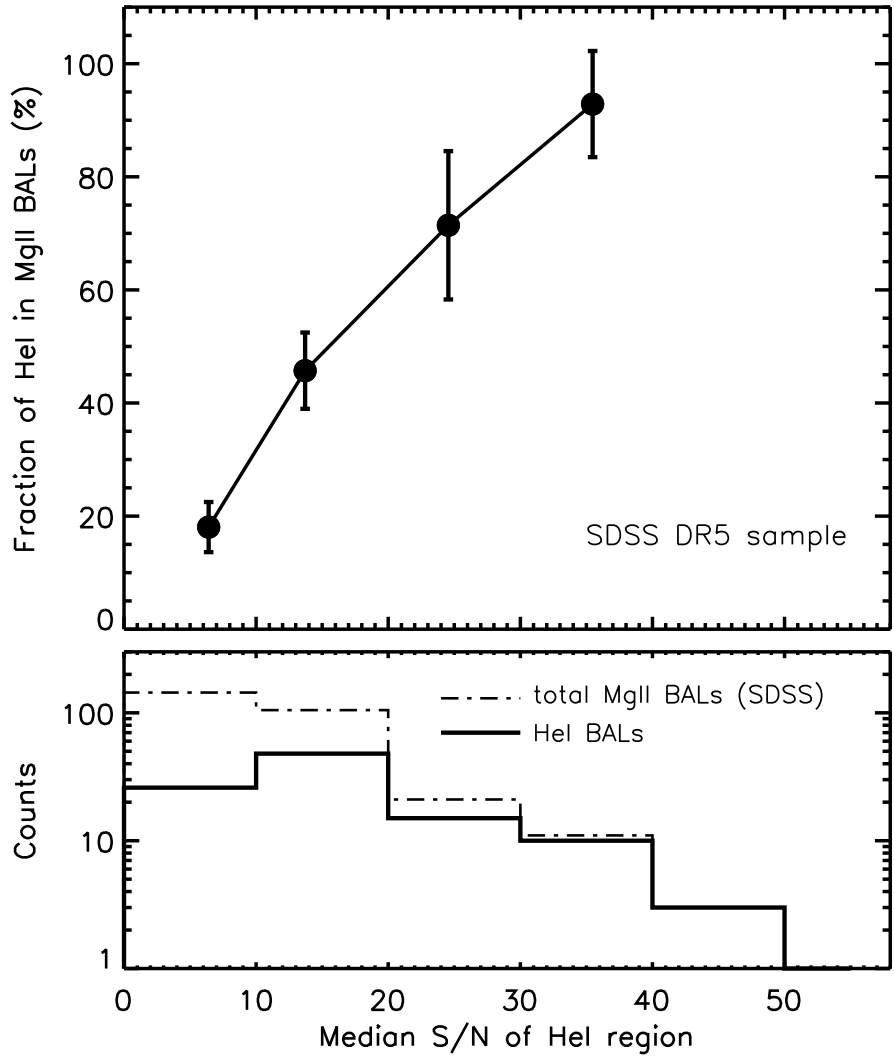}
    \caption{\footnotesize {\it Top}: The detected fraction of \hei\,$\lambda3889$ BALs in the parent \mgii\ BAL sample,
    as function of the spectral S/N.
   Also plotted are the $\pm 1-\sigma$ error.
    {\it Bottom}: Histograms of the spectral S/N of the \hei\ BAL sample (solid line) and the parent \mgii\ sample (dashed line).
    \label{fig:fraction}}
\end{figure}

\begin{figure}[htbp]
    \includegraphics[width=3.2in]{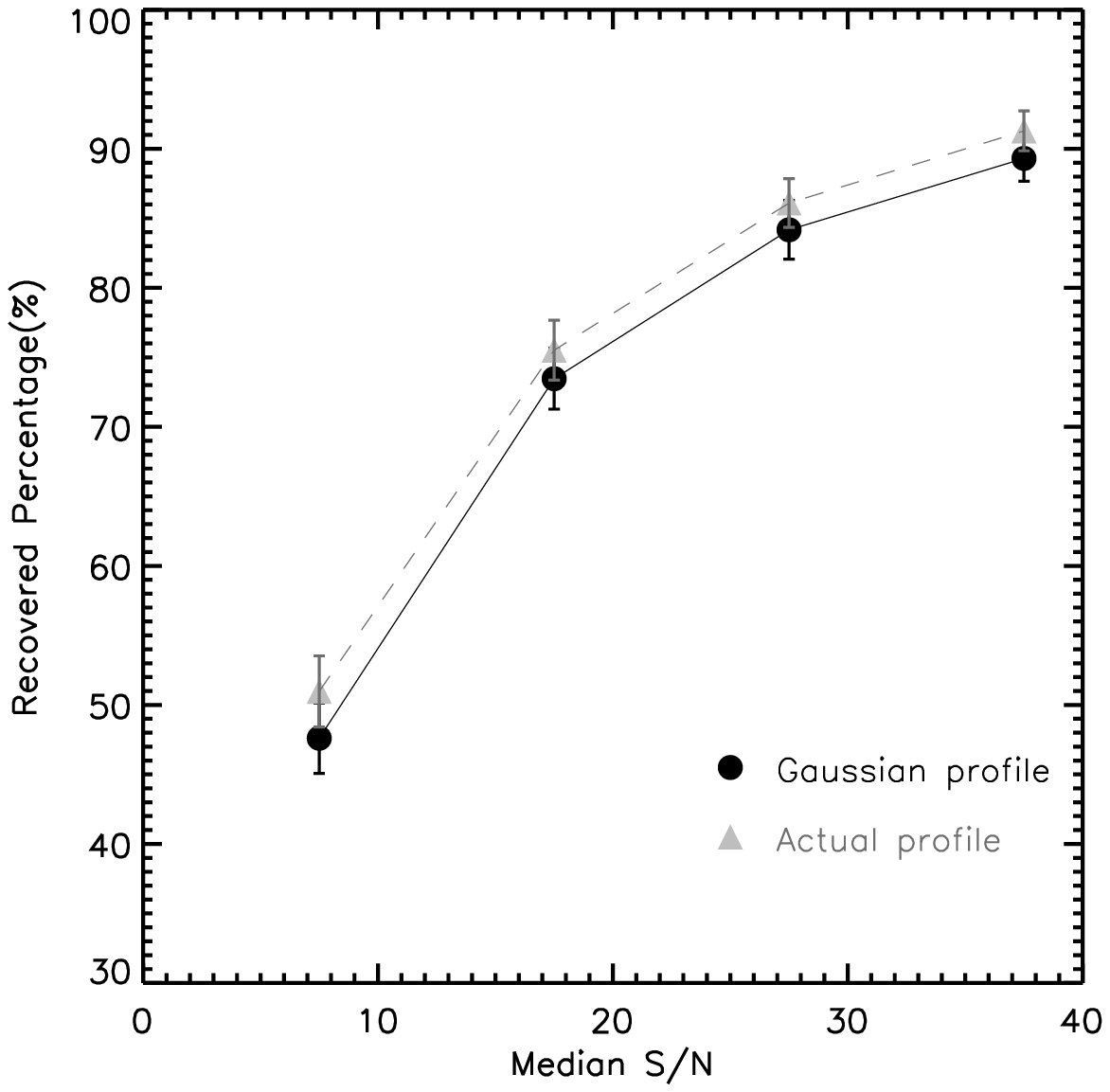}\includegraphics[width=3.2in]{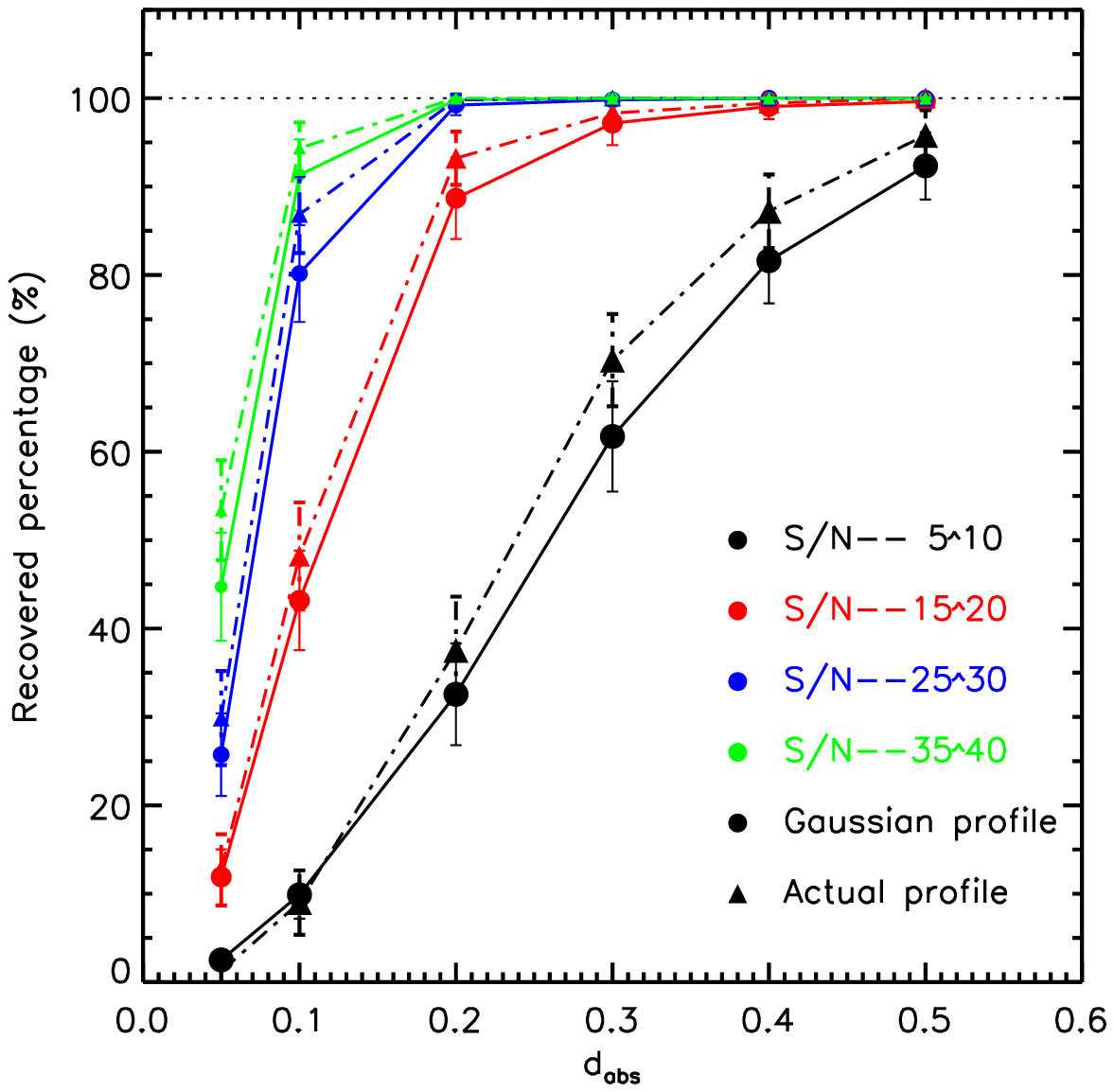}
    \caption{\footnotesize {\it Left}: The recovered fraction of \hei\ BALs as a function 
      of S/N based on the simulated spectra (see Figure~4 and Appendix~\ref{app:Tests}).
      {\it Right}: The recovered fraction of \hei\ BALs as a function 
      of absorption depth, based on simulated spectra of different spectra S/N (see Appendix~\ref{app:Tests} for details).
      \label{fig:modelfrac}}
\end{figure}

\begin{figure}[htbp] 
    \centering
    \includegraphics[width=3.2in]{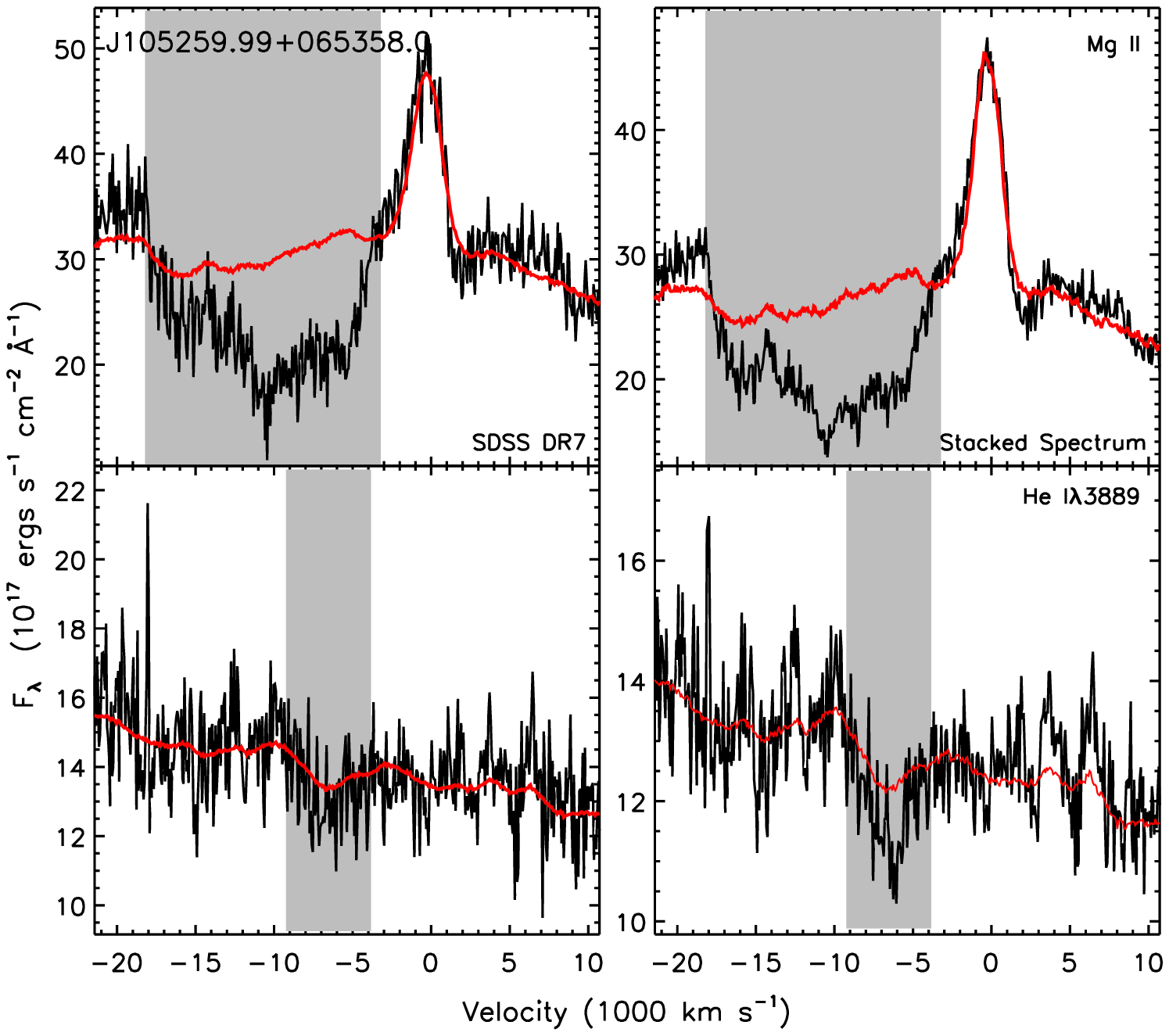}\includegraphics[width=3.2in]{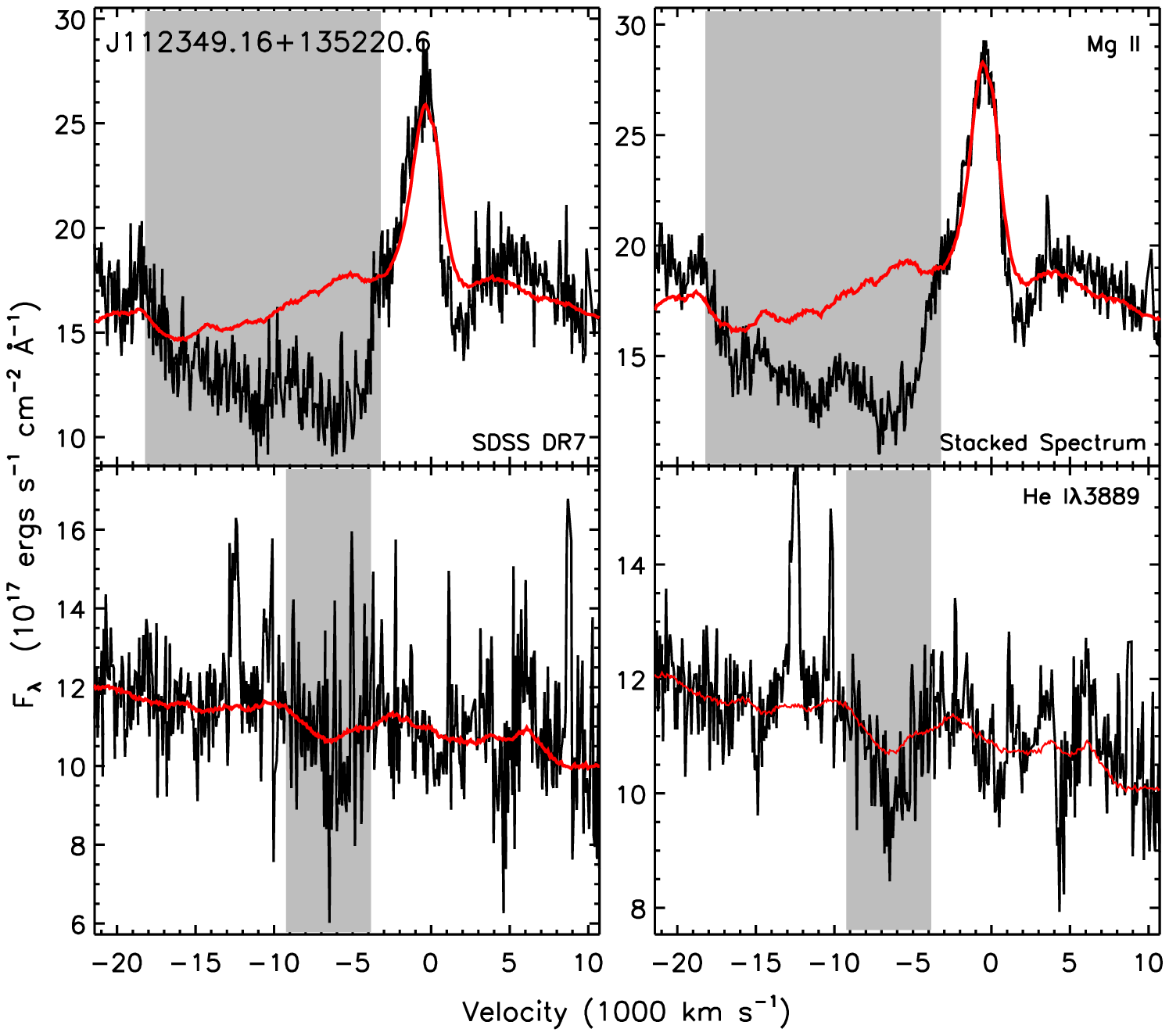}
    \includegraphics[width=3.2in]{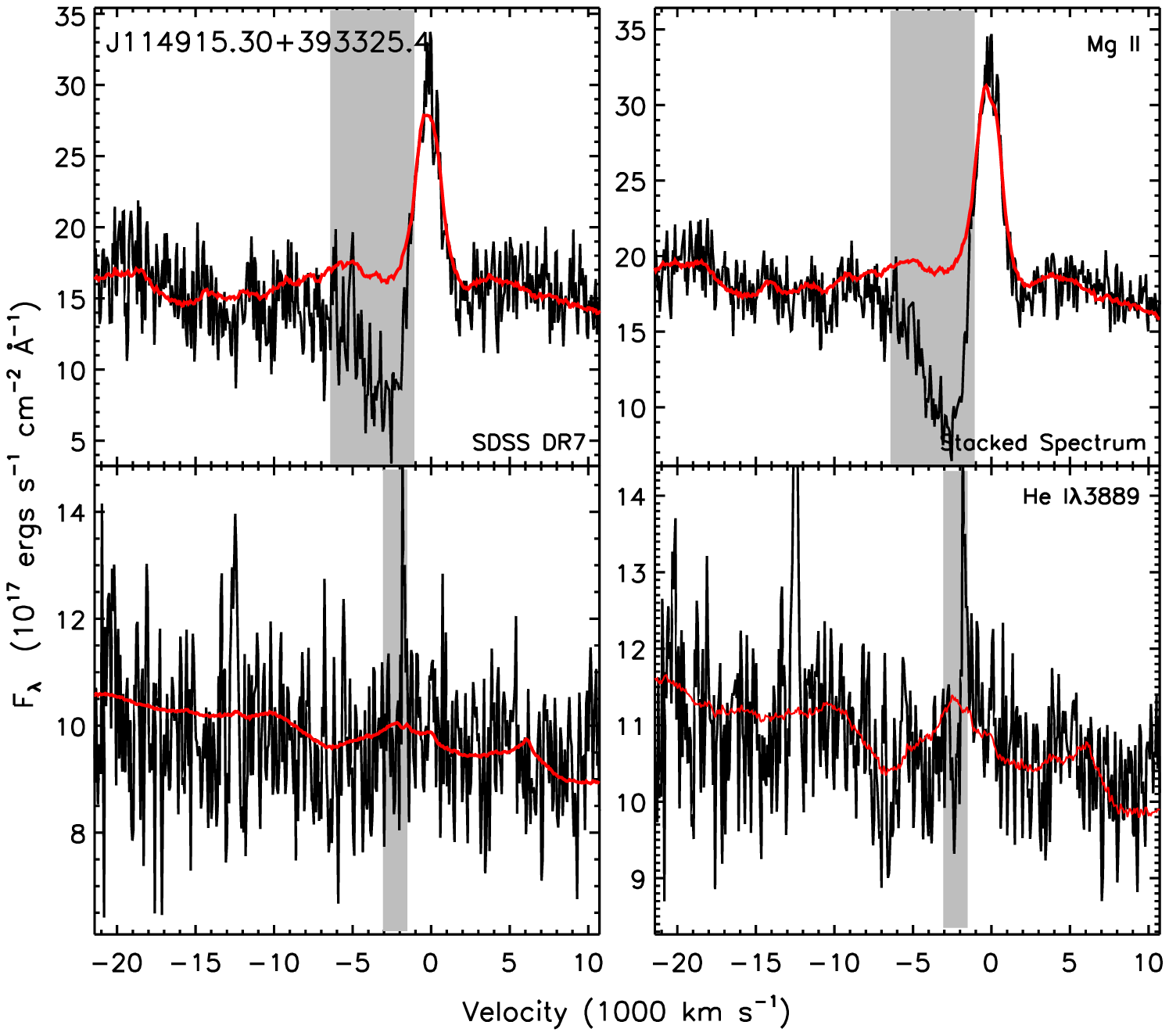}\includegraphics[width=3.2in]{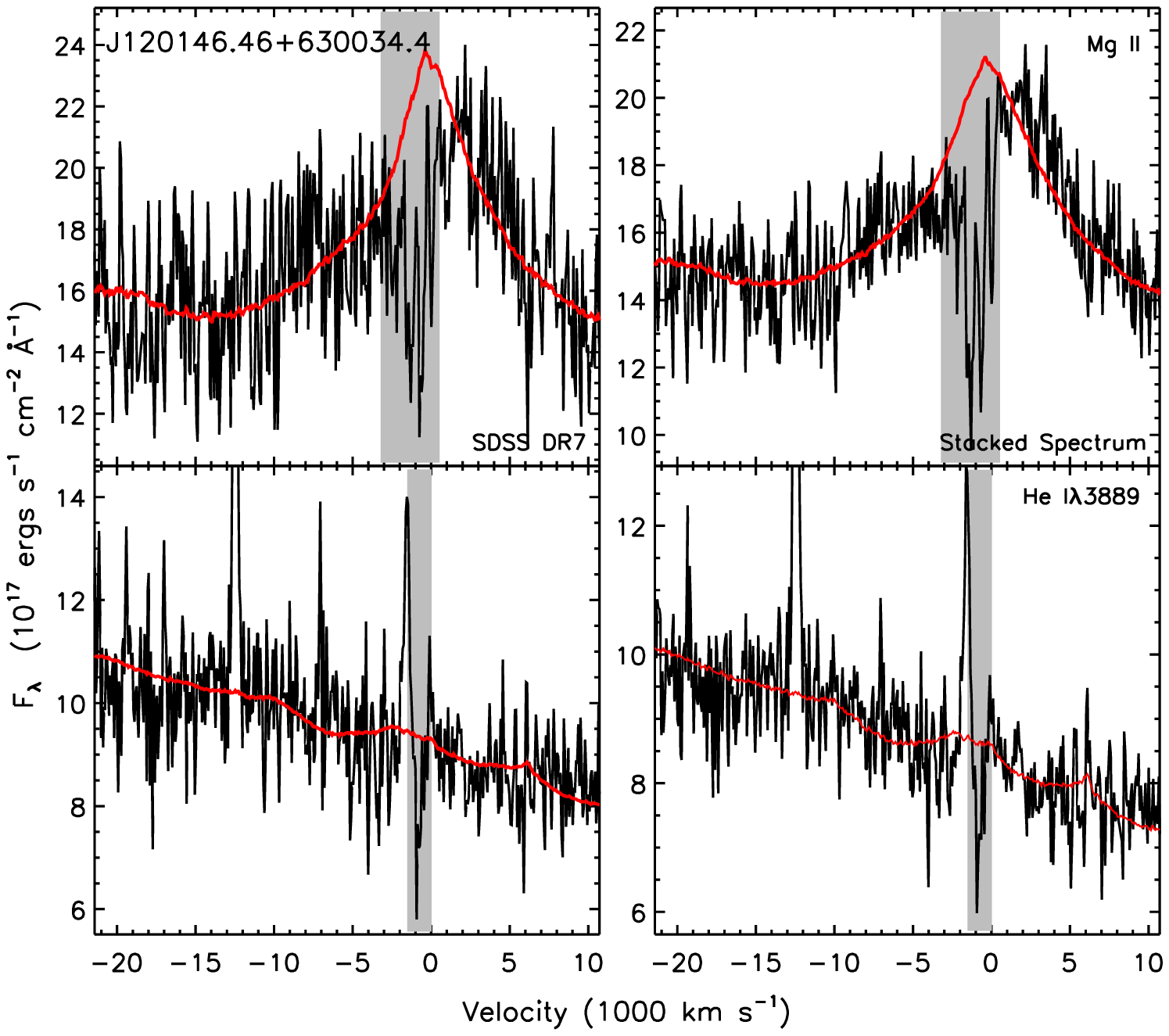}
    \caption{\footnotesize Demonstration of the four sources in the parent \mgii\ BAL sample, whose \hei\,$\lambda3889$ absorption lines
    are not able to be detected in their SDSS DR7 spectra, but are detected from the higher-S/N spectra by
    combining multiple observations in the SDSS and BOSS data set.
    Also shown are the modeled intrinsic spectra by the pair-matching method.
    \label{fig:dr10example}}
\end{figure}
\clearpage

\begin{figure}[htbp]
        \centering
    \includegraphics[width=6.4in]{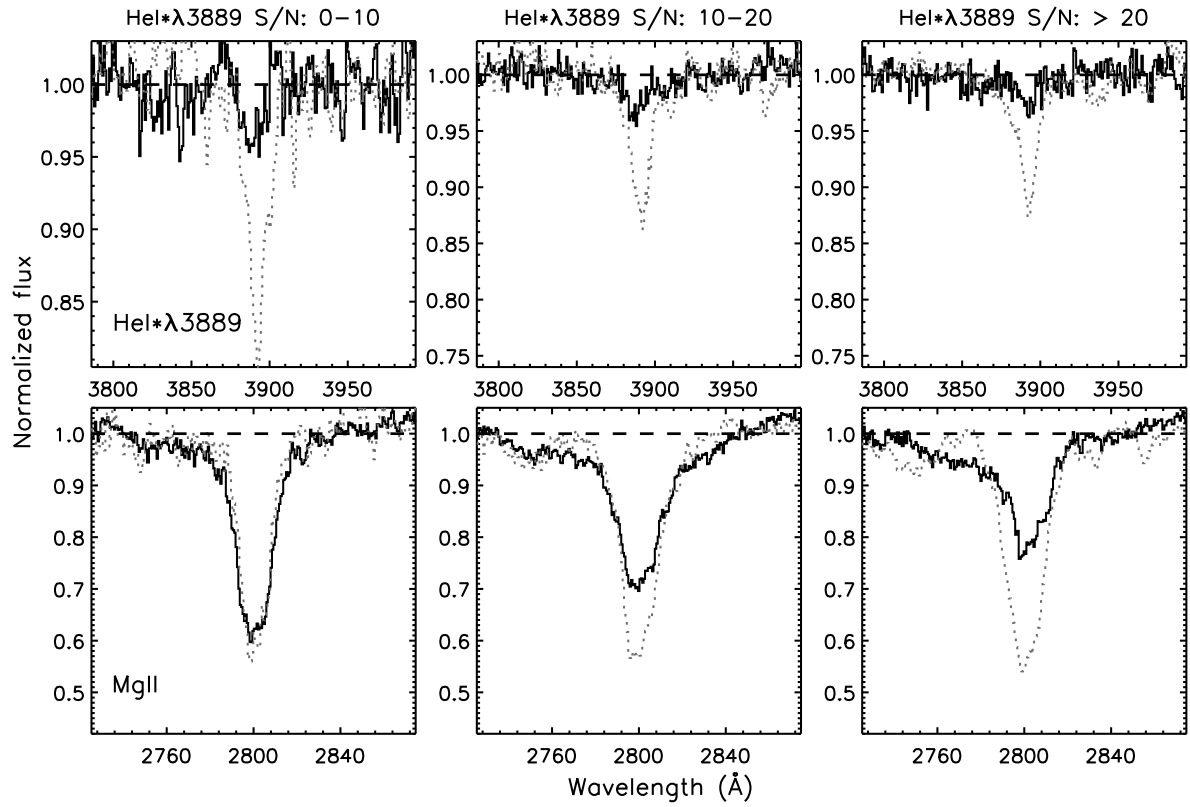}
    \caption{\footnotesize {\it Top}: Stacked spectra of three bins of spectral S/N in the \hei\,$\lambda3889$ region 
    for the sources in the parent sample yet without detected \hei\,$\lambda3889$ absorption (black solid lines)
    and those with (gray dotted lines),  
    {\it Bottom}: The stacked spectra in the \mgii\ region correspondingly. 
    \label{fig:superposition}}
\end{figure}

\begin{figure}[tbp] 
        \centering
    \includegraphics[width=5.5in]{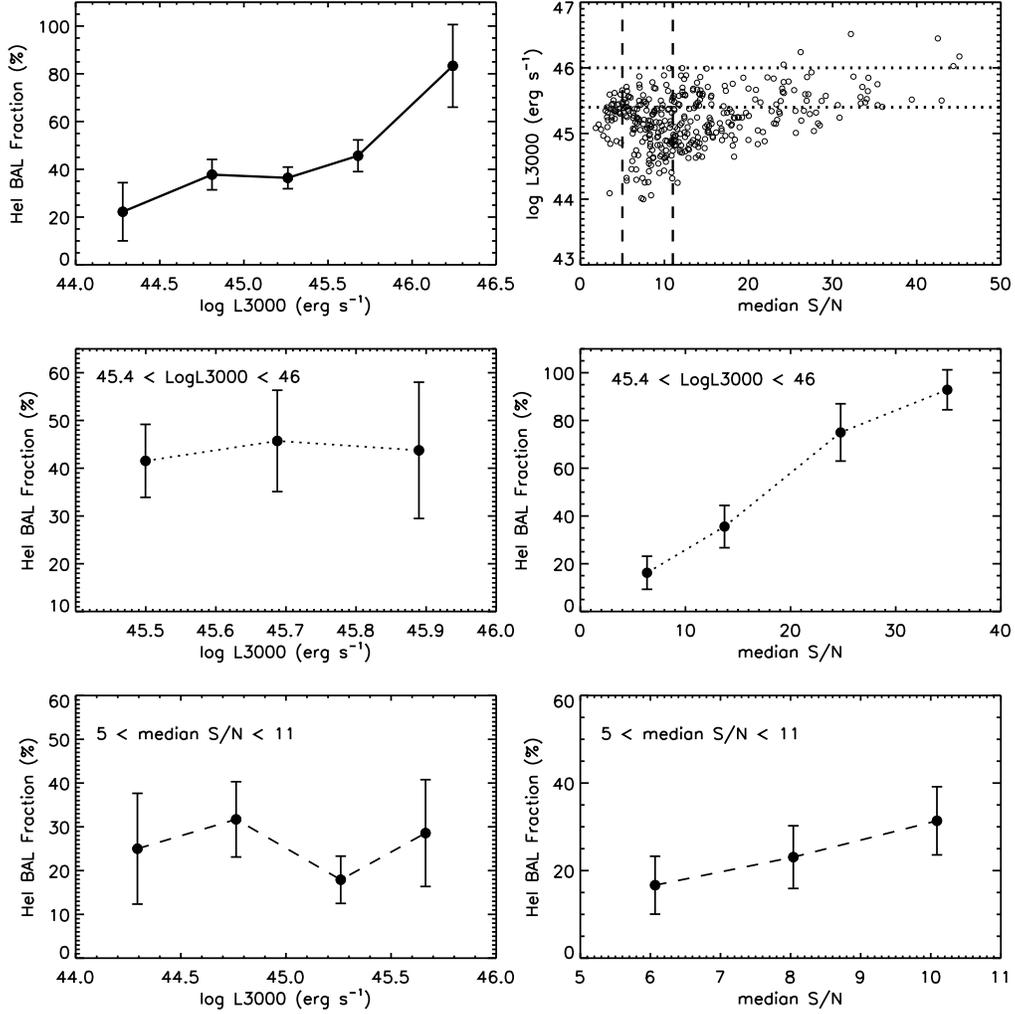}
    \caption{\footnotesize {\it Top-left panel}: Dependence of the \hei\,$\lambda3889$ BAL fraction on the AGN luminosity
    in the parent \mgii\ BAL sample.
    {\it Top-right panel}: Distribution in the plane of luminosity and spectral S/N. 
    {\it Middle panels}: Dependence of the \hei\,$\lambda3889$ BAL fraction on the AGN luminosity (left)
    and on the spectral S/N (right), for the sources in the narrow luminosity bin as denoted by the dotted lines in the top-right panel.
    {\it Bottom panels}: For the sources in the narrow S/N bin as denoted by the dashed lines in the top-right panel. \label{fig:snorlumi}}
\end{figure}
\clearpage

\begin{figure}[htbp]
        \centering
    \includegraphics[width=4.8in]{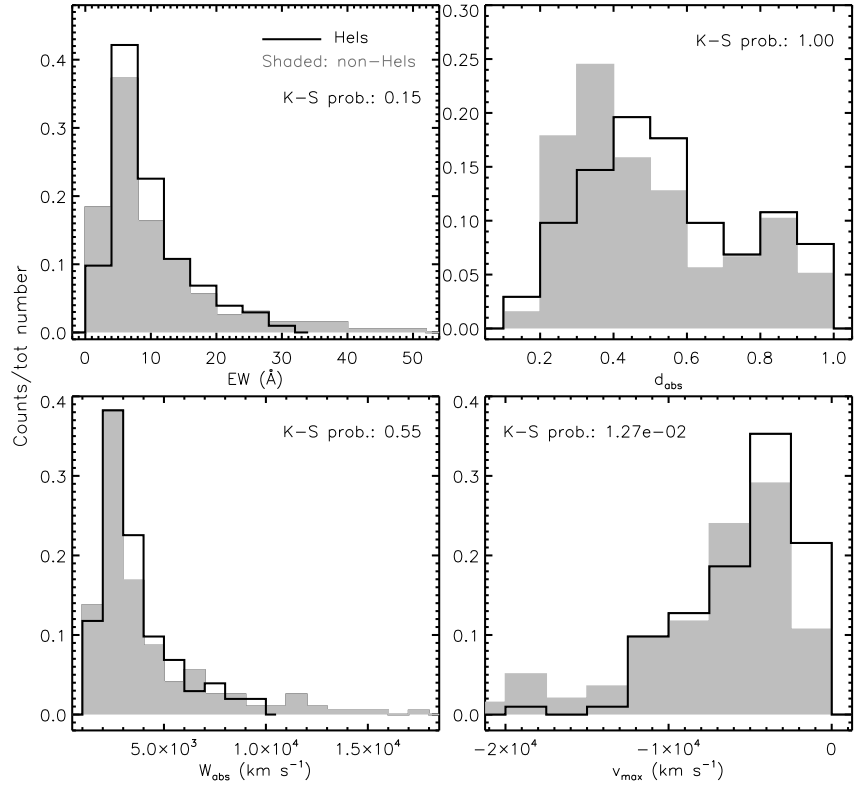}
        \caption{\footnotesize 
        Comparison of the normalized distribution of the \mgii\ BAL parameters between the sources with \hei\,$\lambda3889$ absorption 
        and those without
        in the parent sample.
        Also denoted are the chance probabilities by K-S test, which indicate that the two have the same distribution.
    \label{fig:distrofmg2abs}}
\end{figure}

\begin{figure}[htbp]
        \includegraphics[width=3.1in]{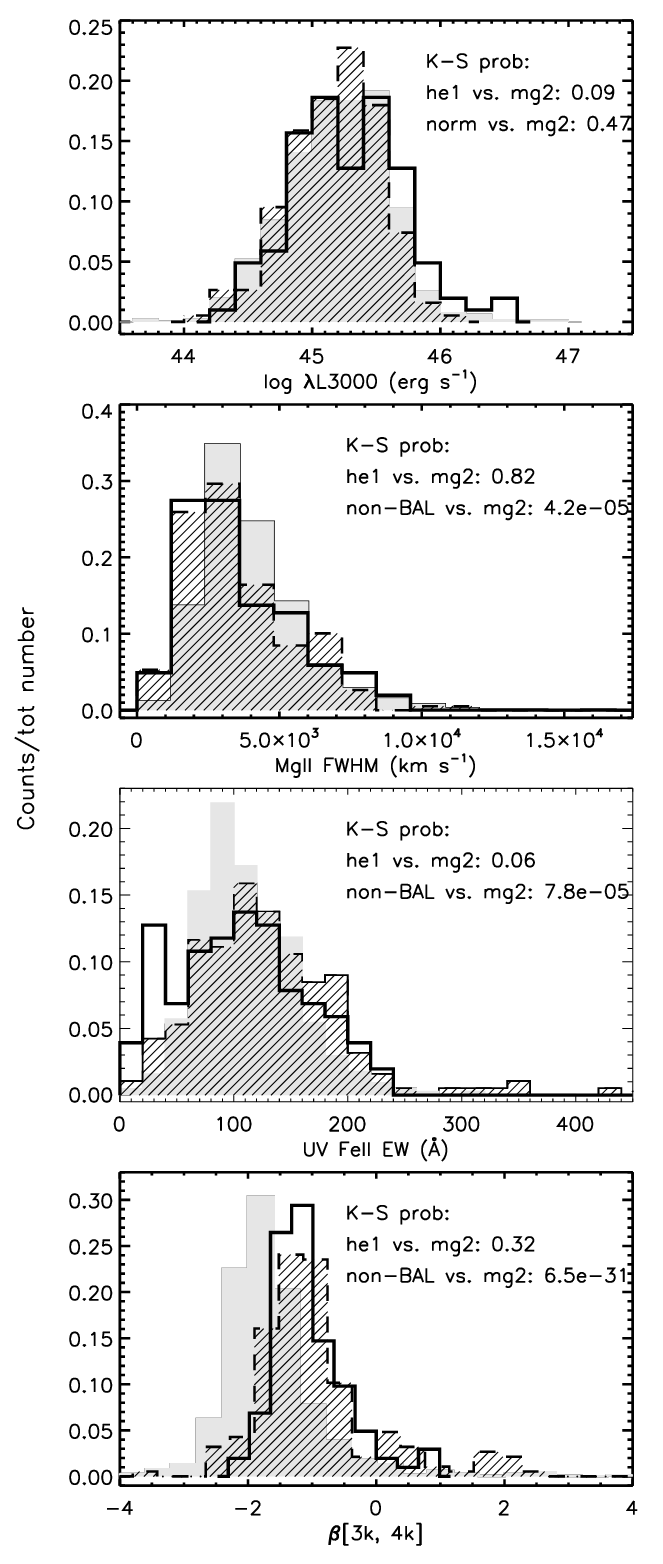}\includegraphics[width=3.1in]{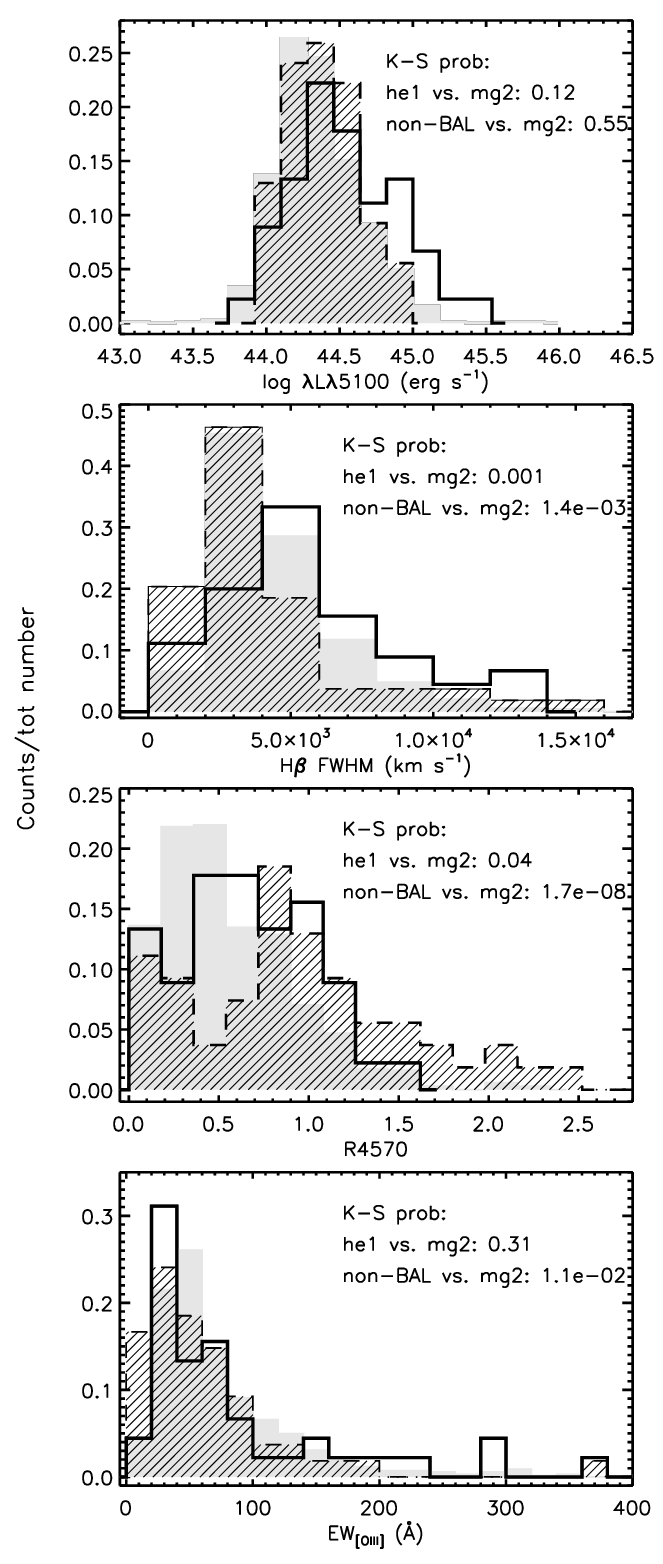}
    \caption{\footnotesize comparison of normalized distributions of continuum and emission-line properties 
    between the \hei\,$\lambda3889$ BAL sample (solid line), the parent \mgii\ BAL sample (filled with tilted lines),
    and non-BAL quasars (gray shaded). Also denoted are the chance probabilities by K-S test between the \hei\ BAL sample and the parent \mgii\ BAL sample,
    and those between the \mgii\ BAL sample and non-BAL quasars.
 \label{fig:distruvopt}}
\end{figure}


\begin{figure}[htbp]
  \centering
  \includegraphics[width=6.4in]{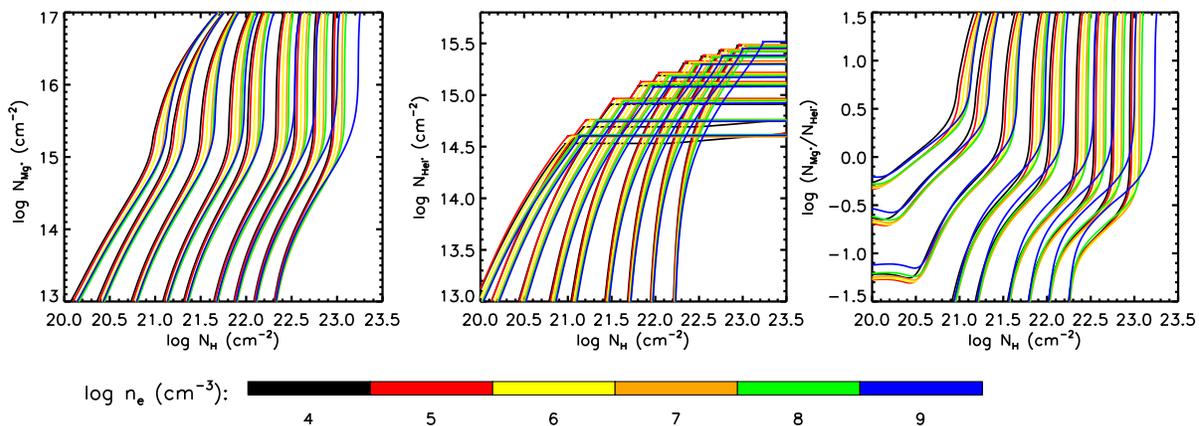}
  \caption{\footnotesize Ionization structure of Mg$^{+}$ (left) and \hei\ (middle), and the ratio of the 
    number of the two species (right) for a cloud slab calculated by CLOUDY. 
    Here the distance from the illuminated surface of the cloud ($r$, x-axis) is represented by \colnh\ 
    that is the total column density integrated from the illuminated surface to $r$. 
    The column densities of \mgii\ and \hei\ (y-axis) are calculated in the same way.
    Models with different hydrogen number densities ($n_{\rm H}$) are marked by different colors.
    Bundles from left to right in each panel denote models with different ionization parameters:
    $\log~U$ = -2.0, -1.8, -1.5, -1.2, -1.0, -0.7, -0.5, -0.3.
    Note that the ionization structure of both species are almost insensitive to $n_{\rm H}$,
    yet fairly sensitive to $U$. \label{fig:modeloverview}}
\end{figure}

\begin{figure}[htbp]
    \centering
    \includegraphics[width=5.2in]{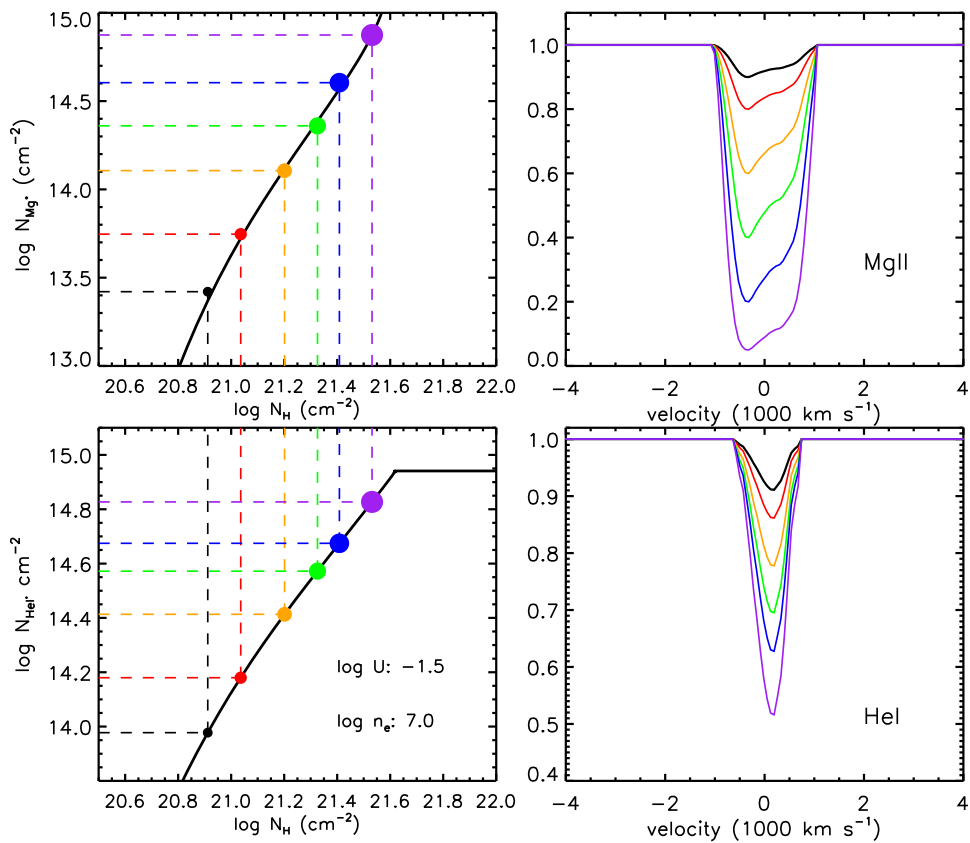}
    \caption{\footnotesize 
    Illustrating  the procedure that generates the corresponding
    \hei\ $\lambda3889$ absorption lines (bottom right panel) from \mgii\ BALs with given absorption depths 
    (top right panel).
    This example is for an absorbing cloud with typical physical parameters: log~$n_{\rm e}= 7.0$ and log~$U=-1.5$.
    The velocity width of MgII absorption line constructed from the sample is $\sim 2000$ km~s$^{-1}$, and 
    the corresponding He\,I* absorption line is $\sim 1000$ km~s$^{-1}$.
    From the observed \mgii\ BAL, the Mg$^{+}$ column density and then  the total column density 
    can be derived according to the CLOUDY calculation (top left panel; cf. Figure~\ref{fig:modeloverview}), 
    then the  \hei\ column density (bottom left panel). 
    Based on these cloud parameters, the \hei\ BAL can finally be generated.
 \label{fig:samplemodeldepths}}
\end{figure} 
\clearpage

\begin{figure}[htbp]
     \centering
     \includegraphics[width=6.4in]{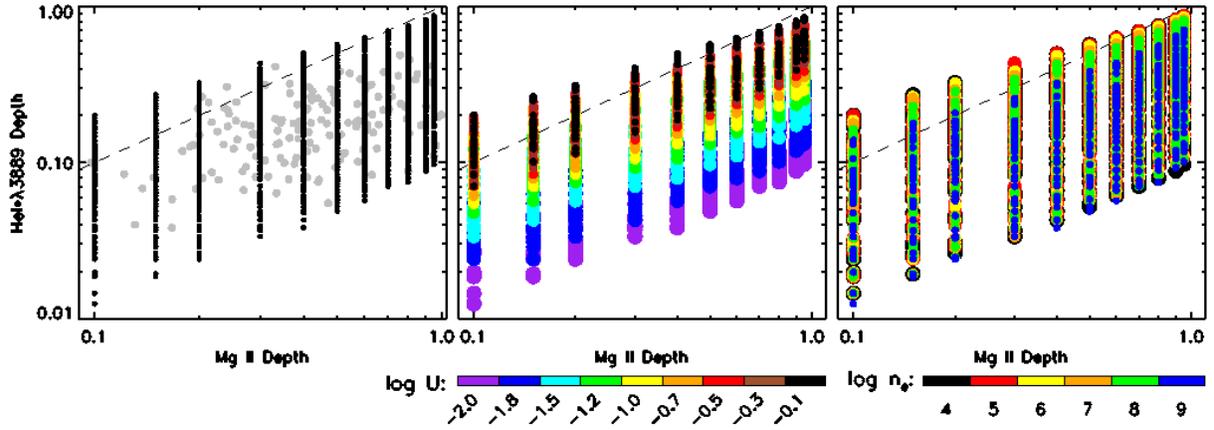}
     \caption{\footnotesize The CLOUDY calculations reveal the relationship of absorption depth between \mgii\ 
       and \hei$\lambda3889$ BALs.
       In left panel, black dots show the relation between \mgii\ and \hei$\lambda3889$ predicted by the calculations.
       Gray dots are the measurements of the \hei\ BAL sample as shown in Figure~\ref{fig:absmg2he1measure}.
       The next two panels are the same with the left panel, but with different colors denoting
       ionization parameters (log$U$) and electron density ($n_{\rm e}$), respectively.
       Note the strong dependence of \hei$\lambda3889$ absorption depth on ionization parameter, yet little on 
       electron density. \label{fig:modeldepth}}
\end{figure} 
\clearpage

\begin{figure}[htbp] 
  \centering
  \includegraphics[width=6.4in]{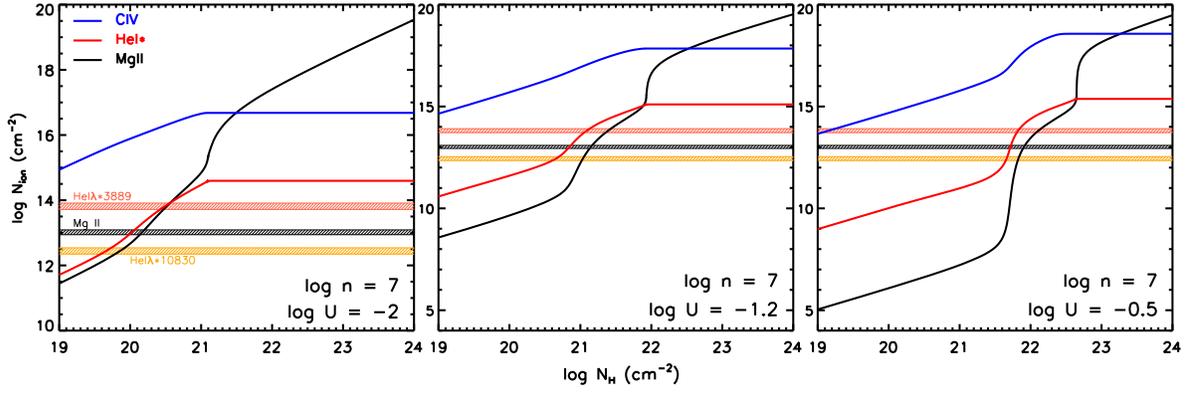}
  \caption{\footnotesize
   Ionization structure in a cloud slab of C\,IV (blue), Mg\,II (black) and HeI* (red) calculated by CLOUDY, 
   which depend strongly on ionization parameter (log~$U$). Horizontal shaded stripes
   represent the ion column density ranges corresponding to the absorption troughs with different widths and with 
   a depth of 0.05, the lower limit by the pair-matching method,
   for He\,I*$\lambda3889$ (red) ,  He\,I*$\lambda10830$ (orange), and Mg\,II (black), respectively.
   \label{fig:modelsequence}}
\end{figure}
\clearpage

\begin{figure}[htbp] 
  \centering
  \includegraphics[width=6.2in]{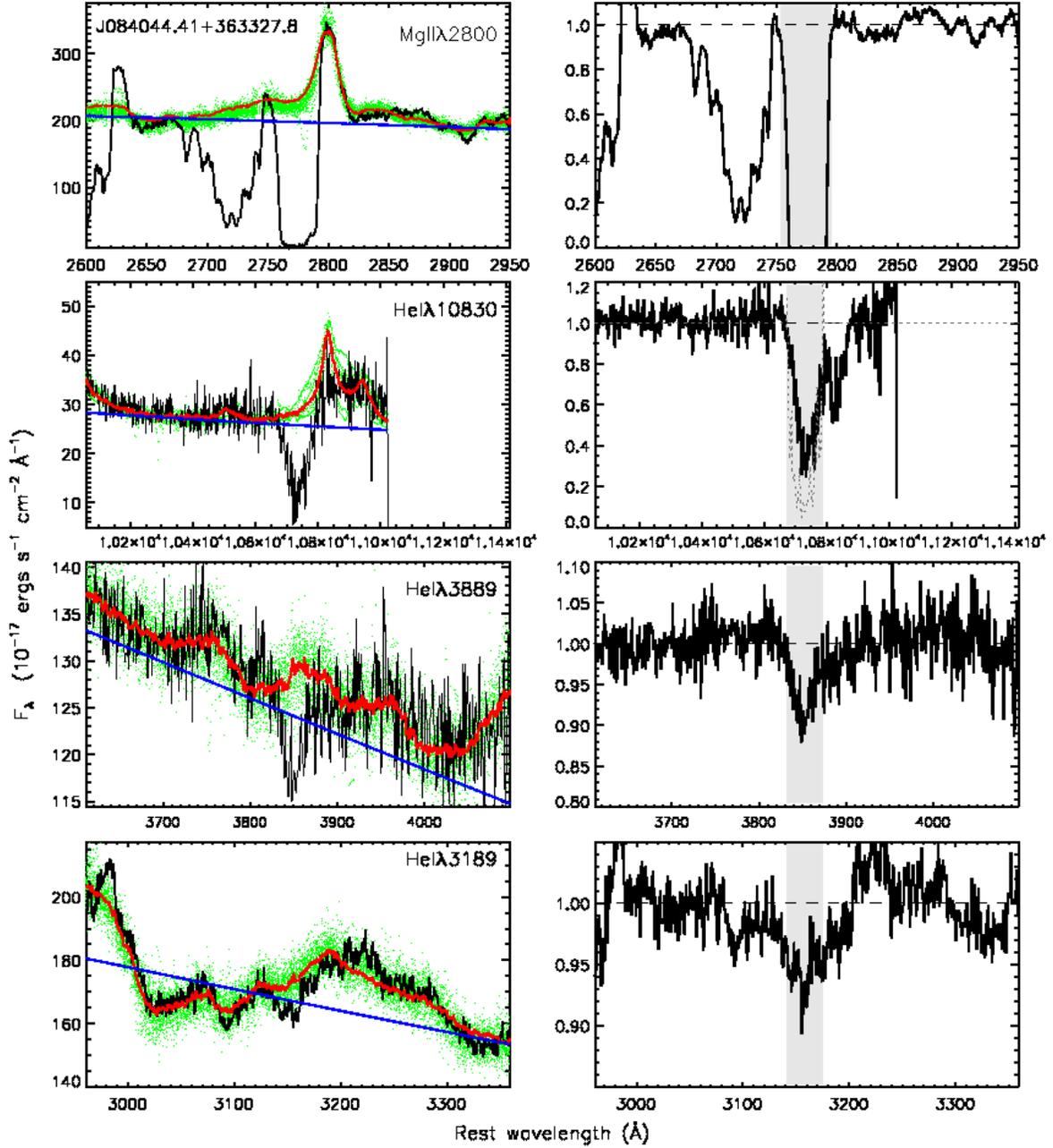}
  \caption{\footnotesize 
  {\it Left}: Demonstration of the absorption troughs in Mg\,II, He\,I*$\lambda\lambda10830,3889,3189$ (black)
  of FBQS~0840+3633, and the acceptable fittings (green) and the mean spectrum (red) by the pair-matching method,
  as well as the intrinsic AGN continuum (blue) determined according to several line-free regions. 
  {\it Right}: Normalized versions corresponding to the left panels.  
    The gray shades indicate the absorption line regions.
    The gray dotted line in the second panel from the top shows the He\,I*$\lambda10830$ absorption profile
    as predicted from the He\,I*$\lambda3889$ absorption trough under the full coverage assumption.\label{fig:J0840abs}}
\end{figure}
\clearpage

\begin{figure}[htbp]
  \centering
  \includegraphics[width=5.2in]{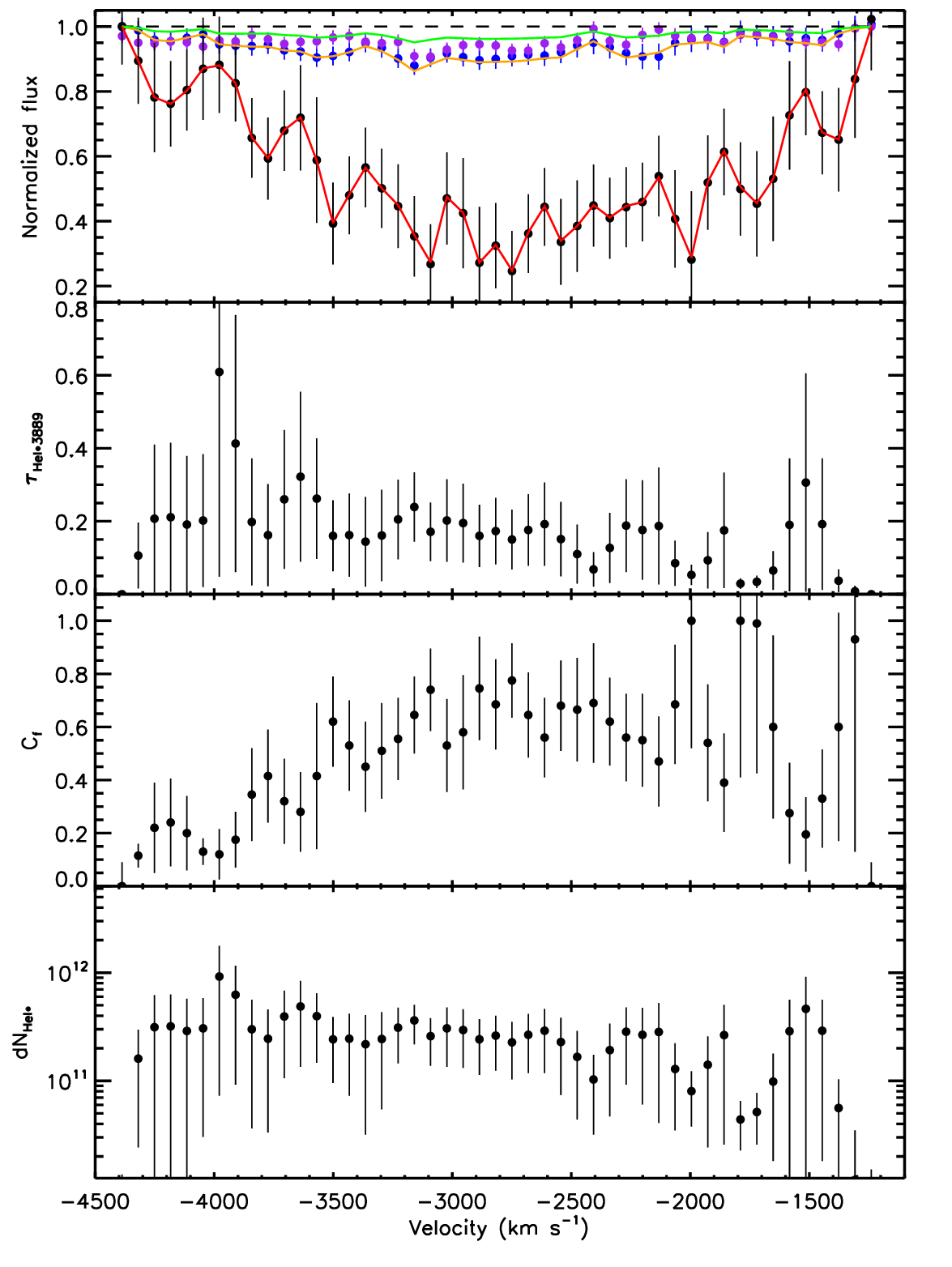}
  \caption{\footnotesize Calculation results (top panel) of the He\,I absorption lines of FBQS~0840$+$3633 
    under partial coverage model.
    The top panel shows the observed profiles with $\pm$1-$\sigma$ error bars of \hei\,$\lambda3189$ (purple dots),
    $\lambda3889$ (blue dots) and $\lambda10830$ (black dots), and their models as reproduced by the model 
    parameters.
    The model parameters (the optical depth, covering fraction and column density of the outflow gas) are 
    shown in the other 3 panels as functions of velocity.\label{fig:J0840partialcover}}
\end{figure} 
\clearpage

\begin{figure}[htbp]
  \centering
  \includegraphics[width=3.2in]{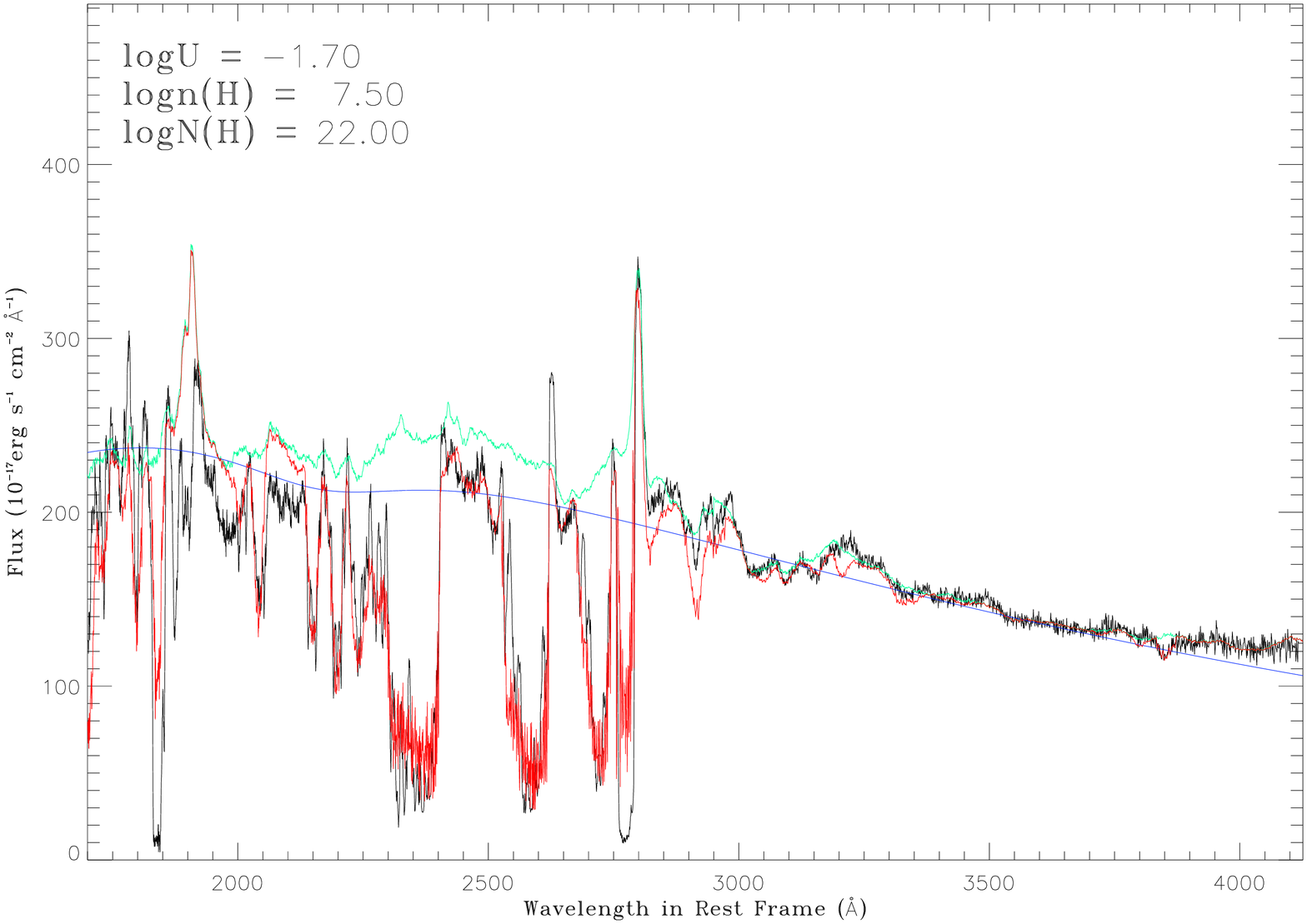}\includegraphics[width=3.2in]{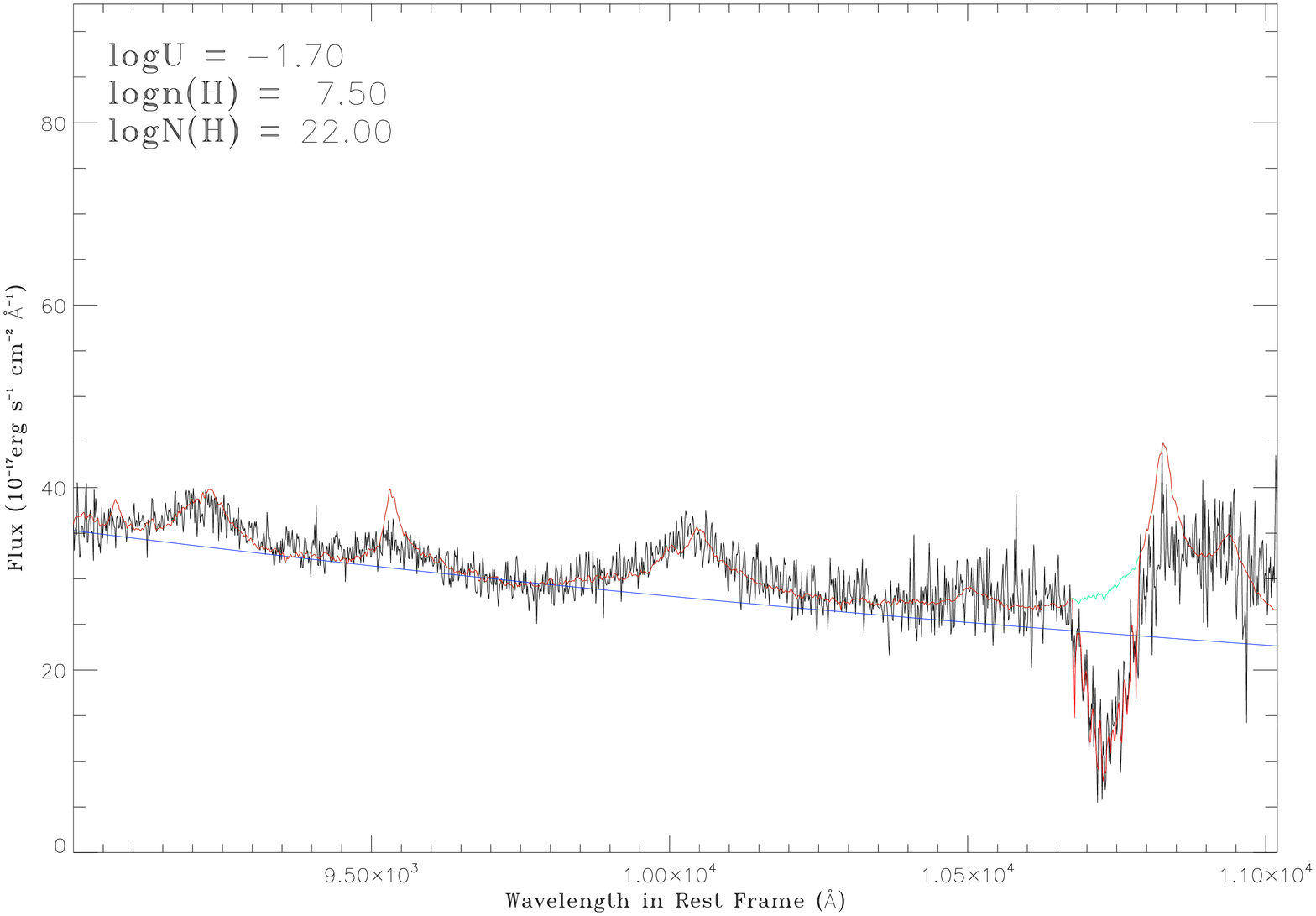}
  \caption{\footnotesize Best-fit model for FBQS~J0840+3633 by the method of synthetic-spectra fitting, with the model spectra synthesized by CLOUDY. \label{fig:J0840cloudymodel}}
\end{figure} 

\clearpage

\begin{deluxetable}{lrrrrrrl}
\tablewidth{\textwidth}
\tablecolumns{8}
\tabletypesize{\small}
\tablecaption{Physical parameters of known \hei\ BALs \label{tab:heiknown}}
\tablehead{
\colhead{Object}      &
\colhead{$z$}           &  \colhead{log$N_{\rm H}$ (cm$^{-2}$)}   & \colhead{log$U$}   &
\colhead{log$n_{\rm e}$}  & \colhead{log$N_{\rm HeI*}$}    &  \colhead{$r$(kpc)} &
\colhead{Ref.}}
\startdata
SDSSJ0300+0048  &0.89   & \nodata     & \nodata &\nodata   & $\leq$14.9   &\nodata           & 1 \\ 
SDSSJ0802+5513  &0.664  & 21$\sim$21.5& -1.8    & 5        & 14.73          & 0.1$\sim$0.25  & 2 \\
SDSSJ1106+1939  &3.038  & 22.1$^{+0.3}_{-0.1}$ & -0.5$^{+0.3}_{-0.2}$ & 4.1$^{+0.14}_{-0.37}$ & 14.68$^{+0.23}_{-0.02}$ & 0.32$^{+0.20}_{-0.14}$   &  3 \\
FBQSJ1151+3822  &0.335  & 21.7$\sim$21.9 & -1.5   & 5.5-8    & 14.9           & 0.0072-0.127   & 4,5\\
LBQSJ1206+1052  &0.396  & 21$\sim$22  & > -1.5  & 6$\sim$8 & 15.01          & \nodata        & 6 \\
Mrk231          &0.042  & 22.7        & -0.5    & 3.75     & 14.96          & $\sim$ 0.1     & 7,8 \\
SDSSJ1512+1119\tablenotemark{a} &  2.106  & 21.9$^{+0.1}_{-0.1}$ & -0.9$^{+0.1}_{-0.1}$ & 5.4$^{+2.70}_{-0.60}$  & 14.84$^{+0.03}_{-0.01}$  & 0.3-0.01 & 3,9 \\ 
AKARIJ1757+5907 &0.615  & > 20.82     & >-2.15  & 3.8      & 14.2           & >3.7           & 10 \\
NVSSJ2359-1241  &0.868  & 20.556      & -2.418  & 4.4      & 14.14$\pm$0.3  & 1.3$\pm$0.4   & 11,12,13 \\
\enddata
\tablerefs{ (1) \citet{2003ApJ....593..189H}; (2) \citet{2014arXiv1412.2881J};
(3) \citet{2013ApJ...762...49B}; (4) \citet{2011ApJ....728....94L}; 
(5) \citet{2014ApJ....783....58L}; (6) \citet{2012RAA.....12..369J} ;
(7) \citet{1985ApJ....288..531R}; (8) \citet{2014ApJ...788..123L};
(9) \citet{2012ApJ...758...69B}; (10) \citet{2011PASJ....63S.457A}; 
(11) \citet{2008ApJ...681..954A}; (12) \citet{2008ApJ...688..108K}; 
(13) \citet{2010ApJ...713...25B}}
\tablenotetext{a}{For SDSSJ1512+1119, we show the physcial properties of component 2 (C2) of absorption 
  trough, which is associated with He\,I* absorption lines (see Figure 4 in \citet{2013ApJ...762...49B}). }
\end{deluxetable}

\clearpage
\begin{deluxetable}{lrrrrrr}
\tablewidth{0.8\textwidth}
\center
\tablecolumns{7}
\tabletypesize{\normalsize}
\tablecaption{Absorption Line information \label{tab:lineinfo}}
\tablehead{
\colhead{Ion}  &  \colhead{$\lambda$ (\AA)}  &  
\colhead{$f_{\rm ik}$}   &  \colhead{$E_{\rm low}$} & \colhead{$E_{\rm up}$} &
\colhead{$g_{\rm low}$}   &  \colhead{$g_{\rm up}$} \\
\colhead{}   & \colhead{Vacuum}  &
\colhead{}   & \colhead{cm$^{-1}$} & \colhead{cm$^{-1}$} &
\colhead{}   & \colhead{}}
\startdata
He I* \ldots\ldots\ldots& 2764.63  & 0.004 & 159856 & 196027 & 3 & 9  \\
He I* \ldots\ldots\ldots& 2829.91  & 0.007 & 159856 & 195193 & 3 & 9  \\
He I* \ldots\ldots\ldots& 2945.96  & 0.012 & 159856 & 193801 & 3 & 9  \\
He I* \ldots\ldots\ldots& 3188.67  & 0.028 & 159856 & 191217 & 3 & 9  \\
He I* \ldots\ldots\ldots& 3889.75  & 0.064 & 159856 & 185565 & 3 & 9  \\
He I* \ldots\ldots\ldots& 10833.1  & 0.539 & 159856 & 169087 & 3 & 9  \\
Mg II \ldots\ldots\ldots& 2796.35  & 0.608 & 0      & 35761  & 2 & 4  \\
Mg II \ldots\ldots\ldots& 2803.53  & 0.303 & 0      & 35669  & 2 & 4  \\
C IV  \ldots\ldots\ldots& 1548.20  & 0.190 & 0      & 64592  & 2 & 4  \\
C IV  \ldots\ldots\ldots& 1550.77  & 0.095 & 0      & 64484  & 2 & 4  \\
\enddata
\tablerefs{These data were taken from 
  Kramida,~A., Ralchenko,~Yu., Reader,~J., and NIST ASD Team (2014),
NIST Atomic Spectra Database (ver. 5.2) [On line], available at http://physics.nist.gov/asd \,[2015, January 22], 
National Institute of Standards and Technology, Gaithersburg, MD, USA.}
\end{deluxetable}

\clearpage

\begin{landscape}

\pagestyle{empty}
\begin{deluxetable}{lrrrrrrrrrrrr}
\tablewidth{0pc}
\tablecolumns{13}
\tabletypesize{\scriptsize}
\tablecaption{The parent sample of \mgii~BAL Quasars in the SDSS DR5\label{tab:dr5mg2table}}
\tablehead{
  \colhead{SDSS Name\tablenotemark{a}}  &
  \colhead{$z$}          &  \colhead{Spectrum\tablenotemark{b}}   &
  \colhead{Mg\,II\tablenotemark{c} }       &  \colhead{Mg\,II\tablenotemark{d}}     & \colhead{Mg\,II\tablenotemark{e}}      &
  \colhead{Mg\,II\tablenotemark{f}}       & \colhead{Mg\,II\tablenotemark{g}}      & \colhead{Mg\,II\tablenotemark{h}} &
  \colhead{Mg\,II\tablenotemark{i}}       & \colhead{Ref.}        \\
  \colhead{}             &
  \colhead{}             & \colhead{}            &
  \colhead{EW (\AA)}     & \colhead{AI (km s$^{-1}$)}& \colhead{$d_{\rm abs}$} &
  \colhead{$v_{\rm max}$ (km s$^{-1}$)}  & \colhead{$v_{\rm min}$ (km s$^{-1}$)} & \colhead{$v_{avg}$ (km s$^{-1}$)} &
  \colhead{S/N}          & \colhead{}
}
\startdata
J000009.26+151754.5 &1.197 &52251-0751-354&10.15$\pm$0.86& 1282&0.80& -4104  & -126 & -1457  & 9.1  & T06,Z10\\
J002623.78+135523.5 &1.319 &52233-0753-002&24.81$\pm$1.71& 2547&0.45& -17685 & -6550& -11943 & 14.0 & T06,G09,Z10\\
J004610.17+000449.7 &0.826 &52199-0691-494&4.92 $\pm$1.20& 391 &0.33& -1846  & 567  & -699   & 10.3 & Z10 \\
J005722.48+010101.8 &1.146 &51783-0395-417&3.75 $\pm$0.66& 420 &0.40& -5249  & -3001& -4204  & 11.3 & T06,Z10\\
J010352.46+003739.7 &0.705 &51816-0396-471&5.34 $\pm$0.44& 556 &0.28& -11571 & -8433& -10071 & 23.7 & T06,G09,Z10\\
J011117.34+142653.6 &1.155 &51821-0423-310&4.51 $\pm$0.51& 510 &0.39& -4427  & -2652& -3614  & 20.3 & T06 \\
J013816.16+140431.6 &0.877 &51882-0426-494&6.25 $\pm$1.08& 657 &0.64& -5167  & -2850& -4050  & 6.6  & T06 \\
J013853.94-101125.7 &1.305 &52145-0663-217&7.86 $\pm$1.07& 702 &0.29& -10090 & -5311& -7844  & 16.8 & T06,G09,T06\\
J014012.07+130241.8 &1.188 &51882-0426-081&8.30 $\pm$1.21& 887 &0.55& -4994  & -2196& -3697  & 5.4  & T06,G09\\
J014534.36+143136.9 &0.636 &51820-0429-372&4.67 $\pm$0.84& 517 &0.50& -1888  & -29  & -1016  & 10.1 & T06 \\
J014950.96-010314.1 &1.082 &51793-0402-260&4.69 $\pm$0.50& 467 &0.25& -9342  & -5841& -7640  & 27.3 & T06,Z10 \\
J015636.03+135212.3a&1.129 &51900-0427-354&7.70 $\pm$0.83& 877 &0.73& -4044  & -2267& -3209  & 10.0 & T06,Z10\\
J015636.03+135212.3b&1.129 &51900-0427-354&5.06 $\pm$0.95& 456 &0.54& -6420  & -4793& -5603  & 10.0 & T06,Z10\\
J020105.14+000617.9 &1.205 &51871-0403-593&6.47 $\pm$0.76& 643 &0.34& -3684  & 715  & -1827  & 23.4 & T06,G09,Z10\\
J023102.49-083141.2 &0.587 &51908-0454-582&5.36 $\pm$0.91& 552 &0.30& -4939  & -1660& -3167  & 12.9 & T06,G09,Z10\\
\enddata
\tablenotetext{a}{The SDSS designation, hhmmss.ss+ddmmss.s (J2000). 
           Sources with the same designation but postfixed with a/b/c denote sources with multiple Mg\,II 
	 absorption components.}
\tablenotetext{b}{The SDSS spectra are designated as mjd-plate-fiberid.} 
\tablenotetext{c}{The total error $\sigma_{\rm total}$ as defined in \S2.2\,.}
\tablenotetext{d}{Absorption Index (AI) measured according to the definition of \citet{2010ApJ....714..367Z}.}
\tablenotetext{e}{Maximum depth of Mg\,II absorption troughs.}
\tablenotetext{f}{Maximum velocity of Mg\,II absorption troughs from the zero velocity.}
\tablenotetext{g}{Minimum velocity of Mg\,II absorption troughs from the zero velocity.}
\tablenotetext{h}{Weighted average velocity of Mg\,II BAL troughs.}
\tablenotetext{i}{Median signal to noise (S/N) in the range 2400--3000~\AA.}

{\it (This table is available in its entirety in a machine-readable form in the online
journal. A portion is shown here for guidance regarding its form and content.)}

\end{deluxetable}
\end{landscape}

\begin{landscape}

\pagestyle{empty}
\begin{deluxetable}{lrrrrrrrrrrrr}
\tablewidth{0pc}
\tablecolumns{13}
\tabletypesize{\footnotesize}
\tablecaption{The \hei\,$\lambda3889$ BAL Quasars \label{tab:dr5he1table}}
\tablehead{
  \colhead{SDSS Source\tablenotemark{a}}  &
  \colhead{$z$}          &  \colhead{Spectrum\tablenotemark{b}}   &
  \colhead{He\,I}       & \colhead{He\,I\tablenotemark{c}}      &
  \colhead{He\,I\tablenotemark{d}}       & \colhead{He\,I\tablenotemark{e}}      & \colhead{He\,I\tablenotemark{f}} &
  \colhead{He\,I\tablenotemark{g}}       \\
  \colhead{}             &
  \colhead{}             & \colhead{}            &
  \colhead{EW (\AA)}      & \colhead{$d_{\rm abs}$}           &
  \colhead{$v_{\rm max}$ (km s$^{-1}$)}  & \colhead{$v_{\rm min}$ (km s$^{-1}$)} &
  \colhead{$v_{\rm avg}$ (km s$^{-1}$)} &
  \colhead{S/N}          & \colhead{}
}
\startdata
000009.26+151754.5 & 1.197 &52251-0751-354 &2.57 $\pm$1.13 &0.35 &-1821  & -445   & -1024  & 4.7  \\
010352.46+003739.7 & 0.705 &51816-0396-471 &0.67 $\pm$0.29 &0.07 &-10548 & -9614  & -10089 & 30.3 \\
011117.34+142653.6 & 1.155 &51821-0423-310 &2.35 $\pm$0.54 &0.17 &-1598  & -290   & -1088  & 16.9 \\
013816.16+140431.6 & 0.877 &51882-0426-494 &3.01 $\pm$1.33 &0.23 &-4734  & -2687  & -3733  & 7.4  \\
013853.94-101125.7 & 1.305 &52145-0663-217 &2.61 $\pm$1.08 &0.25 &-7988  & -6707  & -7281  & 5.4  \\
014950.96-010314.1 & 1.082 &51793-0402-260 &0.51 $\pm$0.21 &0.08 &-7504  & -6897  & -7187  & 20.7 \\
020105.14+000617.9 & 1.205 &51871-0403-593 &1.77 $\pm$0.82 &0.14 &-2702  & -1673  & -2213  & 11.1 \\
023102.49-083141.2 & 0.587 &51908-0454-582 &1.06 $\pm$0.51 &0.13 &-2868  & -1771  & -2377  & 13.5 \\
023153.64-093333.6 & 0.555 &51908-0454-017 &4.00 $\pm$1.43 &0.32 &-1282  & -42    & -774   & 8.5  \\
023445.76-085908.5 & 1.275 &51909-0455-238 &3.20 $\pm$1.36 &0.33 &-8079  & -7001  & -7608  & 4.3  \\
024220.10-085332.7 & 0.800 &51910-0456-291 &2.00 $\pm$0.73 &0.19 &-3894  & -2665  & -3238  & 14.2 \\
073122.84+430241.0 & 0.975 &53312-1865-044 &3.92 $\pm$1.14 &0.30 &-3962  & -2664  & -3164  & 8.8  \\
074554.74+181817.0 & 1.054 &52939-1582-256 &1.03 $\pm$0.46 &0.14 &-1962  & -1207  & -1665  & 11.2 \\
075927.12+363431.5 & 0.983 &52238-0757-474 &1.24 $\pm$0.56 &0.16 &-4160  & -3341  & -3780  & 12.7 \\
080248.19+551328.8 & 0.664 &53384-1871-440 &2.64 $\pm$0.52 &0.24 &-684   &  767   &  52    & 23.0 \\
080934.64+254837.9 & 0.545 &52670-1205-588 &2.00 $\pm$0.31 &0.27 &-40    &  790   &  366   & 20.0 \\
\enddata
\tablenotetext{a}{The SDSS designation, hhmmss.ss+ddmmss.s (J2000). Sources with the same designation but postfixed with a/b
  is corresponding to that of \mgii\ in table \ref{tab:dr5mg2table}.}
\tablenotetext{b}{The SDSS spectrum is designated by its mjd-plate-fiber.}
\tablenotetext{c}{Maximum depth of He\,I$\lambda3889$ absorption trough.}
\tablenotetext{d}{Maximum velocity of He\,I$\lambda3889$ absorption trough from the zero velocity (3889.74 \AA).}
\tablenotetext{e}{Minimum velocity of He\,I$\lambda3889$ absorption trough from the zero velocity (3889.74 \AA).}
\tablenotetext{f}{Weighted average velocity of the He\,I$\lambda3889$ BAL troughs.}
\tablenotetext{g}{Median signal to noise (S/N) of 3500~$\sim$~4000~\AA.}
{\it (This table is available in its entirety in a machine-readable form in the online
journal. A portion is shown here for guidance regarding its form and content.)}

\end{deluxetable}
\end{landscape}

\clearpage

\begin{landscape}

\pagestyle{empty}
\begin{deluxetable}{lrrrrrrrrrrrr}
\tablewidth{0pc}
\tablecolumns{13}
\tabletypesize{\footnotesize}
\tablecaption{\hei$\lambda3189$~BAL Quasars \label{tab:dr5he13189table}}
\tablehead{
  \colhead{SDSS Name\tablenotemark{a}}  &
  \colhead{$z$}          &  \colhead{Spectrum\tablenotemark{b}}   &
  \colhead{He\,I$\lambda3189$}       & \colhead{He\,I$\lambda3189$\tablenotemark{c}}      &
  \colhead{He\,I$\lambda3189$\tablenotemark{d}}    & \colhead{He\,I$\lambda3189$\tablenotemark{e}} &
  \colhead{He\,I$\lambda3189$\tablenotemark{f}} &
  \colhead{He\,I$\lambda3189$\tablenotemark{g}}       \\
  \colhead{}             & 
  \colhead{}             & \colhead{}            &
  \colhead{EW (\AA)}      & \colhead{$d_{\rm abs}$}           &
  \colhead{$v_{\rm max}$ (km s$^{-1}$)}  & \colhead{$v_{\rm min}$ (km s$^{-1}$)} & 
  \colhead{$v_{\rm avg}$ (km s$^{-1}$)} &
  \colhead{S/N}          & \colhead{}
}
\startdata
J000009.26+151754.5 &  1.197 & 52251-0751-354 & 1.65 $\pm$0.68 & 0.24 & -1539 & -437  & -894  & 10.6\\
J020105.14+000617.9 &  1.205 & 51871-0403-593 & 0.76 $\pm$0.34 & 0.07 & -2968 & -1666 & -2335 & 25.0\\
J023102.49-083141.2 &  0.587 & 51908-0454-582 & 1.20 $\pm$0.59 & 0.10 & -3339 & -2107 & -2714 & 14.1\\
J023153.64-093333.6 &  0.555 & 51908-0454-017 & 1.15 $\pm$0.50 & 0.15 & -1343 & -241  & -766  & 9.4\\
J023445.76-085908.5 &  1.275 & 51909-0455-238 & 1.41 $\pm$0.64 & 0.16 & -7937 & -6724 & -7380 & 10.5\\
J024220.10-085332.7 &  0.800 & 51910-0456-291 & 1.20 $\pm$0.56 & 0.19 & -3682 & -2520 & -3153 & 11.4\\
J073122.84+430241.0 &  0.975 & 53312-1865-044 & 1.12 $\pm$0.48 & 0.13 & -4022 & -2725 & -3269 & 11.1\\
J074554.74+181817.0 &  1.054 & 52939-1582-256 & 0.77 $\pm$0.32 & 0.14 & -1886 & -1268 & -1583 & 16.5\\
J075927.12+363431.5 &  0.983 & 52238-0757-474 & 1.64 $\pm$0.67 & 0.14 & -3811 & -2033 & -2932 & 14.6\\
J080934.64+254837.9 &  0.545 & 52670-1205-588 & 1.42 $\pm$0.38 & 0.12 & 106   & 798   & 404   & 19.9\\
J081820.31+200046.1 &  0.986 & 53327-1925-040 & 0.72 $\pm$0.29 & 0.10 & -2126 & -1371 & -1735 & 17.8\\
J082231.53+231152.0 &  0.653 & 53317-1926-546 & 2.53 $\pm$0.35 & 0.15 & -1811 & 256   & -805  & 29.9\\
J083522.77+424258.3 &  0.807 & 52232-0762-085 & 0.59 $\pm$0.17 & 0.10 & -564  & 58    & -249  & 24.7\\
J084044.41+363327.8 &  1.235 & 52320-0864-149 & 1.01 $\pm$0.22 & 0.07 & -4448 & -2262 & -3325 & 43.8\\
J084824.14+034542.3 &  0.699 & 52224-0564-575 & 0.85 $\pm$0.41 & 0.15 & -1104 & -346  & -691  & 12.3\\
J085053.12+445122.4 &  0.542 & 52605-0897-359 & 1.55 $\pm$0.53 & 0.09 & -4716 & -2532 & -3632 & 31.5\\
J085215.65+492040.8 &  0.567 & 51993-0551-274 & 1.67 $\pm$0.51 & 0.22 & -1615 & -514  & -1051 & 10.8\\
J085357.88+463350.6 &  0.549 & 52238-0764-248 & 1.04 $\pm$0.37 & 0.08 & -4098 & -2458 & -3271 & 20.7\\
J093759.60+453801.8 &  0.429 & 52672-1202-330 & 1.13 $\pm$0.53 & 0.08 & -1715 & -131  & -901  & 14.2\\
J094355.00+560649.0 &  1.055 & 52253-0557-299 & 1.62 $\pm$0.52 & 0.18 & -1545 & -443  & -966  & 12.5\\
J102839.11+450009.4 &  0.585 & 52990-1429-401 & 0.49 $\pm$0.21 & 0.13 & -1054 & -503  & -806  & 33.3\\
J102943.75+370127.2 &  1.344 & 53415-1957-601 & 1.98 $\pm$0.78 & 0.37 & -997  & 176   & -452  & 5.1\\
J103255.37+083503.2 &  0.894 & 52734-1240-316 & 0.68 $\pm$0.29 & 0.07 & -3398 & -2782 & -3094 & 23.3\\
J104459.60+365605.1 &  0.701 & 53463-2090-329 & 0.40 $\pm$0.16 & 0.05 & -4413 & -3458 & -3936 & 30.3\\
J104705.08+590728.4 &  0.392 & 52427-0949-326 & 1.28 $\pm$0.49 & 0.18 & -1380 & 690   & -474  & 7.9\\
J104845.83+353110.7 &  1.011 & 53463-2090-131 & 0.95 $\pm$0.31 & 0.08 & -3365 & -2681 & -3030 & 23.8\\
J105404.72+042939.3 &  0.579 & 52338-0579-072 & 1.95 $\pm$0.69 & 0.26 & -1499 & -329  & -870  & 5.7\\
J105638.08+494943.3 &  1.148 & 52669-0876-110 & 0.45 $\pm$0.18 & 0.05 & -11546& -10681& -11134& 36.5\\
J111628.00+434505.8 &  0.801 & 53061-1364-095 & 1.04 $\pm$0.25 & 0.09 & -4636 & -3204 & -3901 & 22.8\\
J114043.62+532438.9 &  0.530 & 52734-1015-085 & 2.45 $\pm$0.67 & 0.15 & -2704 & -850  & -1873 & 11.6\\
J115553.87+012427.6 &  1.010 & 52051-0515-043 & 1.87 $\pm$0.51 & 0.12 & -5277 & -2755 & -4074 & 20.5\\
J120924.07+103612.0 &  0.395 & 52723-1229-489 & 1.77 $\pm$0.19 & 0.10 & -1548 & 383   & -662  & 40.7\\
J122614.97+120925.4 &  0.871 & 53120-1614-145 & 3.17 $\pm$0.40 & 0.28 & -1768 & -392  & -1164 & 18.1\\
J122703.19+505356.2 &  0.765 & 52644-0971-459 & 0.85 $\pm$0.33 & 0.09 & -1276 & -657  & -1001 & 10.2\\
J124300.87+153510.6 &  0.562 & 53502-1769-584 & 1.46 $\pm$0.36 & 0.10 & -1466 & 673   & -423  & 14.6\\
J130952.89+011950.6 &  0.547 & 52295-0525-250 & 0.69 $\pm$0.20 & 0.09 & -630  & 405   & -103  & 28.0\\
J135226.34+024549.4 &  1.222 & 52026-0530-588 & 1.71 $\pm$0.67 & 0.13 & -3879 & -2375 & -3077 & 13.1\\
J142647.47+401250.8 &  0.749 & 52797-1349-348 & 1.40 $\pm$0.55 & 0.14 & -3700 & -2401 & -2974 & 8.8\\
J142927.28+523849.5 &  0.595 & 52781-1327-343 & 2.53 $\pm$0.34 & 0.14 & -5431 & -2226 & -3954 & 23.0\\
J143144.91+391910.2 &  1.091 & 52797-1349-257 & 1.21 $\pm$0.41 & 0.10 & -1560 & -1147 & -1396 & 20.5\\
J150847.41+340437.7 &  0.788 & 53108-1385-173 & 1.46 $\pm$0.22 & 0.17 & -1202 & 109   & -606  & 27.3\\
J151053.63+574055.1 &  1.037 & 52079-0612-424 & 1.15 $\pm$0.31 & 0.09 & -5656 & -3889 & -4706 & 25.6\\
J153036.83+370439.2 &  0.417 & 53144-1401-367 & 2.73 $\pm$0.39 & 0.14 & -4621 & -2025 & -3263 & 19.2\\
J153646.88+515755.2 &  1.132 & 52378-0795-095 & 1.78 $\pm$0.58 & 0.13 & -5613 & -3641 & -4711 & 16.9\\
J155905.39+250047.2 &  0.933 & 53523-1655-085 & 1.15 $\pm$0.42 & 0.17 & -1811 & -159  & -1047 & 9.6\\
J160329.72+502722.2 &  0.638 & 52375-0620-126 & 2.16 $\pm$0.93 & 0.17 & -4599 & -2826 & -3596 & 9.7\\
J163255.46+420407.8 &  0.728 & 52379-0816-569 & 2.30 $\pm$0.76 & 0.18 & -2929 & -526  & -1858 & 11.3\\
J163656.84+364340.4 &  0.852 & 52782-1174-337 & 5.62 $\pm$0.95 & 0.39 & -5919 & -3266 & -4303 & 9.5\\
J171032.23+214451.3 &  0.867 & 53177-1689-069 & 1.46 $\pm$0.44 & 0.18 & -2041 & -45   & -1243 & 12.6\\
J210757.67-062010.6 &  0.644 & 52174-0637-610 & 2.41 $\pm$0.27 & 0.17 & -1741 & 464   & -509  & 30.2\\
J214118.78-070957.4 &  0.869 & 52824-1177-484 & 2.50 $\pm$0.91 & 0.25 & -3853 & -2349 & -3236 & 8.1\\
J232550.73-002200.4 &  1.011 & 51818-0383-142 & 0.89 $\pm$0.30 & 0.09 & -2419 & -1320 & -1846 & 21.9\\
\enddata
\end{deluxetable}
\end{landscape}
\clearpage

\begin{landscape}
\begin{deluxetable}{lrrrrrrrrrr}
\tablewidth{0pc}
\tablecolumns{11}
\tabletypesize{\scriptsize}
\tablecaption{Low-$z$ BAL AGNs \label{tab:lowzbal}}
\tablehead{
  \colhead{SDSS Name}      &  \colhead{$z$}      &  \colhead{Spectrum}      &
  \colhead{He\,I*$\lambda3889$}  & \colhead{He\,I*$\lambda3889$}    &  \colhead{He\,I*$\lambda3889$}  &
  \colhead{He\,I*$\lambda3889$}  & \colhead{He\,I*$\lambda3889$}    &  \colhead{He\,I*$\lambda3889$}  \\
  \colhead{ }       &  \colhead{ }      &  \colhead{ }      &
  \colhead{EW (\AA)}  & \colhead{$d_{\rm abs}$}    &  \colhead{$v_{\rm max}$(km~s$^{-1}$)}  &
  \colhead{$v_{min}$(km~s$^{-1}$)}  & \colhead{$v_{avg}$(km~s$^{-1}$)}    &  \colhead{S/N}  &
  }
\startdata
J013117.14+162535.5& 0.274 &55833-5137-627& 2.54$\pm$0.68 &0.27 &-1805 &-291 &	-897 &	15.19\\ 
J014219.00+132746.5& 0.267 &51820-0429-303& 3.17$\pm$0.35 &0.23 &-621  &553  &	-57  &  26.49\\ 
J075217.84+193542.2& 0.117 &52939-1582-612& 1.78$\pm$0.23 &0.10 &-661  &444  &	-94  &  43.72\\ 
J081527.29+445937.4& 0.268 &51877-0439-034& 3.06$\pm$0.65 &0.14 &-3398 &-515 &	-2180& 	17.89\\ 
J081542.53+063522.9& 0.244 &52934-1295-580& 1.98$\pm$0.72 &0.09 &-5106 &-1485& 	-3303& 	21.14\\ 
J081652.88+241612.5& 0.276 &52962-1585-178& 0.70$\pm$0.25 &0.06 &-7291 &-6278& 	-6790& 	28.11\\ 
J092247.03+512038.0& 0.161 &52247-0766-614& 0.95$\pm$0.32 &0.06 &-7421 &-6611& 	-7035& 	26.16\\ 
J093653.84+533126.8& 0.228 &52281-0768-473& 1.20$\pm$0.22 &0.11 &-1404 &-302 &	-851 &	34.85\\ 
J101325.43+221229.4& 0.275 &53739-2365-389& 0.69$\pm$0.36 &0.07 &-7524 &-6647& 	-7099& 	29.39\\ 
J105311.38+261522.6& 0.249 &53793-2357-388& 1.32$\pm$0.27 &0.16 &-1249 &-422 &	-826 &	20.85\\ 
J113804.88+400118.9& 0.292 &53466-1972-484& 1.48$\pm$0.50 &0.22 &-838  &-218 & 	-503 &	12.01\\ 
J130534.49+181932.8& 0.118 &54479-2603-443& 1.86$\pm$0.13 &0.24 &-1023 &-196 &	-563 &	34.66\\ 
J130712.33+340622.5& 0.148 &53476-2006-628& 2.21$\pm$0.30 &0.16 &-1856 &-66  &   -896& 	21.18\\ 
J134704.91+144137.6& 0.135 &53858-1776-612& 1.39$\pm$0.19 &0.16 &-7510 &-6498& 	-6996& 	41.80\\ 
J140136.63+041627.2& 0.164 &52339-0856-010& 0.84$\pm$0.27 &0.09 &-7479 &-6399& 	-6918& 	24.73\\ 
J153539.25+564406.5& 0.208 &52072-0617-352& 2.82$\pm$0.21 &0.20 &-1939 &-425 &	-984 &	25.86\\ 
J163459.82+204936.0& 0.129 &53224-1659-542& 4.70$\pm$0.49 &0.34 &-4498 &-2860& 	-3638& 	15.53\\ 
J215408.71-002744.4& 0.218 &52078-0371-106& 2.75$\pm$0.58 &0.17 &-2438 &-307 &	-1211& 	17.68\\ 
J222024.58+010931.2& 0.213 &52140-0375-361& 5.44$\pm$0.36 &0.42 &-1672 &-88  &   -724& 	31.15\\ 
\enddata
\end{deluxetable}
\end{landscape}
\clearpage

 \appendix

 \section{Appendix material}
 \subsection{Overview of existing Mg\,II BAL samples \label{app:parentsample}}

   As descried in \S\ref{subsec:parentsample}, we combine the \mgii\ BAL quasars in the T06, G09 and Z10 samples,
   from which a parent sample of 285 objects is built.
   Below we overview and compare those three samples.

   The traditional definition for \civ\ BAL, ``balnicity index'' (BI), was first introduced 
   by \citet{1991ApJ....373....23W}, which is defined as
   \begin{equation}
     BI \equiv \int^{\upsilon_{u} =25,000}_{\upsilon_{l}=3,000} [1 - \frac{f(-\upsilon)}{0.9}]~C\,d\upsilon ~,
   \end{equation} 
   where f($\upsilon$) is the continuum-normalized spectral flux. 
   The dimensionless factor $C$ is initially set to 0, and changes to 1 only when
   an absorption trough continuously dips 10\% or more below the estimated continuum over an interval of 2000 \kms.
   $BI = 0$ means no BAL,  while a positive BI indicates not only the presence
   of one or more BAL troughs but also the strength of the total absorption.
   Using BI-based criteria and spectroscopic data set from the SDSS DR5, G09 identified 5039 BAL quasars.
   In their work, the $\upsilon_{l}$ in the BI definition was simply revised to be 0~km~s$^{-1}$ instead of the
   blueshifted 3000~km~s$^{-1}$, to avoid omitting the BAL features at low outflow velocities.

   T06 identified 4784 quasars in the SDSS DR3 quasar catalog \citet{2005AJ....130..367S}.
   In their work, a different definition of BALs, termed ``intrinsic absorption index'' (AI), was devised; 
   the original definition of AI was first introduced by \citet{2002ApJS..141..267H} to select LoBAL quasars.
   AI is essentially an equivalent width (EW) to measure all the absorption flux within the absorption trough.
   The definition of T06 is as follow,
   \begin{equation}
      AI \equiv \int^{\upsilon_{u}=29,000}_{\upsilon_{l}=0} [1 - f(~-\upsilon)]~C^{'} d\upsilon
   \end{equation}
   where f($\upsilon$) is the normalized template-subtracted flux spectrum. $C^{'}$($\upsilon)=1$ in continuous
   troughs that exceed the minimum depth (10\%) and the minimum width (1000~km~s$^{-1}$); otherwise,
   $C^{'}$($\upsilon)=0$.

   Z10 have searched specifically for \mgii\ BAL quasars in the SDSS DR5 data set, and obtained 68
   \mgii\ BAL quasars at 0.4~$\leq$~z$\leq$~0.8, with a median S/N~$\geq$~7 pixel$^{-1}$ of the SDSS spectra.
   They used an AI-based selection criteria, with setting a maximum velocity of \mgii\ BALs to 20,000~km~s$^{-1}$,
   and a minimum velocity width of 1600~km~s$^{-1}$. The latter is a trade-off between the completeness and
   consistency with respect to the canonical definition (i.e., the velocity width of the trough being $\geq$
   2000~km~s$^{-1}$). 
   Because the published \mgii\ BAL quasars
   in Z10 are only at 0.4~$\leq~z~\leq$~0.8, we use the same pipeline of Z10 to enlarge this sample to 175 \mgii\
   BALs at redshifts of 0.4~$\leq~z~\leq$~1.35 in the SDSS DR5.

\begin{figure}[htbp]
        \centering
        \includegraphics[width=3.0in]{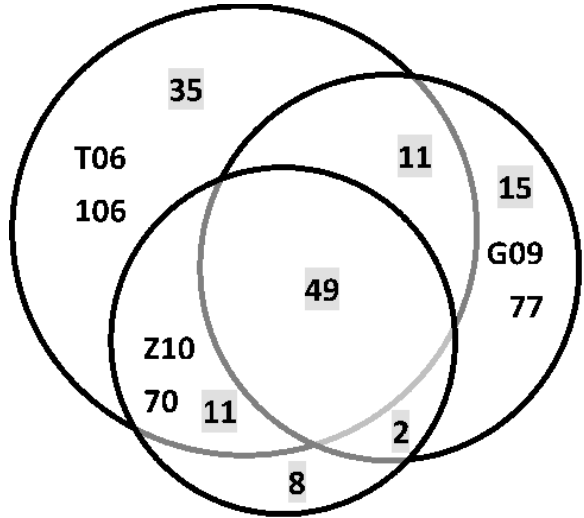}
    \caption{\footnotesize Venn diagram displaying the relationships of the \mgii\ BAL quasars 
    among the T06, G09 and the original Z10 samples. The sources of the three samples used here 
    are reselected according to the same BAL criteria based on our measurements 
    (see \S~\ref{subsec:parentsample}).
     The numbers with a shaded box denote the number of sources in every minimum regions enclosed by the arcs.
     \label{fig:venn}}
   \end{figure}

   For the ease of comparison, we select only the sources in those samples satisfying the following criteria:
      (1) from the SDSS DR3 only, which is common in the three samples;
      (2) with median spectral S/N > 7 pixel$^{-1}$;
      (3) the width of the continuous \mgii\ absorption trough $\geq 2000$ \kms,
      which is the strictest BAL width criterion among the three samples.
   The thus culled samples have 106 (T06), 77 (G09), and 70 (Z10) sources, respectively.
   Forty-nine sources are common in the three samples and the combined sample have 134 sources in total.
   Thirty-three percent objects of T06, 19.5 percent of G09, and 11.4 percent of Z10 are rejected by the 
   other two samples respectively (See Figure~\ref{fig:venn}).
   This indicates the degree of incompleteness of the samples caused by different selection procedures and criteria.

  \subsection{Reliability of the pair-matching method \label{app:Tests}}

     We perform a series of tests to evaluate the pair-matching method.
     The tests aim at three aspects: to identify the major factors influencing the BAL measurements, 
     to assess the efficiency of our selection procedure 
     (particularly the completeness of the thus selection sample),
      and to check the accuracy of our measurements of BAL parameters 
     (namely, assessing the total error $\sigma_{\rm tot}$ and the systematic error 
     $\sigma_{\rm sys}$ described in \S2.2).

    According to the BAL definitions (BI and AI; see Appendix \ref{app:parentsample}),
    the depth ($d_{\rm abs}$) and width ($W_{\rm abs}$) of absorption troughs are the two quantities 
    that impact the BAL measurement.
    Obviously, the spectral S/N is another influencing factor to any measurements.
    So in the following tests, we focus on evaluating the three factors.

    We carry out the tests using Monte Carlo simulations, first for the parent \mgii\ BAL sample and 
    then for the \hei\ BAL sample.
    We generate 200 spectra for each (S/N, $d_{\rm abs}$, $W_{\rm abs}$) grid,
    by randomly combining a BAL spectrum to an unabsorbed quasar spectrum.
    Firstly, for simplicity the BAL spectra are assumed to be of Gaussian profile, with the centroid values
    (namely the blueshifted offset relative to the emission line) following the distribution of the 
    parent \mgii\ BAL sample.
    The distributions in every $d_{\rm abs}$ and $W_{\rm abs}$ intervals also follow the observed 
    distributions of the parent \mgii\ BAL sample.
    The first $d_{\rm abs}$ bin is [0.05, 0.1],  and the rest range from 0.1 to 0.9 with a bin size of 0.1.
    The $W_{\rm abs}$ bins are centered at 1600, 2000, 4000, 6000, 8000 and 10000~km~s$^{-1}$.
    The blueshift of absorption line is fixed at -5000 km s$^{-1}$, since the $v_{\rm avg}$ 
    of \mgii\ and \hei\ absorption roughs are mainly between -10000 $\sim$ 0 km s$^{-1}$.
    We will also adopt actual absorption profiles later.
    The unabsorbed quasar spectra are selected from the SDSS DR7 data set, with decent spectral S/N.
    Gaussian noises may be added to the generated spectra to meet the S/N of every bins.
     The S/N bins are [5,10], [15,20], [25,30] and [35,40], covering the spectral S/N range of our sample.
     
\begin{figure}[htbp]
        \centering
    \includegraphics[width=6.0in]{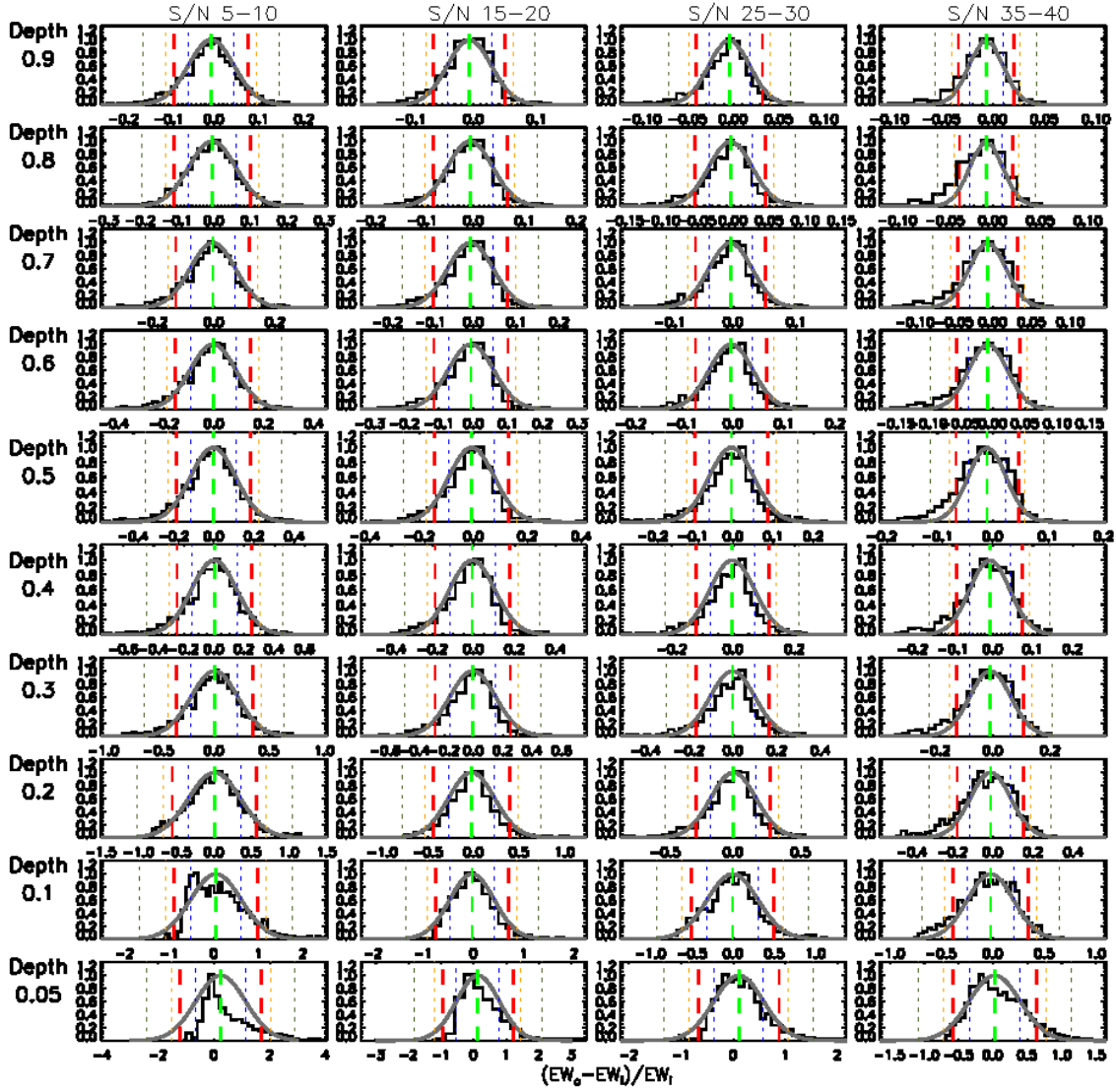}
    \caption{\footnotesize 
    Histograms of the relative difference between input and output EW of the \mgii\ BALs,
    $\frac{\rm EW_o - x_i}{\rm EW_i}$,
    in every 2-dimensional grids of spectral S/N and absorption depth ($d_{\rm abs}$)
    with the absorption blueshift being fixed to -5000~\kms.
    In every panels, the green dashed line denotes the mean value of the distribution, 
    which is almost zero; the blue,
    orange and gray dotted lines denote the measured 
    1-$\sigma$ (standard deviation), 2-$\sigma$ and 3-$\sigma$ of the distribution, respectively;
    the red dashed line denotes the 1.6-$\sigma$, which corresponds to the  90\% confidence level
    and is regarded as the relative total measurement uncertainty (namely $\sigma_{\rm tot}$/EW)
    of the pair-matching method.
    Note that, except for the panel of the shallowest width and poorest S/N, 
    the distributions are close to Gaussian (dark gray lines).
    \label{fig:grid_check}}
    \end{figure}

     Then we apply the pair-matching method to the simulated spectra, using the same procedure as described 
     in \S~\ref{subsec:mg2pairmethod}. Based on the fitting results of all the 
     (S/N, $d_{\rm abs}$, $W_{\rm abs}$) grids, we can evaluate the measurement uncertainties, the detection 
     ability, etc., as functions with the three influencing factors. 
     First of all, we analyze the fitting results of the 200 spectra in every single grids. 
     We consider the relative difference between the input absorption EW (EW$_{i}$) and the recovered 
     one (EW$_{o}$), $\frac{\rm EW_o - EW_i}{\rm EW_i}$. 
     These relative differences in every grids turn out to be normally distributed, which is illustrated 
     in Figure~\ref{fig:grid_check}.
     This figure, with the blueshift of absorption lines being fixed to be -5000~\kms,
     shows the histograms of $\frac{\rm EW_o - EW_i}{\rm EW_i}$ in the 2-dimensional parameter space 
     of S/N and  $d_{\rm abs}$.
     We can see that basically it is normally distributed. The other grids are all a similar situation. 
     The dispersion of $\frac{\rm EW_o - x_i}{\rm EW_i}$ is essentially the relative total error of the 
     absorption EW, namely $\sigma_{\rm tot}$/EW (see \S2.2).
     The 90\% confidence interval is taken as the measurement uncertainty of the pair-matching method,
     which is set by 
     matching the actually calculated $\sigma_{\rm tot}$ for the \mgii\ BAL quasars (see \S2.2 and Table~2).
     Such a confidence interval (namely 1.6-$\sigma$ instead of the commonly used 1-$\sigma$) should be caused by
     the effects of other---albeit minor---factors (e.g., the profile shape of absorption troughs).
     The mean of the derived systematic errors ($\sigma_{\rm sys}$) of the 200 fittings of a grid 
     is regarded as the typical $\sigma_{\rm sys}$ of that grid.  
     The detection probability is straightforward to define as the recovered fraction of BALs out of the 200 
     spectra in every grids.
     
\begin{figure}[htbp]
        \centering
    \includegraphics[width=3.2in]{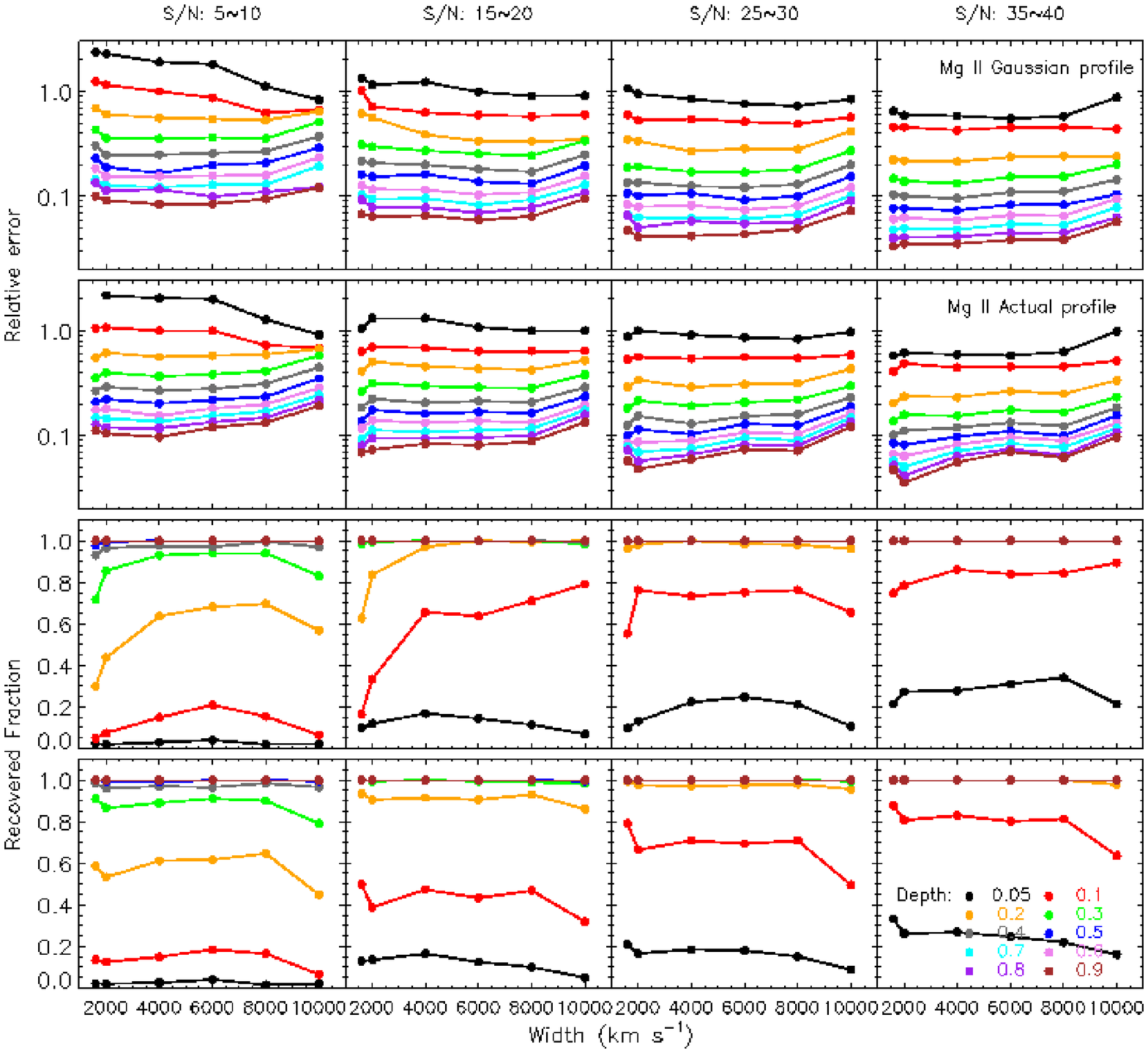}
    \includegraphics[width=3.2in]{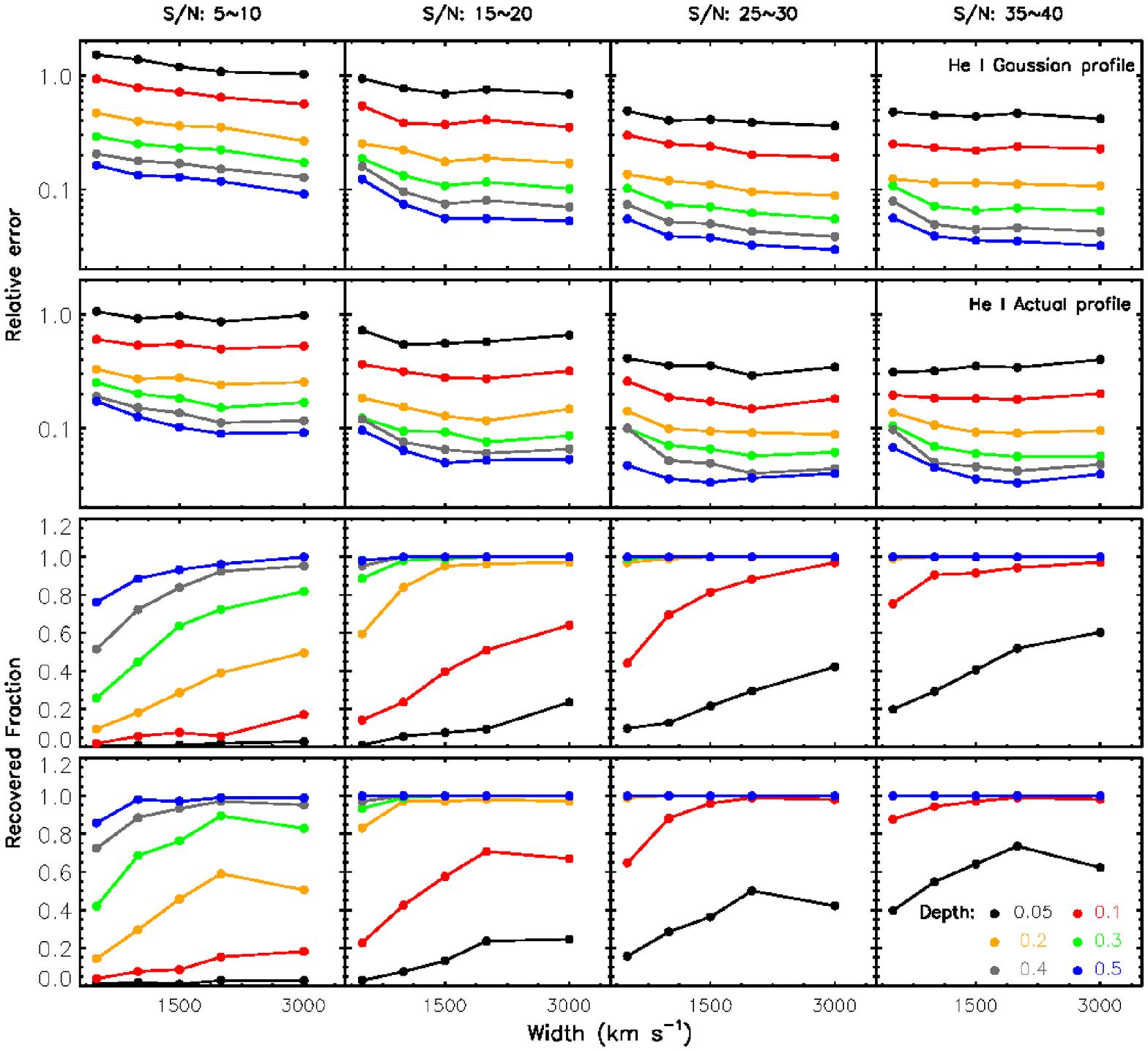}
    \caption{\footnotesize 
	 Relative total errors (measure of $\sigma_{rm tot}$/EW; the top 2 rows) 
	 and the recovered fraction (the bottom 2 rows) as functions of 
	 the spectra S/N, and depth and width of the BALs,
	 for \mgii\ ({\it left}) and for  \hei\,$\lambda3889$ ({\it right}).
	 Rows 1 and 3 from the top are for the tests with simulated BAL of Gaussian profiles;
	 the other 2 rows are of actual BAL profiles.
	 Different absorption depths are denoted with different colors.
      \label{fig:testresults}}
   \end{figure}

     The relative total error and detection probability as functions of S/N, $d_{\rm abs}$ and 
     $W_{\rm abs}$ are shown in Figure~\ref{fig:testresults}  (left panel) for \mgii\ BALs.
     We can clearly see that the relative total error decreases, and the detection probability increases, 
     with both increasing S/N and $d_{\rm abs}$, while the effect of $W_{\rm abs}$ is not so significant.
     Therefore we can reasonably conclude that the spectral S/N and the absorption depth are the two most 
     principal factors influencing the pair-matching method.
     That is why we consider the S/N--$d_{\rm abs}$ plane only, by collapsing the $W_{\rm abs}$ dimension, 
     in Figure~\ref{fig:sndepth}.
     The detection probability of the pair-matching method is almost complete for \mgii\ BALs with S/N $>~35$ or
     $d_{\rm abs} >~0.5$.
     Figure~\ref{fig:testresults} shows 
     $\sigma_{sys}$ and $\sigma_{sys}$/EW as functions with S/N, $d_{\rm abs}$ and $W_{\rm abs}$.
     $\sigma_{sys}$ increases with $W_{\rm abs}$ in every grids of (S/N, $W_{\rm abs}$), yet is less sensitive 
     to S/N and $d_{\rm abs}$. 
     However, similar to what the relative total error, $\sigma_{sys}$/EW bears a positive dependence on 
     $d_{\rm abs}$, although the dependence on S/N is not so significant.

\begin{figure}[htbp]
     \centering
    \includegraphics[width=3.2in]{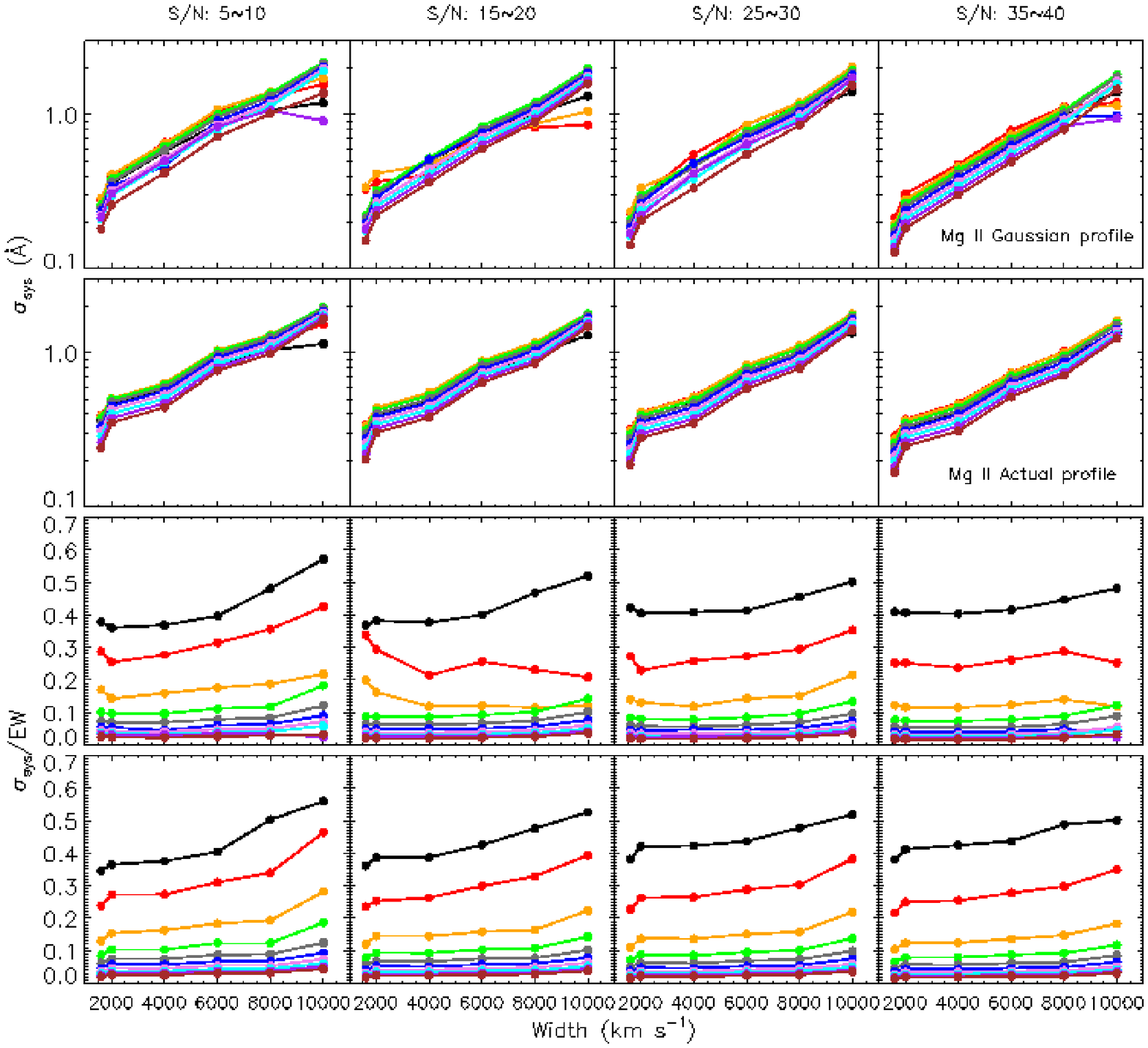}
    \includegraphics[width=3.2in]{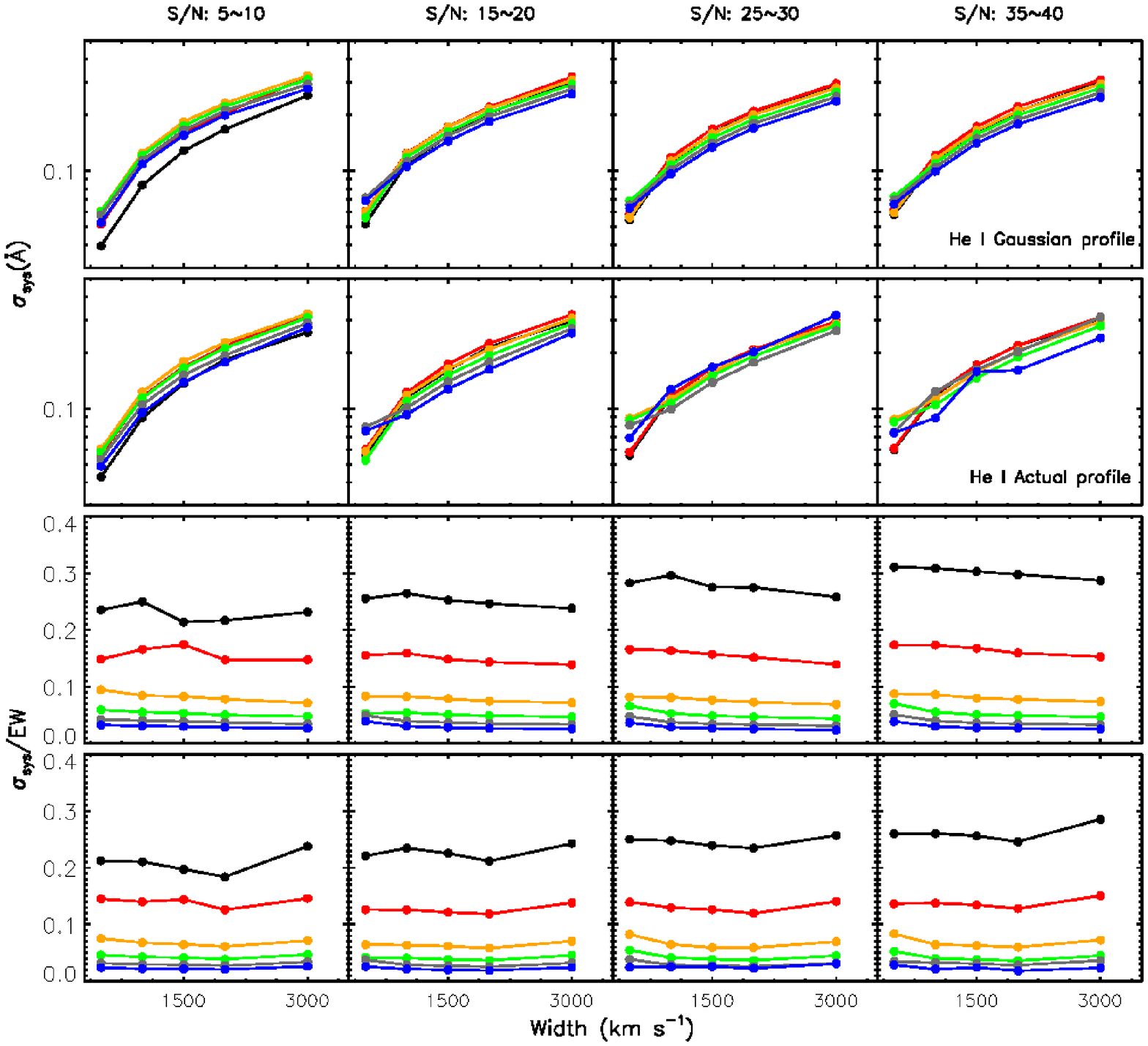}
    \caption{\footnotesize
      The same as Figure~\ref{fig:testresults}, but for the 
      systematic error ($\sigma_{sys}$) and the relative systematic errors ($\sigma_{sys}$/EW) .
      \label{fig:syserror} }
  \end{figure}

\begin{figure}[htbp] 
   \centering
   \includegraphics[width=6.4in]{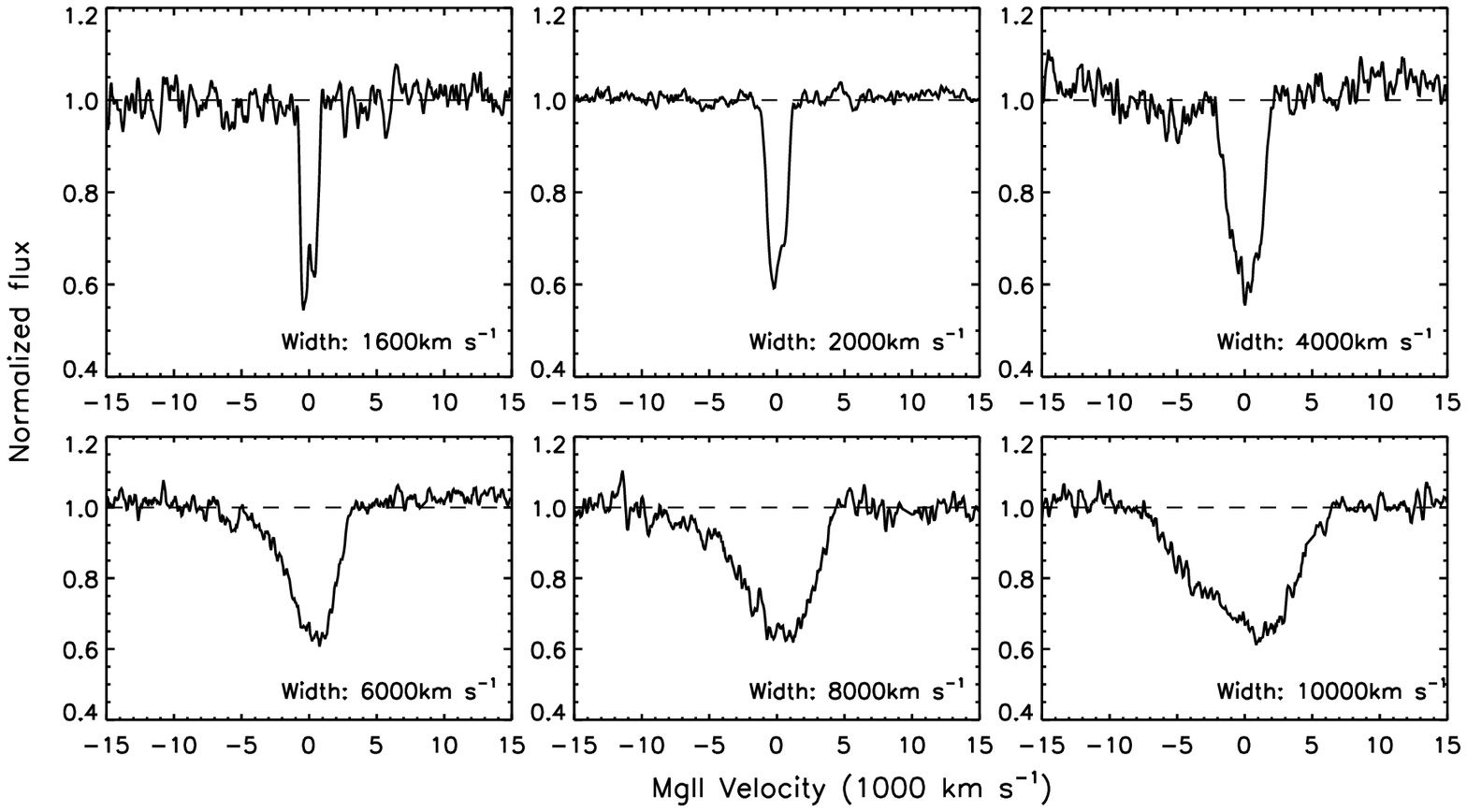}
   \includegraphics[width=6.4in]{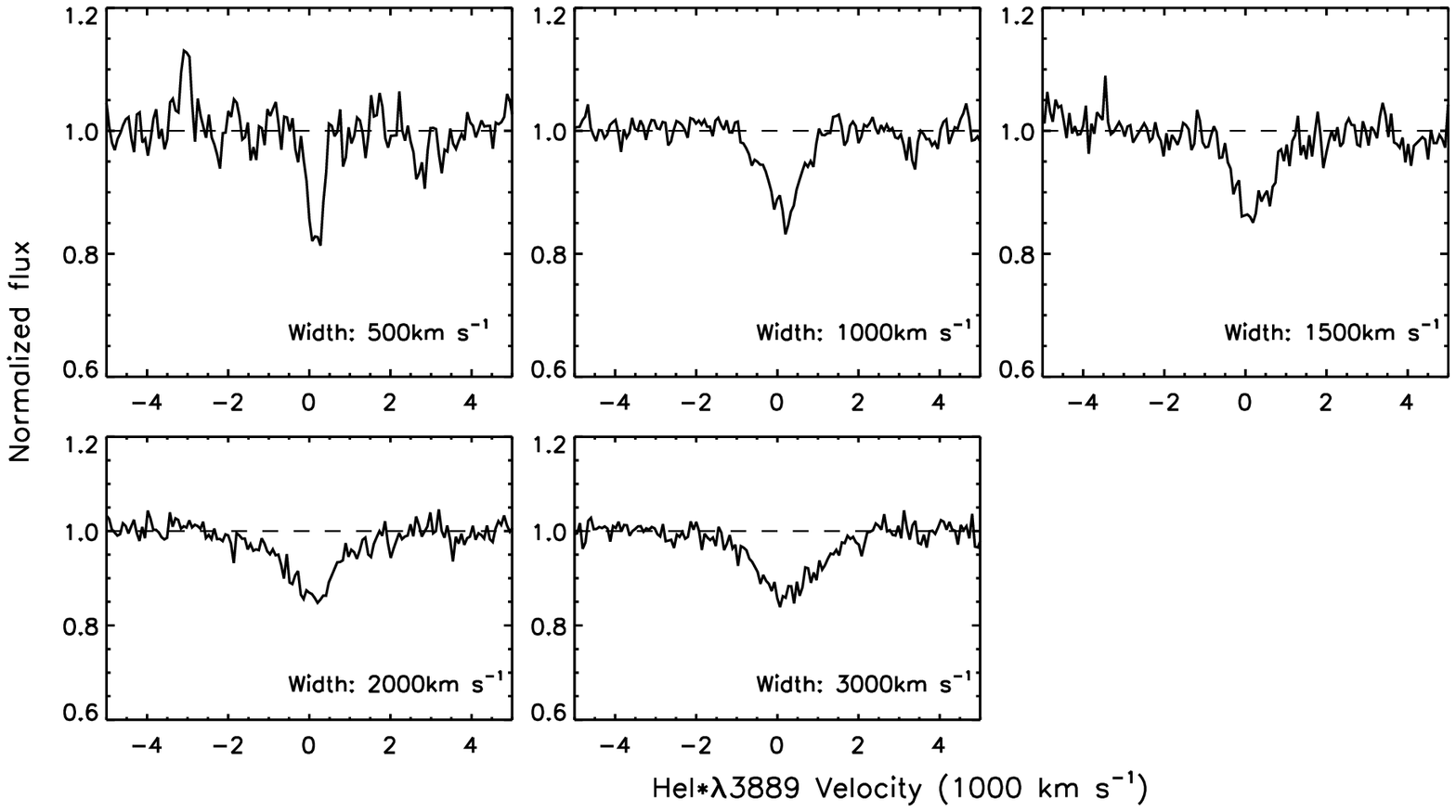}
   \caption{\footnotesize Composite normalized absorption-line spectra for \mgii\ ({\it top}) 
   and \hei$\lambda3889$ ({\it bottom}),
   serving as templates of BAL profiles of different absorption widths in the tests. \label{fig:mg2composite}}
   \end{figure}

     The above conclusions are based on the simulated spectra with BALs assumed to be of Gaussian profile.
     Now we build simulated spectra with actual absorption lines from the parent \mgii\ sample.
     The BAL spectra are normalized and the lines are corrected into zero velocity shift.
     Consistent with the above $W_{\rm abs}$ bins, we 
     categorize the \mgii\ BAL spectra into six subsamples.
     Then we
     build an arithmetic mean composite spectrum of normalized \mgii\ BALs for each,
     using the composite spectrum method of \citet{2010ApJ...721L.143D}.
     The composite spectra  have very high S/N, and serve as the templates of \mgii\ BAL profiles for the 
     six $W_{\rm abs}$ bins (see the upper panel of Figure~\ref{fig:mg2composite}).
     Following the procedure in the case of Gaussian profile, a series of simulated spectra are generated 
     for all the 3-dimensional grids of (S/N, $d_{\rm abs}$, $W_{\rm abs}$), and the pair-matching method 
     is applied to them.
     The results are also shown in Figures~\ref{fig:testresults}, \ref{fig:syserror} and \ref{fig:sndepth}.
     The general trends are almost the same as in the case of Gaussian profile. 
     This confirm that the most principal influencing factors to the pair-matching method are S/N and $d_{\rm abs}$.

     We repeat the above tests for for \hei\,$\lambda3889$ absorption.
     To fully cover the measured parameter space of the \hei\ BALs, 
     the grids in the tests are 
     are 0.05, 0.1, 0.2, 0.3, 0.4, 0.5 for $d_{\rm abs}$,
     500, 1000, 1500, 2000, 3000 \kms\ for $W_{\rm abs}$. 
     The centroid of the \hei\ absorption line is simply fixed to be the blueshifted 5000~\kms, 
     the typical blueshift of the \mgii\ BALs of the parent sample.
     The test results are illustrated in      
     Just like the tests for \mgii, both Gaussian profiles and actual profiles 
     (Figure~\ref{fig:mg2composite}, bottom panel) for the simulated BAL spectra are tried.
     The test results are illustrated in Figures~\ref{fig:sndepth}, \ref{fig:testresults} and \ref{fig:syserror},
     which are similar to the case for  \mgii\ BAL quasars.
     
     As to testing the possibility of false identification of  \hei\,$\lambda3889$ BALs , 
     the most reliable way is to observe other \hei\ absorption troughs at corresponding velocities.
     In our \hei\,$\lambda3889$ BAL sample, the \hei$\lambda3189$ absorption line is detected in over half the sample
     from the SDSS spectra.
     Moreover, in our follow up NIR observations, 5 of 5 \hei\,$\lambda3889$ BAL quasars (100\%) have 
     the expected \hei$\lambda10830$ BALs (see also \S2.3 and Appendix A.6).
     Thus the false identification rate should be very low.
     More NIR observations of \hei$\lambda10830$ are needed to investigate this issue. 
     
  \subsection{Unusual quasars not included in our parent sample \label{app:unusual}}
\begin{figure}[tbp]
    \centering
      \includegraphics[width=6.4in]{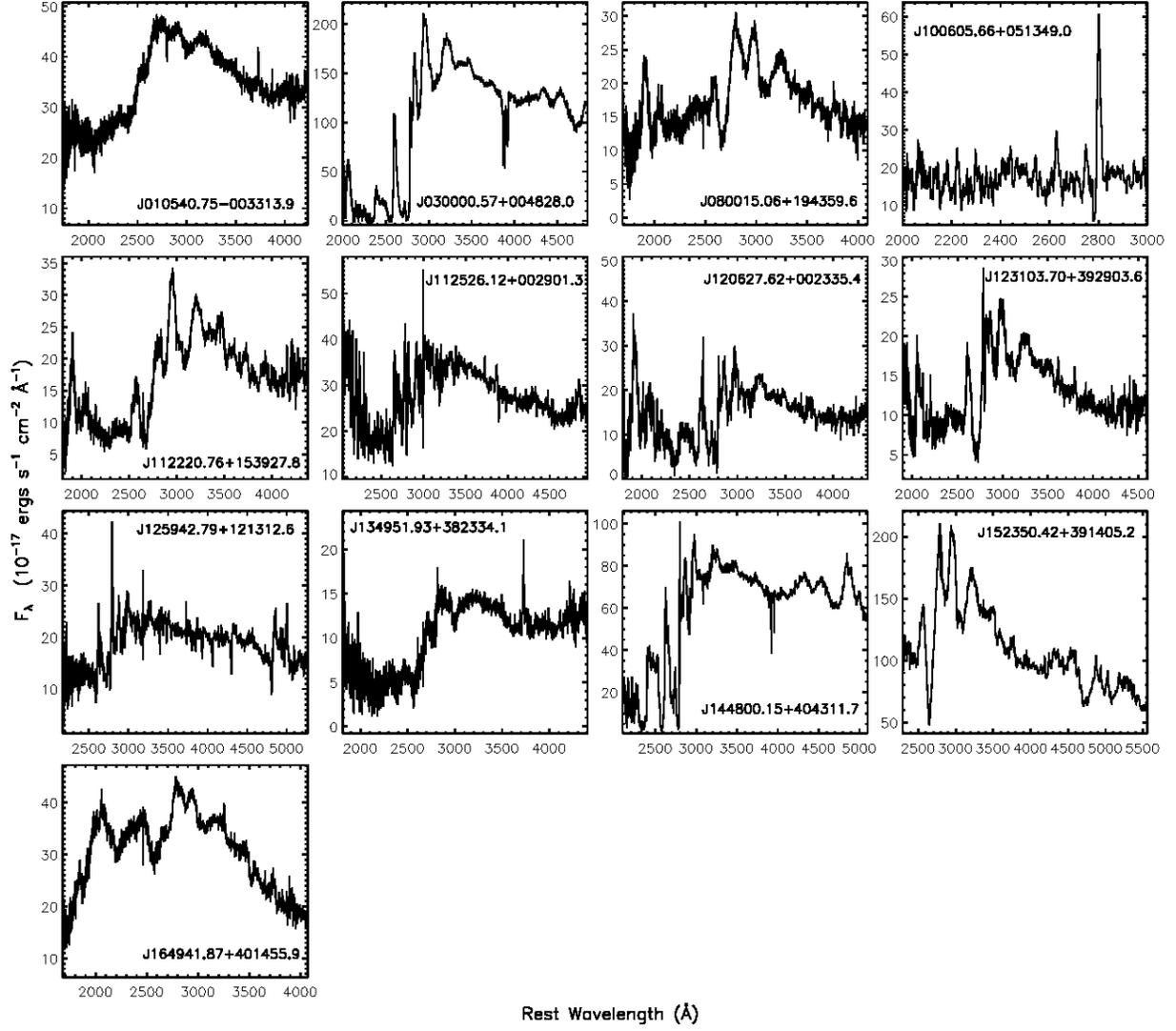}
      \caption{\footnotesize Mg\,II BAL quasars with unusual spectral features. These spectra are smoothed by
        5-pixel boxcar.\label{fig:unusualfig}}
  \end{figure}

  Thirteen sources compiled as \mgii\ BAL quasars  in the T06, G09 or Z10 samples
  show unusual spectra features. It is difficult
  to use our pair-matching method to recover their unabsorbed spectra. 
  Here we list those quasars in Table~\ref{tab:unusual}, and their SDSS spectra are shown in 
  Figure~\ref{fig:unusualfig}.
  Among them, J010540.75$-$003313.9, J030000.57$+$004828.0, and J112526.12$+$002901.3 have 
  been investigated in detail by \citet{2002ApJS..141..267H};
  for J112526.12$+$002901.3,    
  we have studied recently the physical conditions of its outflow gas 
  using the method mentioned in \S4.2 \citet{2014ApJ....578L..31T}.
 
  \begin{deluxetable}{lllllll}
  \centering
  \tablewidth{0pc}
  \tablecolumns{8}
  \tabletypesize{\footnotesize}
  \tablecaption{Unusual Mg\,II BAL quasars not included in the parent sample \label{tab:unusual}}
  \tablehead{
    \colhead{SDSS NAME}  & \colhead{$z$} &
    \colhead{MJD}  & \colhead{Plate} & \colhead{Fiberid} & \colhead{He\,I* Detection} &
    \colhead{Source} 
    }
  \startdata
    J010540.75-003313.9 & 1.179 & 51816 & 0396 & 156 & no  & T06                 \\
    J030000.57+004828.0 & 0.900 & 51816 & 0410 & 621 & yes & T06,G09       \\
    J080015.06+194359.6 & 1.252 & 53315 & 1922 & 474 & no  & G09,Z10        \\
    J100605.66+051349.0 & 0.968 & 52641 & 0996 & 243 & yes & T06,G09        \\
    J112220.76+153927.8 & 1.109 & 53383 & 1753 & 404 & no  & G09                \\
    J112526.12+002901.3 & 0.863 & 51614 & 0281 & 427 & yes & T06,G09        \\
    J120627.62+002335.4 & 1.114 & 51999 & 0286 & 499 & yes & T06,G09         \\
    J123103.70+392903.6 & 1.004 & 53466 & 1992 & 529 & no  & G09                \\
    J125942.79+121312.6 & 0.752 & 53473 & 1695 & 075 & yes & G09,Z10        \\
    J134951.93+382334.1 & 1.094 & 53460 & 2014 & 477 & no  & G09                \\
    J144800.15+404311.7 & 0.808 & 53119 & 1397 & 198 & yes & G09                \\
    J152350.42+391405.2 & 0.658 & 52765 & 1293 & 234 & no  & T06,G09,Z10   \\
    J164941.87+401455.9 & 1.266 & 52050 & 0630 & 482 & no  & T06,G09          \\
  \enddata
  \end{deluxetable}

\subsection{UV and optical measurements for the parent sample \label{app:emissionline}}
 The analysis for the near-UV spectra is performed with the routine described detailedly in 
 \citet{2009ApJ....707.1334W}.
 Here the pseudocontinuum consists a local power-law continuum, an Fe\,II template, but also
 an additional component for the Balmer continuum, which is assumed to be produced in partially optically
 thick clouds with a uniform temperature.
 The Fe\,II emission is modeled with the semi-empirical template for I~Zw~1 generated by 
 \citet{2006ApJ...650...57T}.
 To match the width and possible velocity shift of the Fe\,II lines, the template is convolved with a Gaussian
 and is shifted in velocity space.
 Each of the lines of the Mg\,II doublet is modeled with two components, one broad and the other narrow.
 The broad component is fit with a truncated, five-parameter Gauss-Hermite series; a single Gaussian is used
 for the narrow component. 
 The fitting results for the entire \mgii\ BAL sample are summarized in Table~\ref{tab:uvpar}.

 For the optical spectra of the objects at $z<0.8$, the fitting is performed with the routine 
 described detailedly in \citet{2008MNRAS.383..581D}.
 In the spectra of these BAL quasars the starlight is negligible.
 As the broad emission lines, particularly the Fe\,II multiplets, are so broad and strong that they are in
 blend and essentially leave no line-free wavelength regions, so we fit the nuclear continuum, the Fe\,II
 multiplets, and other emission lines simultaneously.
 The AGN continuum is approximated by a single power law with a free index.
 The optical Fe\,II emission is modeled with two separate sets of analytic spectral templates, one for the
 broad-line system and the other for the narrow-line system, constructed from measurements of I~Zw~1 by
 \citet{2004A&A...417..515V}. Within each system, the respective set of Fe\,II lines is assumed to have no
 relative velocity shifts and the same relative strengths as those in I~Zw~1.
 Emission lines are modeled as multiple Gaussians. Following \citet{2011ApJ...736...86D}, we assume that the broad
 Fe\,II lines have the same profile as the broad \hb.
 The results of the optical fittings for the \mgii\ BAL quasars at $z < 0.8$ are summarized in
 Table~\ref{tab:opticalpar}.

\begin{landscape}
\pagestyle{empty}
\begin{deluxetable}{lrrrrrrrrrr}
\centering
\tablewidth{0pc}
\tablecolumns{11}
\tabletypesize{\footnotesize}
\tablecaption{UV continuum and emission-line parameters of the entire parent sample \label{tab:uvpar}}
\tablehead{
  \colhead{SDSS Name}  & \colhead{$z$} &
  \colhead{$\beta_{[3k-4k]}\tablenotemark{a}$}  & \colhead{log$L3000$} & \colhead{log$F_{2500}$} & 
  \colhead{log$F$(UV Fe\,II)} &
  \colhead{FWHM(Mg\,II$^{b}$)}\tablenotemark{b} & \colhead{log$F$(Mg\,II$^{b})$}\tablenotemark{c} & 
  \colhead{log$F$(Mg\,II$^{n})$}\tablenotemark{d} &
  \colhead{log$M_{BH}$}\tablenotemark{e} & \colhead{$L/L_{Edd}$} \\
  \colhead{}            & \colhead{}     &
  \colhead{}            & \colhead{erg~s$^{-1}$} & \colhead{ergs~s$^{-1}$~cm$^{-2}$ \AA$^{-1}$} &
  \colhead{ergs~s$^{-1}$~cm$^{-2}$}&
  \colhead{km~s$^{-1}$} & \colhead{ergs~s$^{-1}$~cm$^{-2}$} & \colhead{ergs~s$^{-1}$~cm$^{-2}$} &
  \colhead{$\msun$} & \colhead{} }
\startdata
 J000009.26+151754.5 & 1.197 & -1.30 & 45.4  & -15.93 & -14.40 & 2435 & -14.68 & -15.61 & 8.38 & 0.51\\
 J002623.78+135523.5 & 1.319 & 5.03  & 45.39 & -15.95 & -13.47 & 2812 & -14.38 & -15.92 & 8.50 & 0.46\\
 J004610.17+000449.7 & 0.826 & -1.65 & 44.96 & -15.93 & -14.48 & 7771 & -14.39 &        & 9.16 & 0.03\\
 J005722.48+010101.8 & 1.146 & -1.27 & 45.52 & -15.74 & -13.91 & 3239 & -14.36 & -15.94 & 8.69 & 0.33\\
 J010352.46+003739.7 & 0.705 & -0.75 & 45.43 & -15.34 & -13.08 & 2747 & -13.66 & -15.77 & 8.50 & 0.39\\
 J011117.34+142653.6 & 1.155 & -0.25 & 45.76 & -15.64 & -13.77 & 4131 & -14.19 & -15.91 & 9.03 & 0.20\\
 J013816.16+140431.6 & 0.877 & -1.17 & 44.93 & -15.98 & -13.97 &      &        & -16.69 &      &    \\
 J013853.94-101125.7 & 1.305 & -1.56 & 45.47 & -15.90 & -13.69 & 2081 & -14.39 & -15.53 & 8.28 & 0.84\\
 J014012.07+130241.8 & 1.188 &       & 45.20 & -15.98 & -13.90 & 3036 & -14.65 & -15.76 & 8.47 & 0.35\\
 J014534.36+143136.9 & 0.636 & -0.94 & 44.63 & -15.99 & -13.93 & 1984 & -14.93 &        & 7.80 & 0.35\\
 J014950.96-010314.1 & 1.082 & -1.44 & 45.85 & -15.26 & -13.14 & 2622 & -14.01 &        & 8.68 & 0.90\\
 J015636.03+135212.3 & 1.129 & -1.40 & 45.40 & -15.90 & -13.59 & 3055 & -14.25 & -16.12 & 8.58 & 0.29\\
 J020105.14+000617.9 & 1.205 & -2.23 & 45.88 & -15.40 & -13.66 & 6453 & -13.88 & -16.33 & 9.48 & 0.14\\
 J023102.49-083141.2 & 0.587 & -1.70 & 44.83 & -15.65 & -13.49 & 2755 & -14.36 & -15.60 & 8.19 & 0.25\\
 J023153.64-093333.6 & 0.555 & -0.74 & 44.55 & -15.96 & -14.20 & 4724 & -14.83 & -16.22 & 8.51 & 0.05\\
\enddata
\tablenotetext{a}{Continuum slope between $\sim$ 3000 \AA\ and $\sim$ 4000\AA.}
\tablenotetext{b}{Line widths (FWHM) of broad component of \mgii\ emission line.}
\tablenotetext{c}{Flux of broad component of \mgii\ emission line.}
\tablenotetext{d}{Flux of narrow component of \mgii\ emission line.}
\tablenotetext{e}{Mass of SMBH derived by width of \mgii\ emission lines, and the formula we used is the 
  equation 10 in \citet{2009ApJ....707.1334W}.}
{\it (This table is available in its entirety in a machine-readable form in the online
journal. A portion is shown here for guidance regarding its form and content.)}

\end{deluxetable}
\end{landscape}
\clearpage

\begin{landscape}
\pagestyle{empty}
\begin{deluxetable}{lrrrrrrrrrr}
\centering
\tablewidth{0pc}
\tablecolumns{11}
\tabletypesize{\footnotesize}
\tablecaption{Optical continuum and emission-line parameters of the $z<0.8$ objects in the parent sample \label{tab:opticalpar}}
\tablehead{
  \colhead{SDSS NAME}  & \colhead{$z$} &
  \colhead{log$L5100$} & \colhead{$FWHM(\rm H\beta^{b}$)}\tablenotemark{a} & 
  \colhead{log$F(\rm H\beta^{b})$}\tablenotemark{b} & \colhead{log$F(\rm H\beta^{n})$}\tablenotemark{c} &
  \colhead{FWHM([O\,III])}     & \colhead{log$F(\rm [O\,III])$}   & \colhead{log$F(\rm Optical Fe\,II)$} &
  \colhead{log$M_{\rm BH}$}\tablenotemark{d} & \colhead{$L/L_{Edd}$} \\
  \colhead{}  &  \colhead{} &
  \colhead{erg~s$^{-1}$} & \colhead{km~s$^{-1}$} & \colhead{ergs~s$^{-1}$~cm$^{-2}$} &
  \colhead{ergs~s$^{-1}$~cm$^{-2}$} &
  \colhead{km~s$^{-1}$} & \colhead{ergs~s$^{-1}$~cm$^{-2}$} & \colhead{ergs~s$^{-1}$~cm$^{-2}$} &
  \colhead{$\msun$} & \colhead{}}
\startdata
 J010352.46+003739.7 & 0.705 & 44.97 & 4964 & -13.57 & -15.54 & -13.91 & 2913 & -13.66 & 8.63 & 0.15\\
 J014534.36+143136.9 & 0.636 & 44.27 & 7005 & -14.37 & -15.81 & -15.03 & 803  & -14.60 & 8.45 & 0.05\\
 J023102.49-083141.2 & 0.587 & 44.18 & 3812 & -14.23 & -16.36 & -14.78 & 1058 & -14.16 & 8.11 & 0.08\\
 J023153.64-093333.6 & 0.555 & 44.13 & 5557 & -14.27 & -16.08 & -15.23 & 707  & -14.50 & 8.27 & 0.05\\
 J024220.10-085332.7 & 0.800 & 44.43 & 5939 & -14.08 & -15.37 & -14.40 & 354  & -14.50 & 8.45 & 0.07\\
 J025026.66+000903.3 & 0.597 & 44.45 & 8448 & -14.28 & -16.15 & -14.53 & 521  & -14.58 & 8.62 & 0.05\\
 J033438.28-071149.0 & 0.635 & 44.92 & 4620 & -13.53 &        & -14.18 & 527  & -13.65 & 8.58 & 0.16\\
 J080248.19+551328.8 & 0.664 & 44.75 & 2264 & -14.09 & -14.98 & -14.34 & 782  & -13.84 & 8.15 & 0.28\\
 J080934.64+254837.9 & 0.545 & 44.54 & 8778 & -13.88 & -15.38 & -14.62 & 719  & -14.20 & 8.69 & 0.05\\
 J081655.34+074311.5 & 0.645 & 44.37 & 3964 & -14.19 & -16.07 & -14.56 & 992  & -14.23 & 8.23 & 0.10\\
 J082231.53+231152.0 & 0.653 & 44.86 & 9090 & -13.76 & -15.05 & -13.81 & 598  & -15.24 & 8.87 & 0.07\\
 J083000.35+343238.7 & 0.740 & 44.44 & 2030 & -14.52 &        & -15.24 & 241  &        & 7.94 & 0.22\\
 J083525.98+435211.3 & 0.568 & 44.77 & 2518 & -14.20 & -15.01 & -14.84 & 747  & -13.67 & 8.21 & 0.26\\
 J083613.23+280512.1 & 0.743 & 44.33 & 2004 & -14.37 &        & -15.25 & 388  & -14.54 & 7.88 & 0.20\\
 J084716.03+373218.0 & 0.453 & 44.62 & 2539 & -13.63 & -14.88 & -13.74 & 484  & -14.21 & 8.14 & 0.22\\
\enddata

{\it (This table is available in its entirety in a machine-readable form in the online
journal. A portion is shown here for guidance regarding its form and content.)}
\tablenotetext{a}{Line widths (FWHM) of broad component of \hb\ emission line.}
\tablenotetext{b}{Flux of broad component of \hb\ emission line.}
\tablenotetext{c}{Flux of narrow component of \hb\ emission line.}
\tablenotetext{d}{Mass of SMBH derived by width of \hb\ emission lines, and the formula we used is the 
  equation 11 in \citet{2009ApJ....707.1334W}.}

\end{deluxetable}
\end{landscape}

\subsection{Sources with multiple-epoch spectroscopic observations in SDSS and BOSS \label{app:multiobs}}
\begin{figure}[htbp]
  \centering
  \includegraphics[width=6.4in]{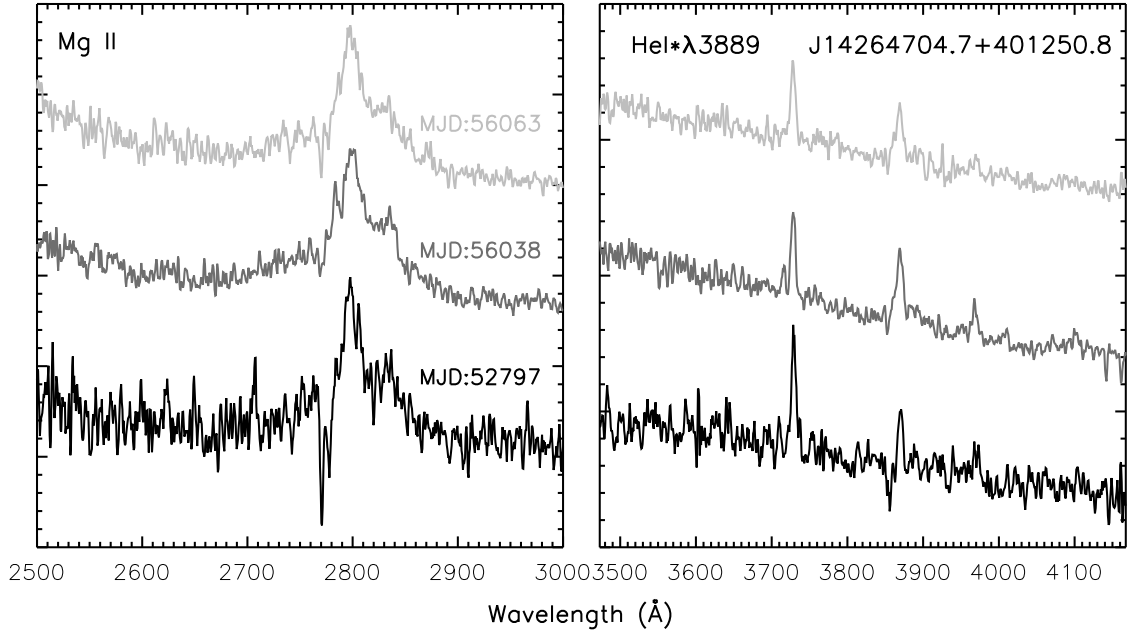}
  \caption{\footnotesize Multi-epoch spectra of SDSS~J14264704.7$+$401250.8 in the rest frame,
  in the Mg\,II region (left) and the He\,I*$\lambda3889$ region.
    Spectra are smoothed with a two-pixel boxcar.
    Note that both the Mg\,II and He\,I* absorption become ever weaker with time. \label{fig:j1426}}
\end{figure}

\begin{figure}[htbp] 
    \centering
    \includegraphics[width=6.4in]{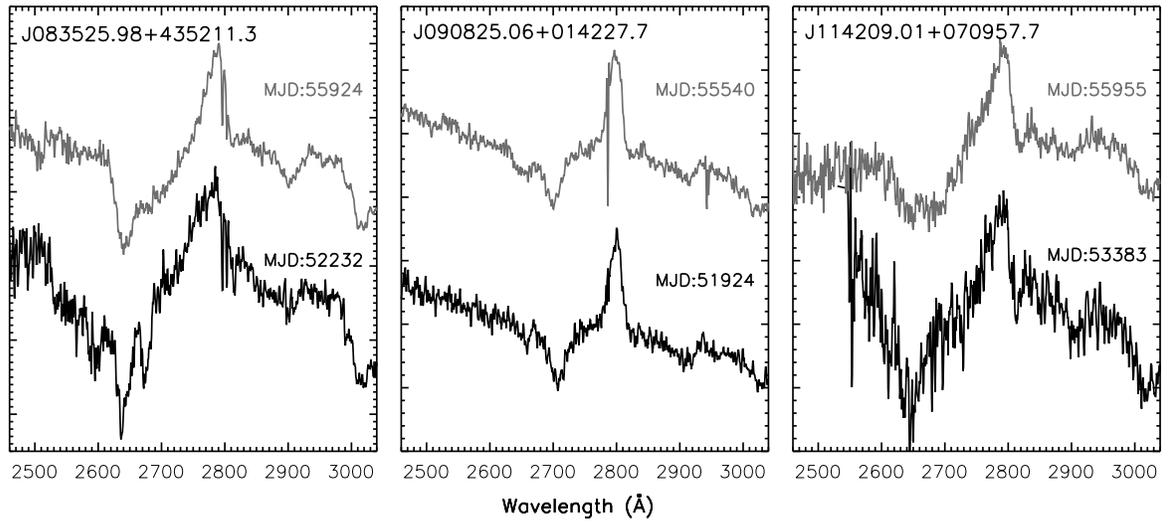}
    \caption{\footnotesize SDSS spectra of the three sources displaying significant changes in 
      their Mg\,II absorption line. \label{fig:mg2vari}}
\end{figure}

  In our parent \mgii\ BAL sample, 61 sources have repeated spectroscopic observations in the SDSS Legacy programme
  and/or in the SDSS DR10 (SDSS-III/BOSS).
  The detailed information of these sources are summarized in Table~\ref{tab:multiobserve}.
  We find that four of them show obvious variability in the absorption profile;
  they are J083525.98$+$435211.3, J090825.06$+$014227.7, J114209.01$+$070957.7, and
  J14264704.7$+$401250.8. The most extreme variability occurs in J14264704.7+401250.8, with 
  both its \mgii\ and \hei\ absorption lines being weakening evidently (Figure~\ref{fig:j1426}). 
  The other three sources, as Figure~\ref{fig:mg2vari} shows, 
  the changes of their \mgii\ absorption line profile are evident.
  Such changes in the absorption may result from the nuclei continuum, the 
  covering factor of the outflow, or the physical conditions of the outflow gas, 
  which is interesting to investigate further.

\begin{deluxetable}{lrccrccc}
\centering
\tablewidth{\textwidth}
\tablecolumns{8}
\tabletypesize{\footnotesize}
\tablecaption{Duplicate spectroscopic observations \label{tab:multiobserve}}
\tablehead{
  \colhead{Object}    & \colhead{$z$}  & \colhead{MJD} & \colhead{Plate} &\colhead{Fiberid}  &
  \colhead{Observation Date} & \colhead{Source}  & \colhead{HeI* Detection\tablenotemark{a}} }
\startdata
J010352.46+003739.7 & 0.705 & 51816&0396&471   &  2000-09-29 &   SDSS & 1  \\
                    &       & 53726&2313&384   &  2005-12-22 &   SDSS & 0  \\
                    &       & 55214&3736&0522  &  2010-01-18 &   BOSS & 0  \\
                    &       & 55475&4226&0600  &  2010-10-06 &   BOSS & 0  \\
J011117.34+142653.6 & 1.155 & 51821&0423&310   &  2000-10-04 &   SDSS & 1  \\
                    &       & 55835&5131&0054  &  2011-10-01 &   BOSS & 0  \\
J014950.96-010314.1 & 1.082 & 51793&0402&260   &  2000-09-06 &   SDSS & 1  \\
                    &       & 55447&4232&0202  &  2010-09-08 &   BOSS & 0  \\
J020105.14+000617.9 & 1.205 & 51871&0403&593   &  2000-11-23 &   SDSS & 1  \\
                    &       & 52179&0701&465   &  2001-09-27 &   SDSS & 1  \\
                    &       & 55201&3609&0620  &  2010-01-05 &   BOSS & 1  \\
J025026.66+000903.3 & 0.597 & 51871&0409&553   &  2000-11-23 &   SDSS & 0  \\
                    &       & 52175&0708&394   &  2001-09-23 &   SDSS & 0  \\
                    &       & 52177&0707&583   &  2001-09-25 &   SDSS & 0  \\
                    &       & 55450&4241&0855  &  2010-09-11 &   BOSS & 0  \\
J025204.17+010710.5 & 1.223 & 51816&0410&321   &  2000-09-29 &   SDSS & 0  \\
                    &       & 51877&0410&325   &  2000-11-29 &   SDSS & 0  \\
J074554.74+181817.0 & 1.054 & 52939&1582&256   &  2003-10-27 &   SDSS & 1  \\
                    &       & 53437&2074&517   &  2005-03-08 &   SDSS & 0  \\
                    &       & 54497&2915&543   &  2008-02-01 &   SDSS & 0  \\
                    &       & 55565&4492&0798  &  2011-01-04 &   BOSS & 0  \\
J075325.52+414842.9 & 1.349 & 51882&0435&408   &  2000-12-04 &   SDSS & 0  \\
                    &       & 51885&0434&011   &  2000-12-07 &   SDSS & 0  \\
J075927.12+363431.5 & 0.983 & 52238&0757&474   &  2001-11-25 &   SDSS & 1  \\
                    &       & 55509&3801&0688  &  2010-11-09 &   BOSS & 0  \\
J080957.38+181804.4 & 0.969 & 53319&1923&023   &  2004-11-10 &   SDSS & 1  \\
                    &       & 55585&4493&0632  &  2011-01-24 &   BOSS & 0  \\
J081312.61+432640.1 & 1.090 & 51959&0547&242   &  2001-02-19 &   SDSS & 0  \\
                    &       & 52205&0546&449   &  2001-10-23 &   SDSS & 0  \\
                    &       & 52207&0547&274   &  2001-10-25 &   SDSS & 0  \\
J083522.77+424258.3 & 0.807 & 52232&0762&085   &  2001-11-19 &   SDSS & 1  \\
                    &       & 55924&4683&0338  &  2011-12-29 &   BOSS & 0  \\
J083525.98+435211.3 & 0.568 & 52232&0762&175   &  2001-11-19 &   SDSS & 0  \\
                    &       & 55924&4683&0683  &  2011-12-29 &   BOSS & 0  \\
J090154.96+380534.4 & 1.038 & 52705&0936&498   &  2003-03-07 &   SDSS & 0  \\
                    &       & 55973&4608&0844  &  2012-02-16 &   BOSS & 0  \\
J090825.06+014227.7 & 1.002 & 51924&0471&499   &  2001-01-15 &   SDSS & 0  \\
                    &       & 55540&3819&0234  &  2010-12-10 &   BOSS & 0  \\
J092035.59+524006.2 & 0.793 & 51999&0553&612   &  2001-03-31 &   SDSS & 0  \\
                    &       & 52252&0767&321   &  2001-12-09 &   SDSS & 0  \\
J092157.62+103539.0 & 0.548 & 53050&1740&170   &  2004-02-15 &   SDSS & 1  \\
                    &       & 55926&5306&0196  &  2011-12-31 &   BOSS & 0  \\
J094225.42+565613.0 & 0.833 & 51991&0556&545   &  2001-03-23 &   SDSS & 0  \\
                    &       & 52253&0557&359   &  2001-12-10 &   SDSS & 0  \\
J094443.13+062507.4 & 0.695 & 52710&0993&535   &  2003-03-12 &   SDSS & 1  \\
                    &       & 55926&4873&0470  &  2011-12-31 &   BOSS & 0  \\
J095057.56+542919.4 & 1.194 & 52282&0769&507   &  2002-01-08 &   SDSS & 0  \\
                    &       & 54530&0769&504   &  2008-03-05 &   SDSS & 0  \\
J095914.91+131639.5 & 1.009 & 53055&1744&362   &  2004-02-20 &   SDSS & 0  \\
                    &       & 55982&5328&0528  &  2012-02-25 &   BOSS & 0  \\
J102943.75+370127.2 & 1.344 & 53415&1957&601   &  2005-02-14 &   SDSS & 1  \\
                    &       & 53432&1973&235   &  2005-03-03 &   SDSS & 0  \\
J103255.37+083503.2 & 0.894 & 52734&1240&316   &  2003-04-05 &   SDSS & 1  \\
                    &       & 55924&5344&0796  &  2011-12-29 &   BOSS & 0  \\
J104459.60+365605.1 & 0.701 & 53463&2090&329   &  2005-04-03 &   SDSS & 1  \\
                    &       & 55615&4635&0704  &  2011-02-23 &   BOSS & 0  \\
J104845.83+353110.7 & 1.011 & 53463&2090&131   &  2005-04-03 &   SDSS & 1  \\
                    &       & 55615&4635&0136  &  2011-02-23 &   BOSS & 0  \\
J105259.99+065358.0 & 0.724 & 52670&1001&080   &  2003-01-31 &   SDSS & 0  \\
                    &       & 55685&4854&0598  &  2011-05-04 &   BOSS & 1  \\
J111628.00+434505.8 & 0.801 & 53061&1364&095   &  2004-02-26 &   SDSS & 1  \\
                    &       & 56013&4686&0594  &  2012-03-27 &   BOSS & 0  \\
J112349.16+135220.6 & 0.928 & 53383&1753&093   &  2005-01-13 &   SDSS & 0  \\
                    &       & 56003&5370&0688  &  2012-03-17 &   BOSS & 1  \\
J112828.31+011337.9 & 0.893 & 51614&0281&523   &  2000-03-11 &   SDSS & 0  \\
                    &       & 51992&0512&123   &  2001-03-24 &   SDSS & 0  \\
                    &       & 55630&4730&0172  &  2011-03-10 &   BOSS & 0  \\
J112901.71+050617.0 & 1.282 & 52376&0836&626   &  2002-04-12 &   SDSS & 0  \\
                    &       & 52642&0837&400   &  2003-01-03 &   SDSS & 0  \\
J113349.81+361027.3 & 1.158 & 53468&2113&608   &  2005-04-08 &   SDSS & 0  \\
                    &       & 55618&4615&0762  &  2011-02-26 &   BOSS & 0  \\
J114111.61-014306.6 & 1.266 & 52282&0328&417   &  2002-01-08 &   SDSS & 1  \\
                    &       & 55207&3775&0820  &  2010-01-11 &   BOSS & 0  \\
J114209.01+070957.7 & 0.497 & 53383&1621&306   &  2005-01-13 &   SDSS & 0  \\
                    &       & 55955&4848&0546  &  2012-01-29 &   BOSS & 0  \\
J114915.30+393325.4 & 0.630 & 53386&1970&137   &  2005-01-16 &   SDSS & 0  \\
                    &       & 55659&4654&0080  &  2011-04-08 &   BOSS & 1  \\
J115852.87-004301.9 & 0.983 & 51663&0285&184   &  2000-04-29 &   SDSS & 0  \\
                    &       & 51930&0285&189   &  2001-01-21 &   SDSS & 0  \\
J120049.55+632211.8 & 0.887 & 52320&0777&609   &  2002-02-15 &   SDSS & 0  \\
                    &       & 52337&0778&377   &  2002-03-04 &   SDSS & 0  \\
J120146.46+630034.4 & 0.659 & 52337&0778&439   &  2002-03-04 &   SDSS & 0  \\
                    &       & 54525&0778&477   &  2008-02-29 &   SDSS & 0  \\
J122614.97+120925.4 & 0.871 & 53120&1614&145   &  2004-04-25 &   SDSS & 1  \\
                    &       & 55979&5403&0524  &  2012-02-22 &   BOSS & 0  \\
J125507.12+634423.8 & 1.067 & 52316&0601&043   &  2002-02-11 &   SDSS & 0  \\
                    &       & 52320&0782&445   &  2002-02-15 &   SDSS & 0  \\
J131323.25+151309.6 & 1.262 & 53089&1772&627   &  2004-03-25 &   SDSS & 1  \\
                    &       & 56033&5424&0206  &  2012-04-16 &   BOSS & 0  \\
J131823.73+123812.5 & 0.589 & 53142&1697&600   &  2004-05-17 &   SDSS & 0  \\
                    &       & 56001&5427&0574  &  2012-03-15 &   BOSS & 0  \\
J132401.53+032020.6 & 0.926 & 52312&0526&609   &  2002-02-07 &   SDSS & 0  \\
                    &       & 55633&4761&0136  &  2011-03-13 &   BOSS & 0  \\
J133603.65+511733.6 & 1.327 & 53433&1668&593   &  2005-03-04 &   SDSS & 0  \\
                    &       & 53433&1669&339   &  2005-03-04 &   SDSS & 0  \\
J140025.53-012957.0 & 0.584 & 52443&0915&380   &  2002-06-18 &   SDSS & 0  \\
                    &       & 55363&4038&0558  &  2010-06-16 &   BOSS & 0  \\
J142010.28+604722.3 & 1.345 & 52365&0606&110   &  2002-04-01 &   SDSS & 0  \\
                    &       & 52368&0607&293   &  2002-04-04 &   SDSS & 0  \\
J142647.47+401250.8 & 0.749 & 52797&1349&0348  &  2003-06-07 &   SDSS & 1  \\
                    &       & 56038&5171&0532  &  2012-04-21 &   BOSS & 0  \\
                    &       & 56063&5170&0928  &  2012-05-16 &   BOSS & 0  \\
J142927.28+523849.5 & 0.595 & 52764&1326&057   &  2003-05-05 &   SDSS & 1  \\
                    &       & 52781&1327&343   &  2003-05-22 &   SDSS & 0  \\
J143144.91+391910.2 & 1.091 & 52797&1349&257   &  2003-06-07 &   SDSS & 1  \\
                    &       & 56038&5171&0340  &  2012-04-21 &   BOSS & 0  \\
J143826.73+642859.8 & 1.222 & 51988&0499&623   &  2001-03-20 &   SDSS & 0  \\
                    &       & 54533&2947&557   &  2008-03-08 &   SDSS & 0  \\
J144436.58+425508.6 & 1.101 & 52734&1289&087   &  2003-04-05 &   SDSS & 0  \\
                    &       & 53112&1396&612   &  2004-04-17 &   SDSS & 0  \\
J145333.01+002943.6 & 1.291 & 51666&0309&424   &  2000-05-02 &   SDSS & 0  \\
                    &       & 51994&0309&433   &  2001-03-26 &   SDSS & 0  \\
                    &       & 52029&0538&202   &  2001-04-30 &   SDSS & 0  \\
J150847.41+340437.7 & 0.788 & 53108&1385&173   &  2004-04-13 &   SDSS & 1  \\
                    &       & 55691&4720&0591  &  2011-05-10 &   BOSS & 0  \\
J152438.79+415543.0 & 1.230 & 53433&1678&106   &  2005-03-04 &   SDSS & 0  \\
                    &       & 56067&5164&0894  &  2012-05-20 &   BOSS & 0  \\
J163656.84+364340.4 & 0.852 & 52782&1174&337   &  2003-05-23 &   SDSS & 1  \\
                    &       & 56048&5195&0930  &  2012-05-01 &   BOSS & 0  \\
J170341.82+383944.7 & 0.554 & 52071&0632&632   &  2001-06-11 &   SDSS & 1  \\
                    &       & 54232&2192&058   &  2007-05-12 &   SDSS & 0  \\
J212017.00+004841.7 & 1.288 & 52523&0987&408   &  2002-09-06 &   SDSS & 0  \\
                    &       & 55469&4192&0874  &  2010-09-30 &   BOSS & 0  \\
                    &       & 55825&5142&0341  &  2011-09-21 &   BOSS & 0  \\
J220931.92+125814.5 & 0.814 & 52519&0735&501   &  2002-09-02 &   SDSS & 1  \\
                    &       & 55749&5041&0390  &  2011-07-07 &   BOSS & 1  \\
J223424.10+005227.1 & 0.884 & 52143&0376&615   &  2001-08-22 &   SDSS & 0  \\
                    &       & 52201&0674&370   &  2001-10-19 &   SDSS & 0  \\
J224028.14-003813.1 & 0.658 & 53261&1901&298   &  2004-09-13 &   SDSS & 1  \\
                    &       & 55470&4204&0232  &  2010-10-01 &   BOSS & 0  \\
J232550.73-002200.4 & 1.011 & 51818&0383&142   &  2000-10-01 &   SDSS & 1  \\
                    &       & 52199&0681&269   &  2001-10-17 &   SDSS & 0  \\
J233635.75-010733.7 & 1.303 & 51821&0384&011   &  2000-10-04 &   SDSS & 0  \\
                    &       & 52525&0682&202   &  2002-09-08 &   SDSS & 0  \\
\enddata
\tablenotetext{a}{Detection of He\,I*$\lambda3889$ absorption line. '1' denotes a spectrum with He\,I*$\lambda3889$
absorption line. '0' denotes a spectrum failed to detect He\,I*$\lambda3889$ absorption line. }
\end{deluxetable}

\subsection{He\,I*$\lambda3889$ BAL quasars with supportive NIR and UV spectra \label{app:nirspectra}}

  We have performed followup near infrared (NIR) spectroscopic observations 
  for sources the \hei$\lambda3889$ BAL sample to check their expected \hei$\lambda10830$ BAL features.
  The NIR observations were made by the Palomar P200 telescope with the Triple-Spec spectroscopy 
  on 2012 April 15--16th,
  2013 February 22--23th, and 2014 January 17--19th.
  All spectra were taken in an A-B-B-A dithering mode, 
  in the primary configuration of the instrument
  (a spectral resolution of R~$\sim$~3500 through a 1$\arcsec$ slit). 
  The telluric standard stars were taken quasi-simultaneously.
  The data were reduced with the IDL program
  SpexTool \citep{2004PASP..116..362C}. The flux calibration and telluric correction were performed 
  with the IDL program that adopts the methods as described in \citet{2003PASP..115..389V}.
 
  On the other hand, some sources have UV spectra in the HST or IUE archives, 
  where C\,IV BAL is present, confirming our detection of their \hei$\lambda3889$ absorption line.

  \subsubsection{Four sources in the \hei\,$\lambda3889$ BAL sample}

  {\bf J074554.74+181817.0}

  J0745$+$1818 is a FeLoBAL quasar at $z=1.054$, which is first identified by G09.
  In its rest frame UV spectrum, broad \mgii\ and Fe\,II absorption lines are seen. 
  The \mgii\ absorption trough, spans from -613 \kms to -3971 \kms, with a weighted average velocity of -1887 \kms.
  The associated \hei$\lambda\lambda3889,3189$ absorption lines are first identified in this work, the detections of 
  which are also confirmed by the NIR observation using TripleSpec on Palomar P200 telescope.
  This NIR observation is performed on 2014 January 19th, and four 240s exposures were taken in A-B-B-A dithering 
  mode. One telluric standard stars was taken quasi-simultaneously.

  {\bf J080248.18$+$551328.9 \citep{2014arXiv1412.2881J}}

  J0802+5513 is identified as a LoBAL quasar by G09 according to detection of \mgii\ absorption lines. Plenty of 
  absorption lines are detected, including \hei, Fe\,II* and Ni\,II* that arise from metastable or excited levels, 
  as well as resonant lines in \mgii, Fe\,II, Mn\,II, Ca\,II and Mg\,I. 
  All these absorption lines are associated with the same profile of $\sim$ 2000 \kms width centered at a 
  common redshift of the quasar emission lines. \citet{2014arXiv1412.2881J} detailedly studied this target and
  determined that the absorber has a density of $n_{\rm e}\sim10^{5}~\rm cm^{-3}$ and a column density of
  $N_{\rm H}\sim10^{21}-10^{21.5} \rm cm^{-2}$, and is located at $\sim100-250$~pc from the central SMBH.
  J0802+5513 is also included in our \hei$\lambda3889$ absorption line sample.

  {\bf J084044.41+363327.8}

  This FBQS quasar was first reported by \citet{1997ApJ...479L..93B} as a radio-moderate LoBAL.
  A spectropolarimetry study by \citet{1997ApJ...487L.113B} reveals that it is a highly polarized BAL quasar.
  The spectrum of this quasar in the rest frame wavelength range is dominated by absorption lines of Fe\,II, and it 
  also contains lines of other singly ionized iron group elements, Si\,II, Mg\,II, Al\,III, and Fe\,III. 
  According to \citet{2002ApJ....570..514D}, the outflow gas of this quasar covers a range of velocity 
  from -700 \kms to -3500 \kms, with two component centered at -900 and -2800 km/s. 
  The low-velocity gas is determined as a 
  low-density gas ($n_{e}~<~500$~cm$^{-3}$) and is $\sim~230$~pc away from the AGN nuclei. 
  The high-velocity gas shows a high density and $\sim~1$~pc away from the AGN nuclei. 
  This distance is determined by the absence of detectable \hei$\lambda2830$ absorption line. 
  In our sample, this quasar is classified as a \hei$\lambda3889$ BAL, which is also confirmed by the NIR observation 
  for \hei$\lambda10830$ absorption line. The total column density and location of the absorption gas can be better 
  determined via the \hei\ lines join with other lines. The NIR observation is carried out on 2014-01-17 using
  TripleSpec at P200 telescope, and 4$\times$180 s exposures were taken in A-B-B-A mode.

  {\bf J120924.07+103612.0}

  LBQS~1206+1052 is reported by \citet{2012RAA.....12..369J} as a rare Balmer broad absorption line quasar, 
  and it also displays \hei\, \mgii\ absorption lines in SDSS spectrum taken on March 24, 2003.
  \citet{2011MNRAS.412.2717J} reported the significant variations over their observational run of $\sim$~4~h.
  We performed a series of follow up observations for this quasar. On May 7th, 2012, LBQS~1206+1052 was taken by
  Yunnan Faint Object Spectrograph and Camera (YFOSC) mounted on Lijiang GMG 2.4m telescope in a rest wavelength 
  range of 2500--5400\AA. Compare with the SDSS spectrum, the absorption troughs fade during nine years.
  On April 23, 2014, this quasar was taken by Double Spectrograph (DBSP) at P200 telescope in a rest wavelength range 
  of 2250--4180 and 5620--7450 \AA. The states of the absorption lines show little variability relative to the YFOSC 
  observation.
  A NIR observation was carried out on February 22, 2013 to obtain \hei$\lambda10830$ absorption line profile, 
  which show a narrow faded \hei$\lambda10830$ absorption trough. This target is detailed studied in 
  Sun \etal\ (2015, in preparation).
  \begin{figure}[htbp]
   \centering
    \includegraphics[width=3.2in]{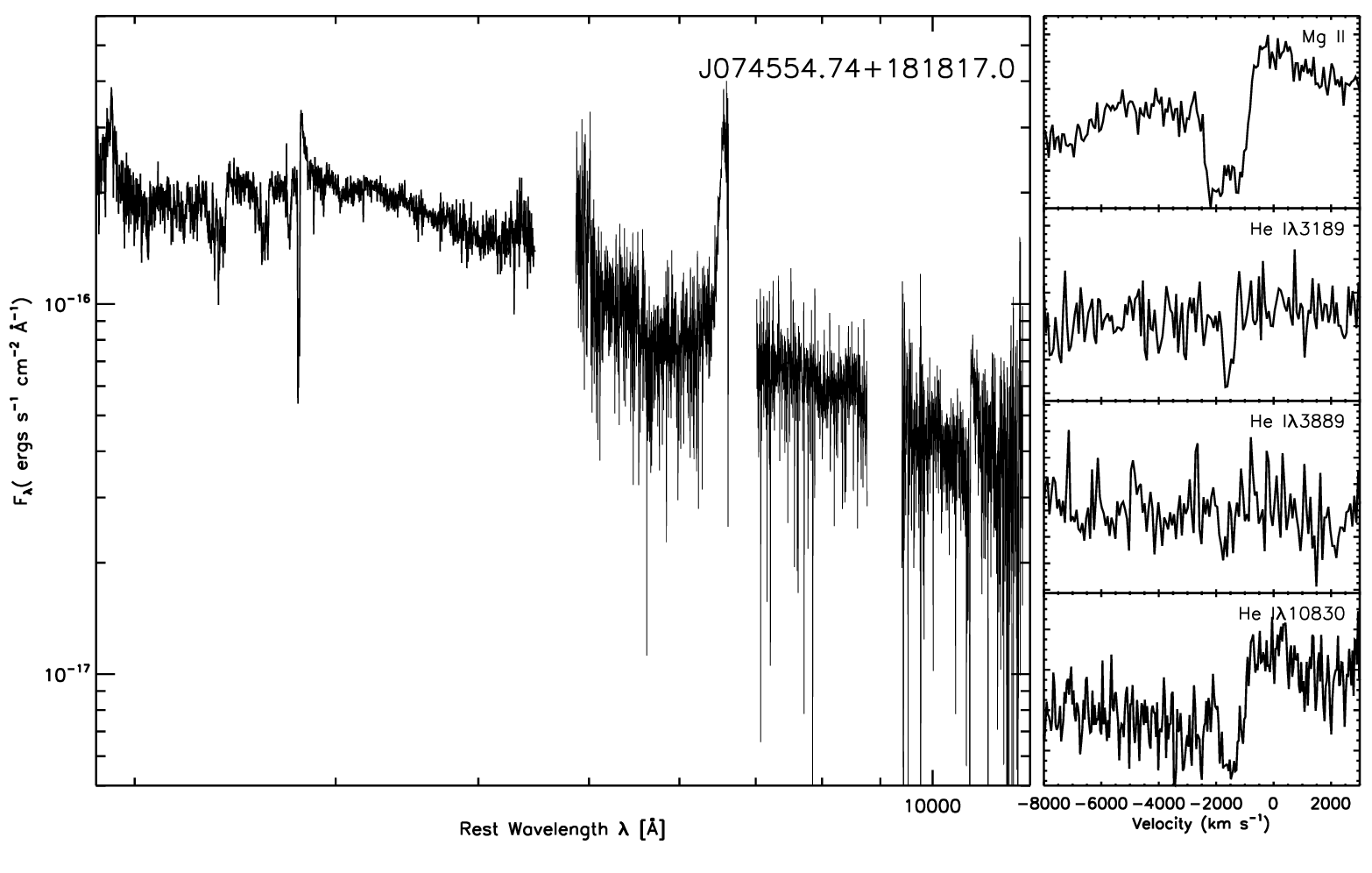}
    \includegraphics[width=3.2in]{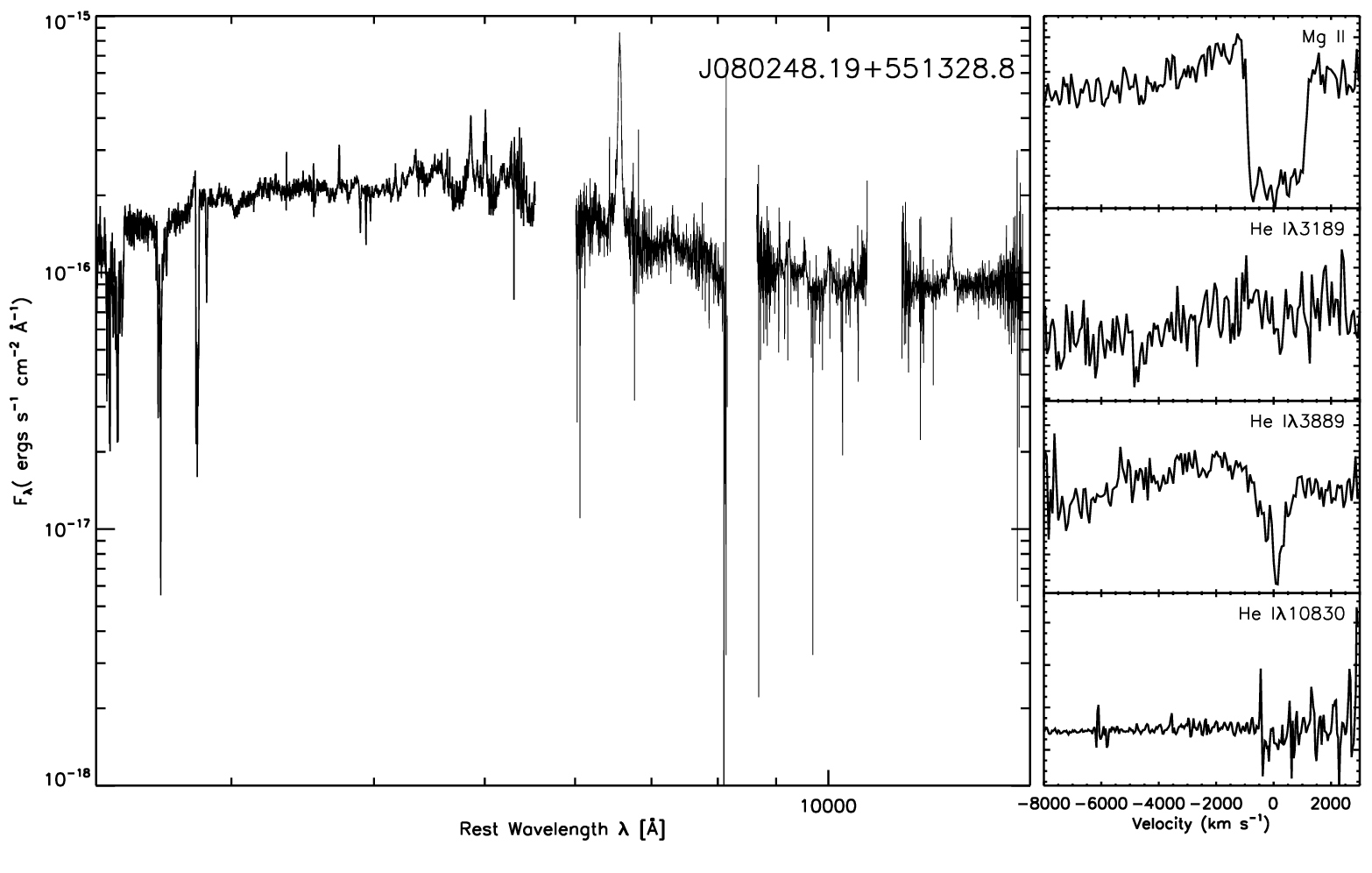}
    \includegraphics[width=3.2in]{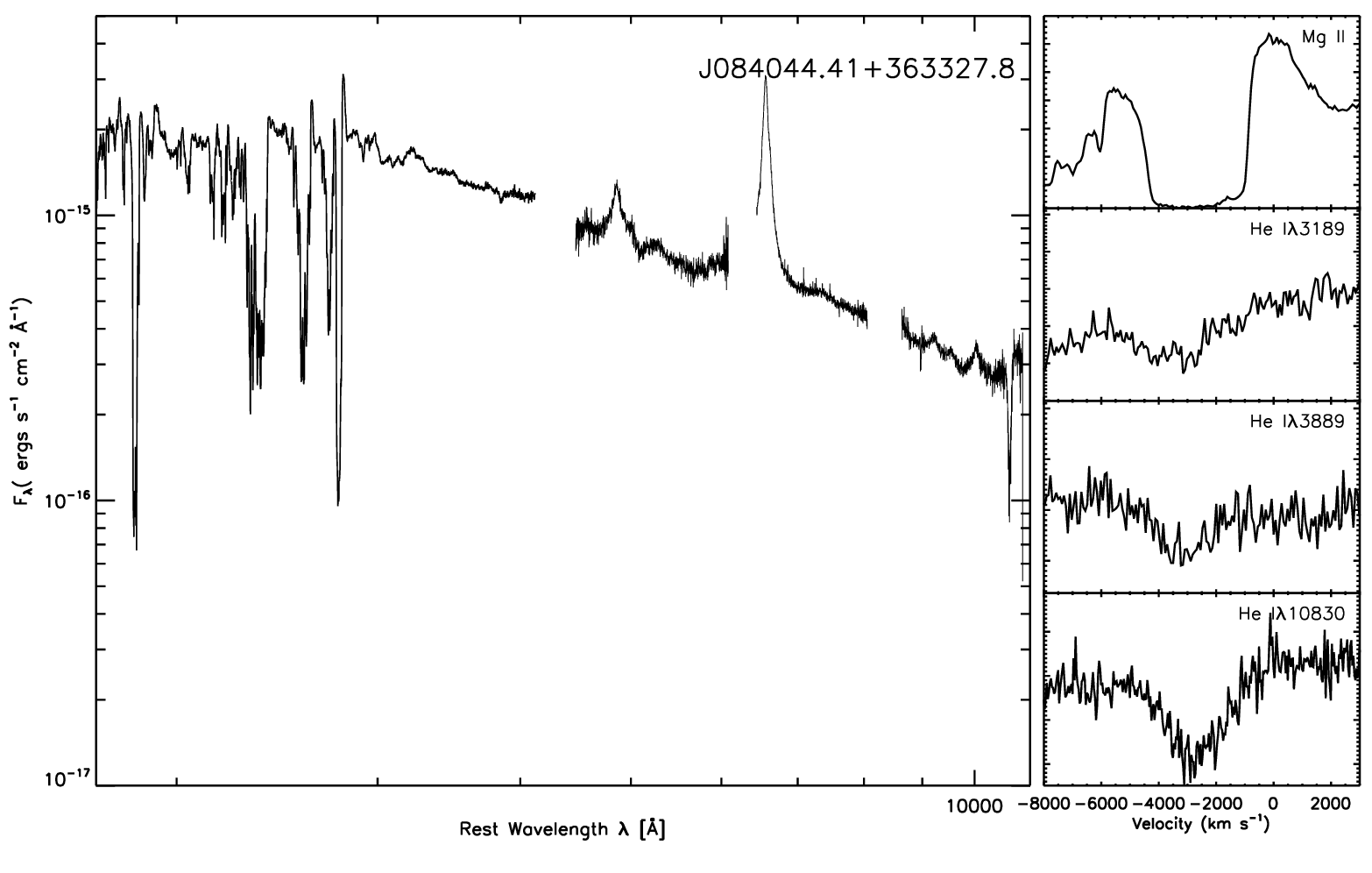}
    \includegraphics[width=3.2in]{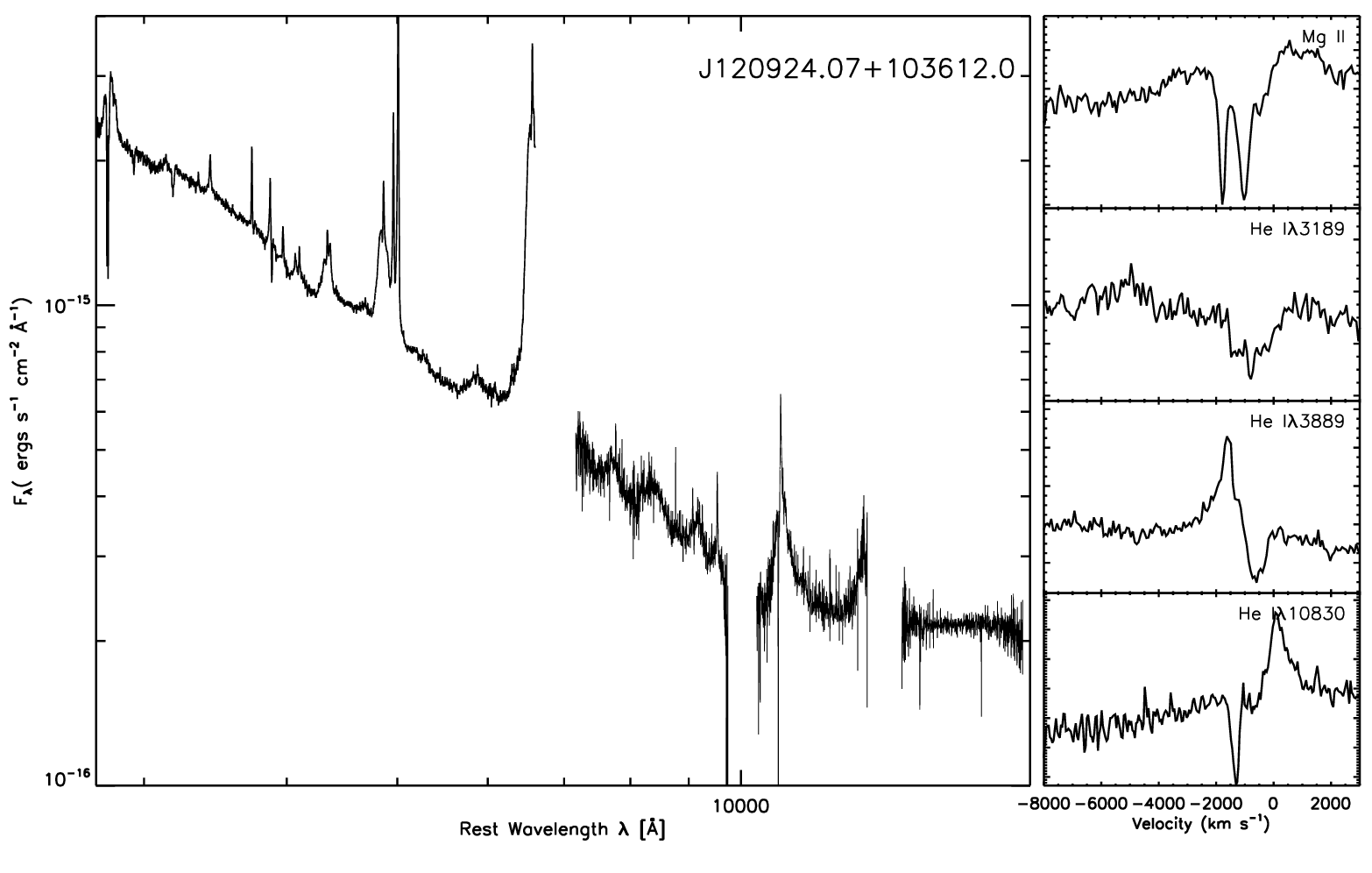}
   \caption{Sources in our main sample that have NIR spectra. (continue)}
 \end{figure}

  \subsubsection{Two example C\,IV BAL quasars with evident \hei$\lambda3889$ absorption yet weak \mgii\ absorption}

  {\bf J035230.55-071102.3}

  J0352-0711 is known as 3C094, which is a steep-spectrum radio-loud quasar.
  The FUV spectrum for this quasar, which was obtained with Faint Object Spectrograph (FOS) on board the HST by 
  \citet{1995ApJ...438..420T} using G270H grating, shows a broad C\,IV absorption line. 
  We detected the \hei$\lambda3889$ absorption line in SDSS spectrum. 
  However, this quasar is not in our parent \mgii\ BAL sample because of the weakness 
  of \mgii\ absorption line. We observe this quasar using TripleSpec at P200 telescope on February 23, 2012, and 
  4$\times$180 s exposures were taken in A-B-B-A mode. 
  The significant \hei$\lambda10830$ absorption trough are seen in the NIR spectrum.

  {\bf J141348.33+440014.0}

  PG 1411+442 is one of nearest HiBAL quasars ($z_{em}$ = 0.0896). 
  It shows broad Ly$\alpha$, N\,V, Si\,IV, C\,IV absorption features in UV spectrum 
  \citep{1987ApJ...322..729M,1999ApJ...519L..35W}. It is also known as a X-ray quiet quasar due to substantial 
  intrinsic absorption with $N_{H} > 10^{23}$~cm$^{-2}$ \citep[e.g.,][]{1999A&A...345...43B}.
  PG 1411+442 is also a luminous infrared quasar with a total infrared luminosity of log$L_{IR}$=11.6 \lsun 
  \citep{2012ApJ...761..184W}.
  On January 17, 2014, this quasar was observed by TripleSpec at P200 telescope, and 4$\times$180 s exposures were 
  taken in A-B-B-A mode. 
  \hei$\lambda10830$ absorption trough are seen in the NIR spectrum, which share the common blueshift with the
  C\,IV absorption line.
  On April 23, 2014, this quasar was observed by DBSP at P200 telescope in a rest wavelength range of 3000--5400 
  and 7200--9500 \AA, and 2 $\times$ 300s exposure were taken. 
  \hei$\lambda3889$ absorption trough is detected in the NUV spectrum.

  \begin{figure}[tbp]
        \centering
        \includegraphics[width=3.2in]{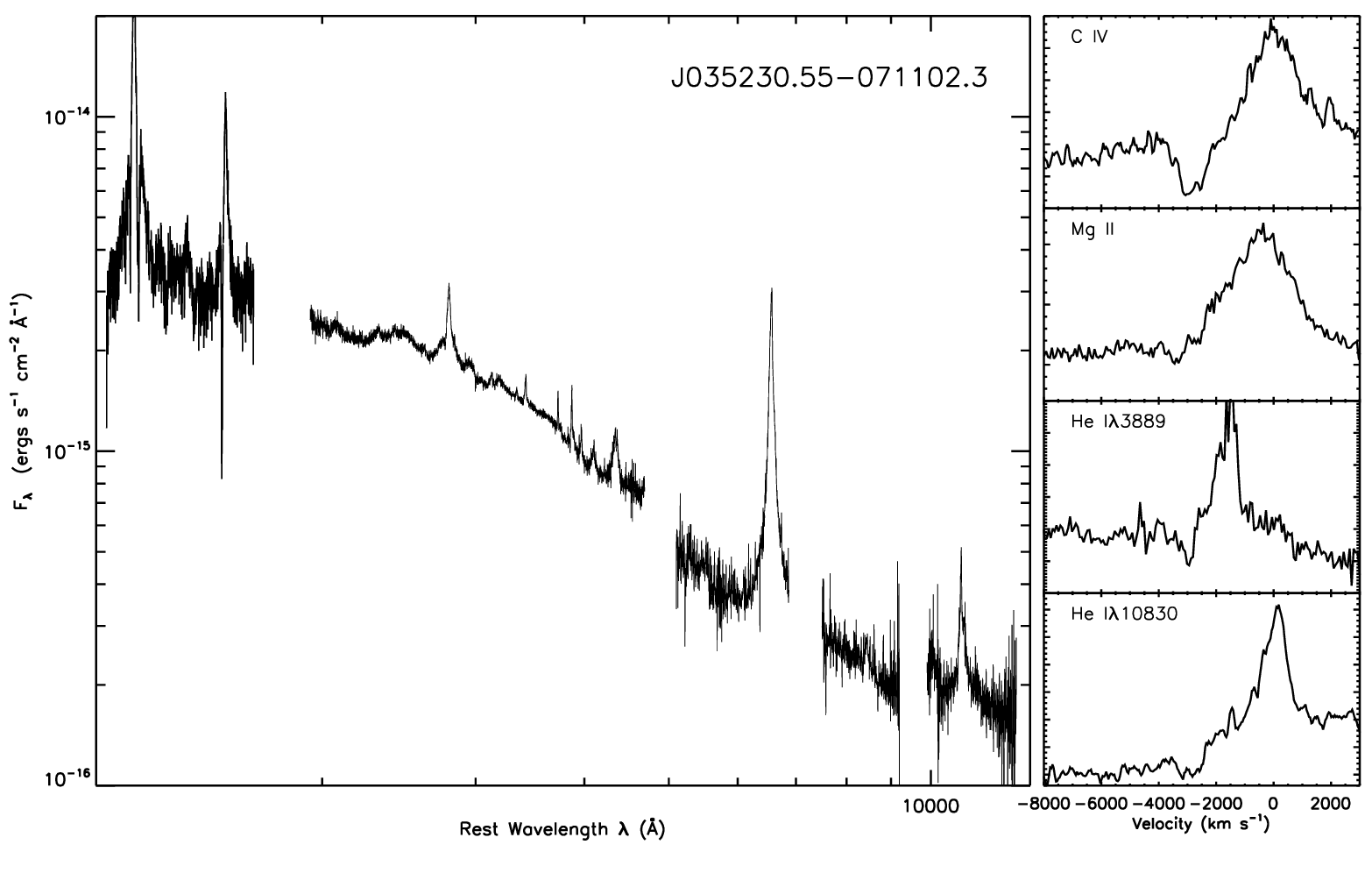}
        \includegraphics[width=3.2in]{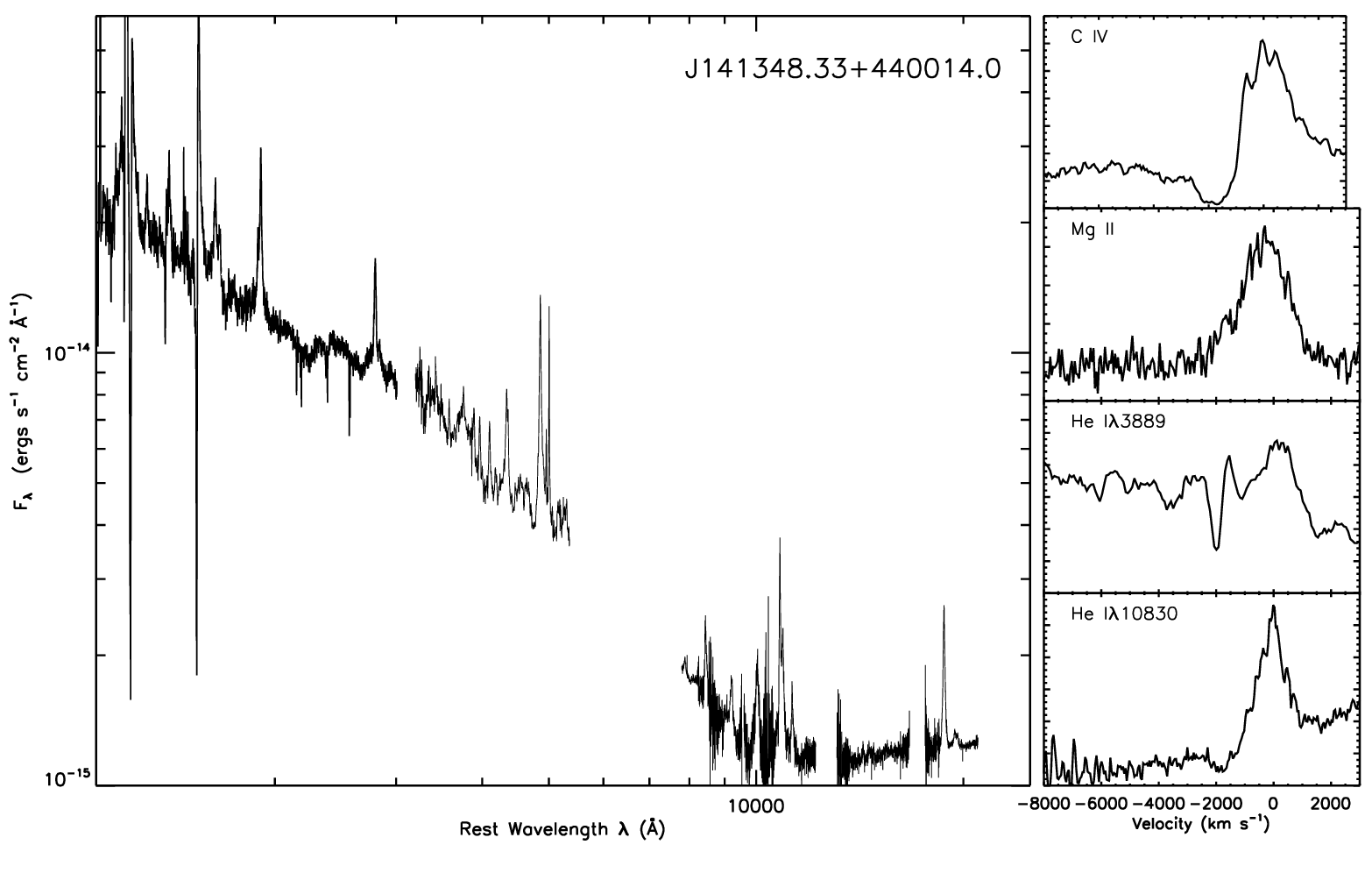}
    \caption{ C\,IV BALs that have obvious \hei\ absorption line and have weak \mgii\ 
      absorption line \label{fig:J0352J1413}}
  \end{figure}

  \subsubsection{Six sources in the low-$z$ BAL sample identified with \hei$\lambda3889$}

  {\bf J075217.84+193542.2}

  J075217.84+193542.2 is a Seyfert I galaxy at $z=0.117$, which is the most nearest targets in our low-$z$ \hei\ 
  sample. We observed this target using TripeSpec at P200 telescope on February 23th, 2013, and 4$\times$120s 
  exposures were taken in A-B-B-A mode. The \hei$\lambda10830$ absorption line is detected in the NIR spectrum.  

  {\bf J093653.81+533127.0}

  J0936+5331 is a Seyfert 1 galaxy at $z=0.227$.  
  It is a low-$z$ C\,IV HiBAL according to the detection of C\,IV absorption trough and absence of \mgii\ absorption 
  feature in UV observations by HST.
  FUV and NUV spectrum for this quasar were obtained with Cosmic Origins Spectrograph (COS) G140L grating and
  Space Telescope Imaging Spectrograph (STIS) G430L grating on board the HST respectively by 
  \citet{2014MNRAS.441..316L}.
  We detected the \hei$\lambda\lambda3189,3889$ lines in its SDSS spectrum, which have the common blueshift with that 
  of C\,IV.
  The NIR observation was carried out using TripleSpec at P200 telescope on February 23th, 2013, and 4$\times$180s 
  exposures were taken in A-B-B-A mode. 
  The expected \hei$\lambda10830$ absorption line are seen in the NIR spectrum.

  {\bf J130534.49+181932.8}

  J130534.49+181932.8 is a low-$z$ ($z = 0.118$) quasar. We detect the \hei$\lambda3889$ absorption line in its SDSS 
  spectrum. This source is also observed by COS G140L grating and STIS G430L grating on board the HST telescope in 
  2011 by \citet{2014MNRAS.441..316L}. 
  C\,IV and Mg\,II absorption lines are detected in the FUV and NUV spectrum, and 
  they have common blueshift with the \hei$\lambda3889$ absorption line. 

  {\bf J153539.25+564406.5}

  J1535+5644 is a Seyfert I galaxy at $z=0.208$. \hei$\lambda\lambda3189,3889$ and rare \ha\ absorption line are 
  detected in the SDSS spectrum. 
  The NIR observation was carried out using TripleSpec at P200 telescope on Feburay 23th, 2013, and 4$\times$180s
  exposures were taken in A-B-B-A mode. The expected \hei$\lambda10830$ absorption line are seen in the NIR spectrum.

  {\bf J163459.82+204936.0}

  J1634+2049 is a nearby ($z=0.1295$) ultra-luminous infrared galaxy with a total infrared luminosity of
  $\sim~1\times~10^{12}~\lsun$ according to the photometry of IRAS.
  The star forming rates (SFR) gets from $L_{IR}$ is $\sim~120~\msun$, but the [OII] shows the SFR is normal.
  This quasar shows a red broad band spectra energy distribution (SED) relative to that of composite
  spectra of quasars. The reddening calculated from decrement for narrow lines is much lager than that 
  for broad lines, with $E(B-V)_{narrow}$ = 1.13 and $E(B-V)_{broad}$ = 0.75. 
  These narrow lines are mainly from HII region according to the BPT diagram. 
  These properties suggest this quasar is an obscured starburst quasar.
  \hei$\lambda3889$ is detected in the SDSS spectrum, and interestingly broad Na\,D 
  absorption line is also detected, which indicate a large column density of the outflow gas. 
  The NIR observation is carried out on April 15, 2012 using TripleSpec at P200 telescope, and 4$\times$120s 
  exposures were taken in A-B-B-A dithering mode. The expected \hei$\lambda10830$ absorption line is seen in the 
  NIR spectrum, and strong P$\alpha$, P$\gamma$ emission lines are also detected. 
  We will detailedly investigate this quasar in later paper (Liu \etal\ 2015, in preparation).

  {\bf J222024.58+010931.2}

  J2220+0109 is a type I quasar in the SDSS Stripe 82, and it shows significant variability 
  \citep{2011A&A...525A..37M}.
  We detected deep and broad \hei$\lambda\lambda3189,3889$, and weak Balmer absorption lines in the SDSS spectrum.
  The NIR observation was carried out using TripleSpec at P200 telescope on October 21th, 2011, and 4$\times$300s 
  exposures were taken in A-B-B-A mode. 
  The expected \hei$\lambda10830$ absorption line are seen in the NIR spectrum.
  \clearpage

 \begin{figure}[htbp]
   \centering
   \includegraphics[width=3.2in]{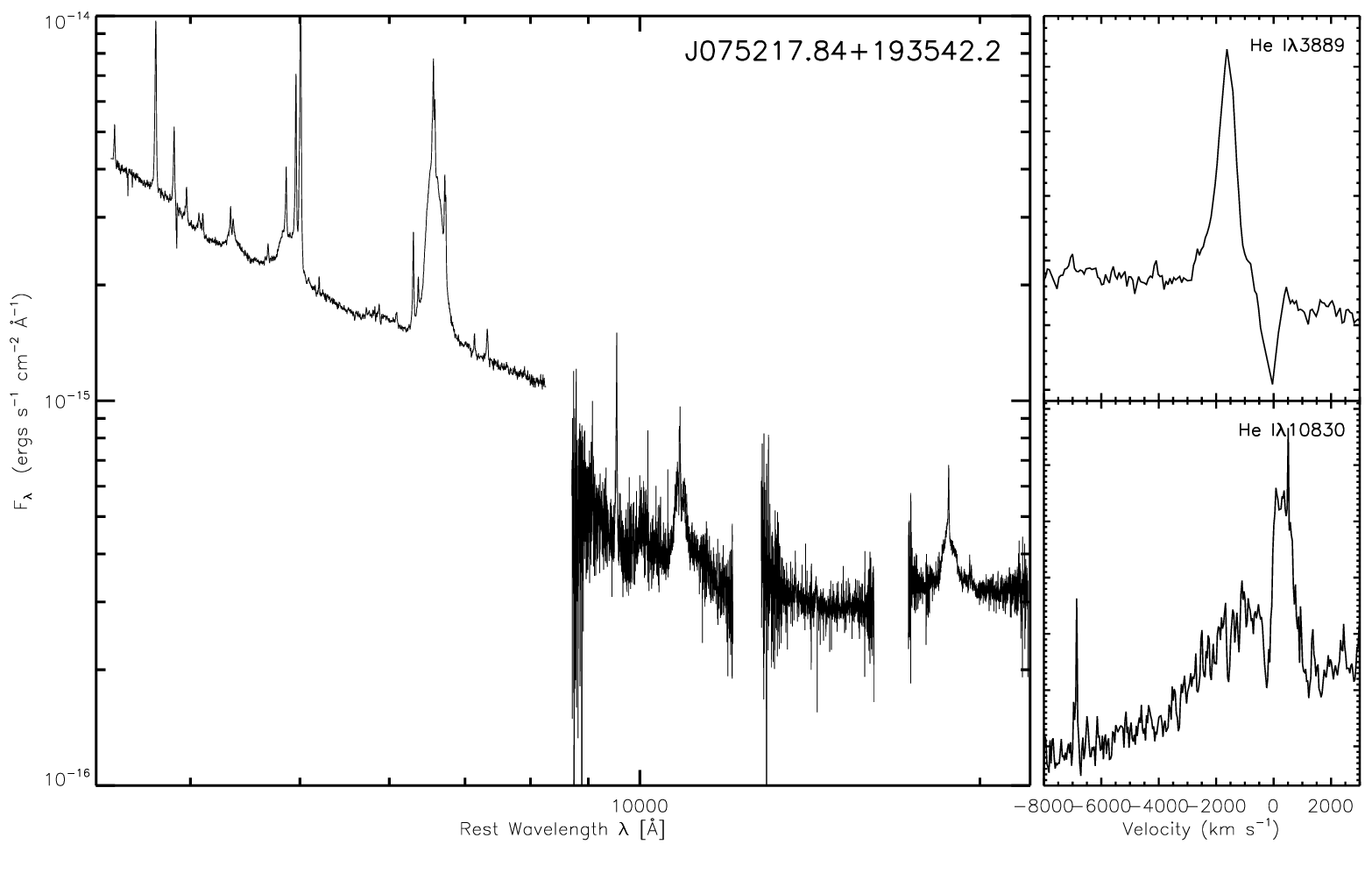}
   \includegraphics[width=3.2in]{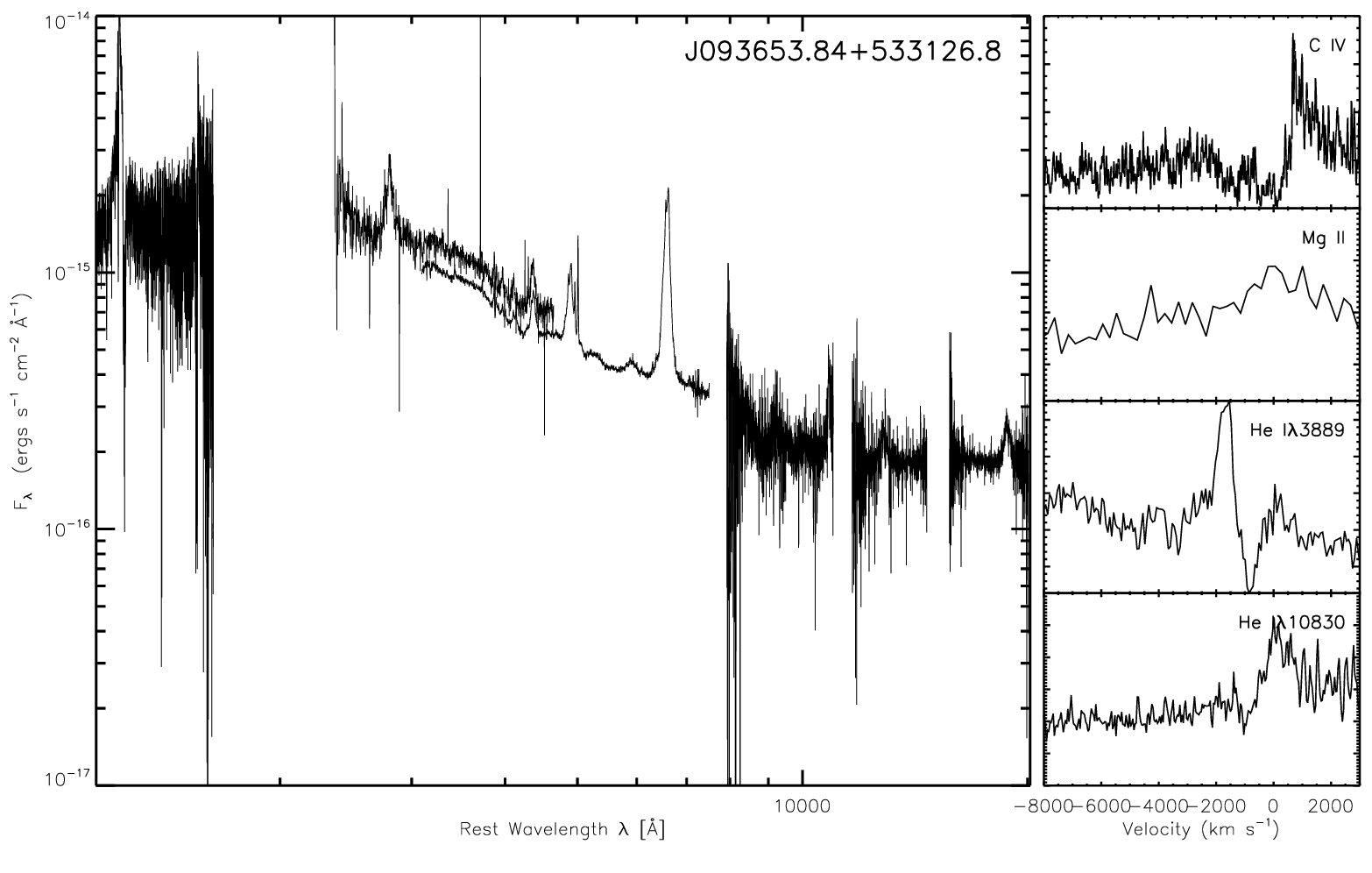}
   \includegraphics[width=3.2in]{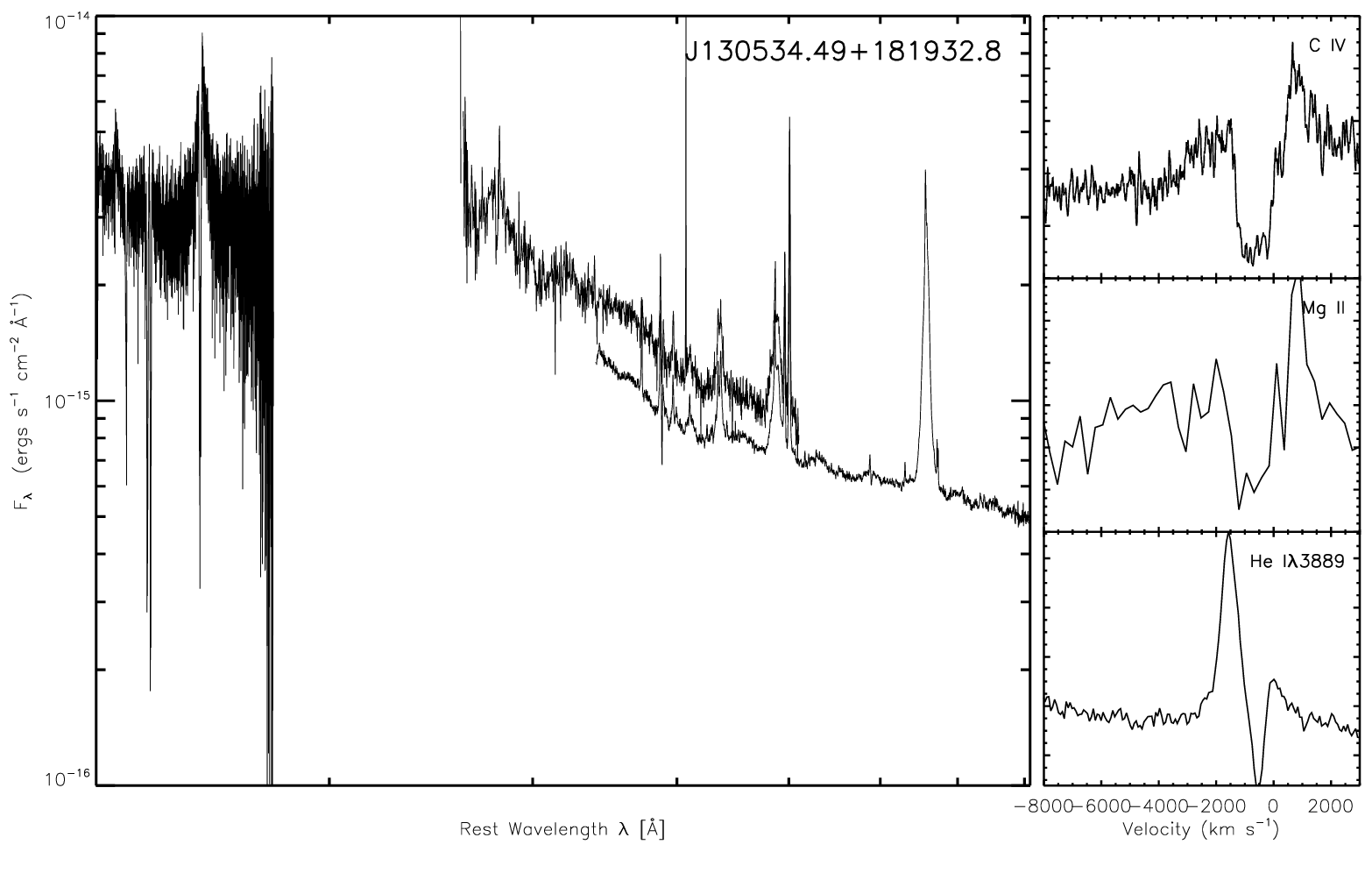}
   \includegraphics[width=3.2in]{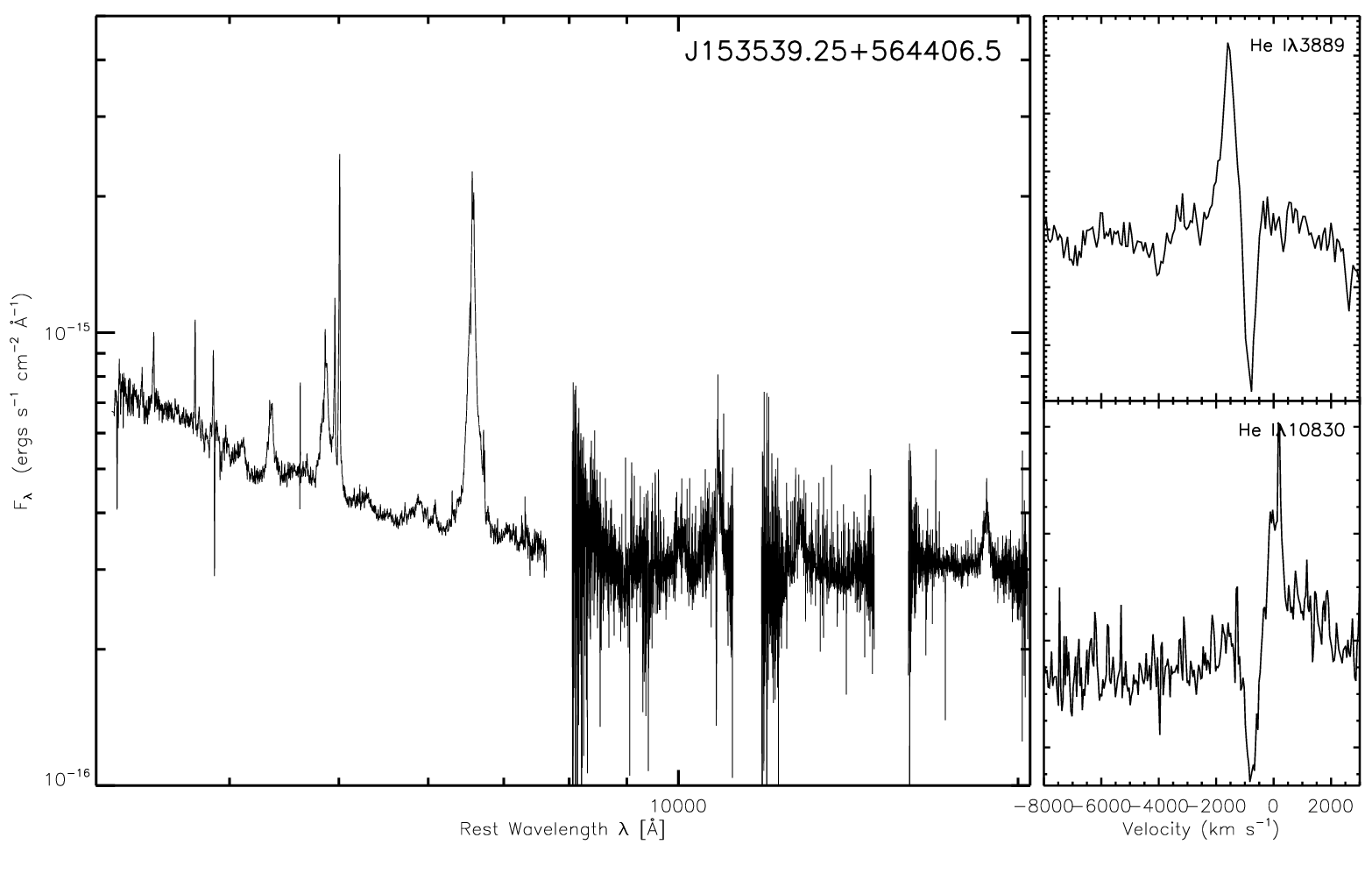}
   \includegraphics[width=3.2in]{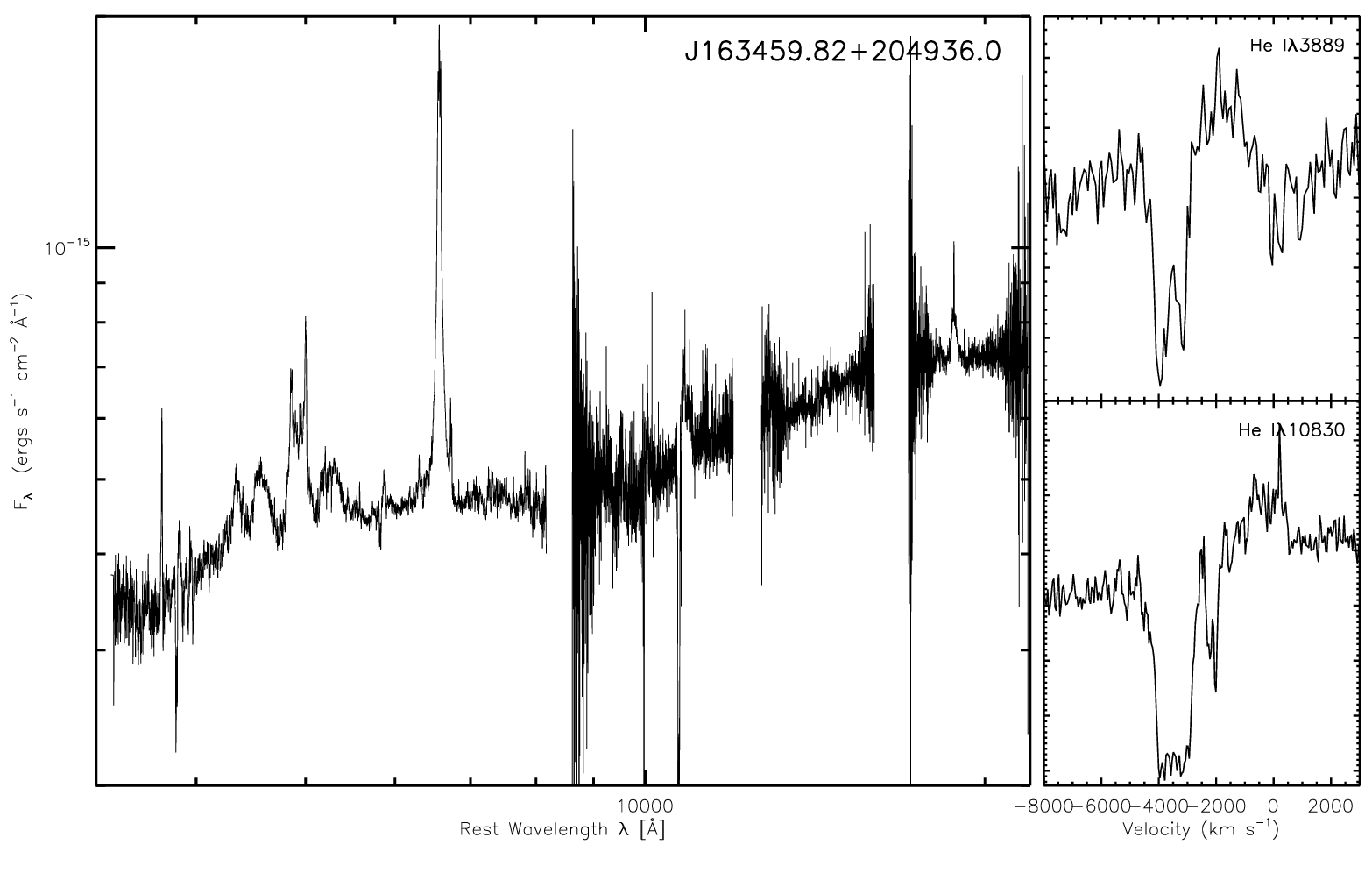}
   \includegraphics[width=3.2in]{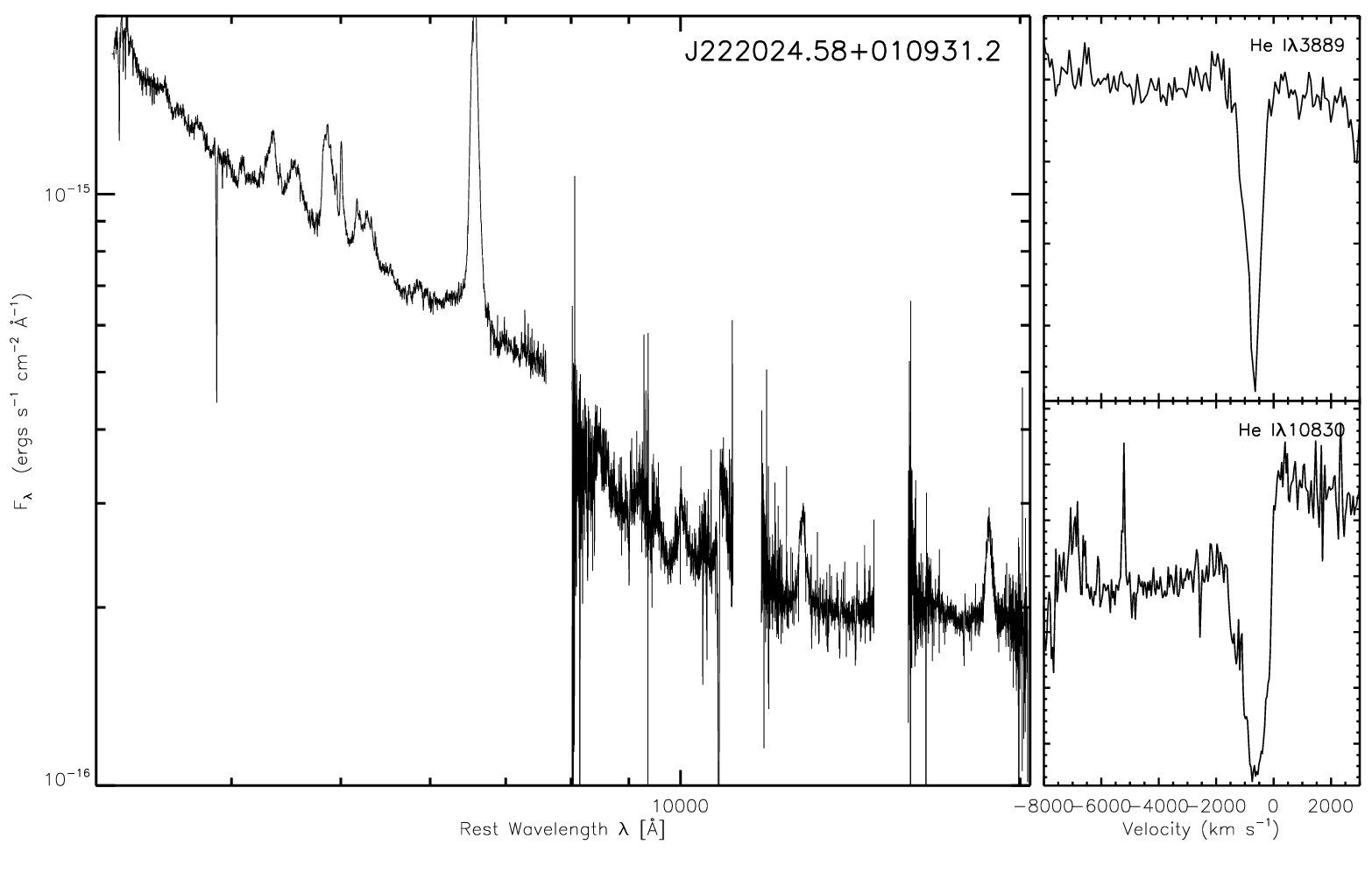}
   \caption{Low-z \hei\ BALs that have UV or NIR spectra.}
 \end{figure}

 \subsection{Demonstration of fitting the \hei$\lambda3889$ BAL quasars \label{app:fitresult}}

 \subsubsection{Sources also with \hei$\lambda3189$ absorption line}
  \begin{figure}[htbp]
   \centering
   \includegraphics[width=5.6in]{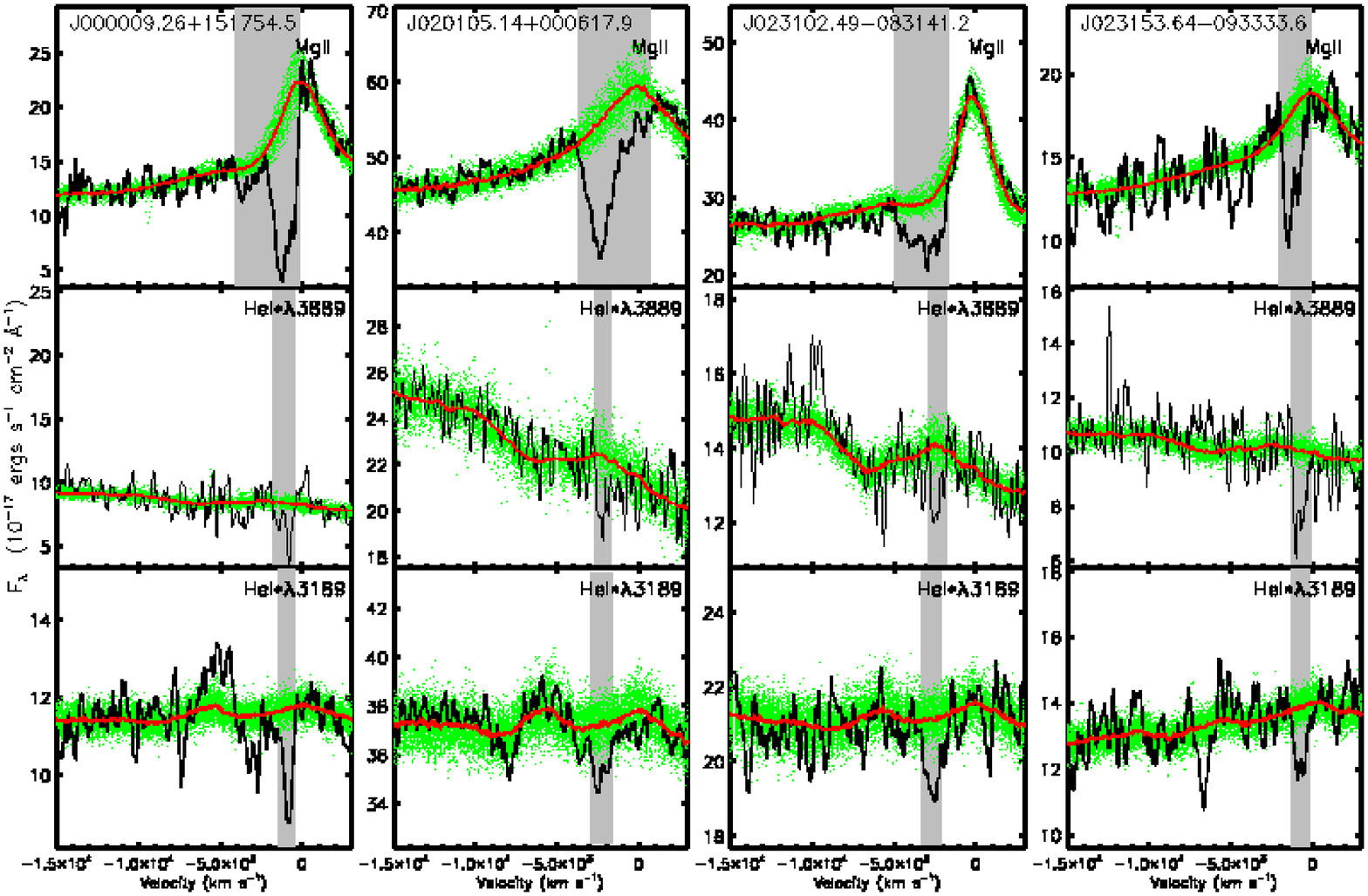}
   \includegraphics[width=5.6in]{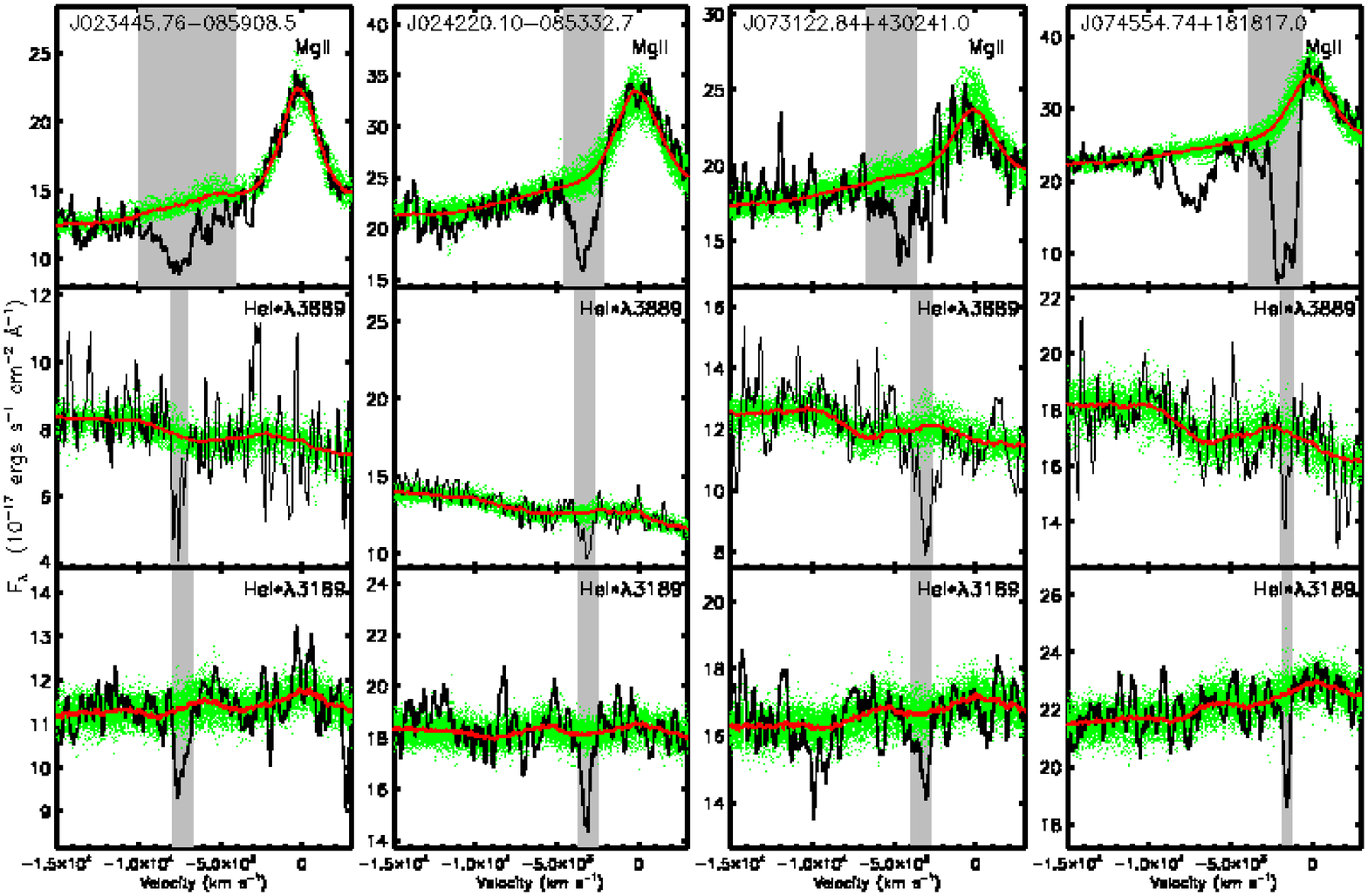}
   \caption{\footnotesize Demonstration of the spectral fitting for all the 52 sources that have Mg\,II, 
     He\,I*$\lambda3889$ and He\,I*$\lambda3189$ absorption troughs. 
     The acceptable fittings of the unabsorbed spectrum of an object are denoted as green dotted lines,
     and their mean spectrum denoted as a red solid line.   \label{fig:fitresult}  }
 \end{figure}

  \begin{figure}[htbp]
   \centering
   \includegraphics[width=5.8in]{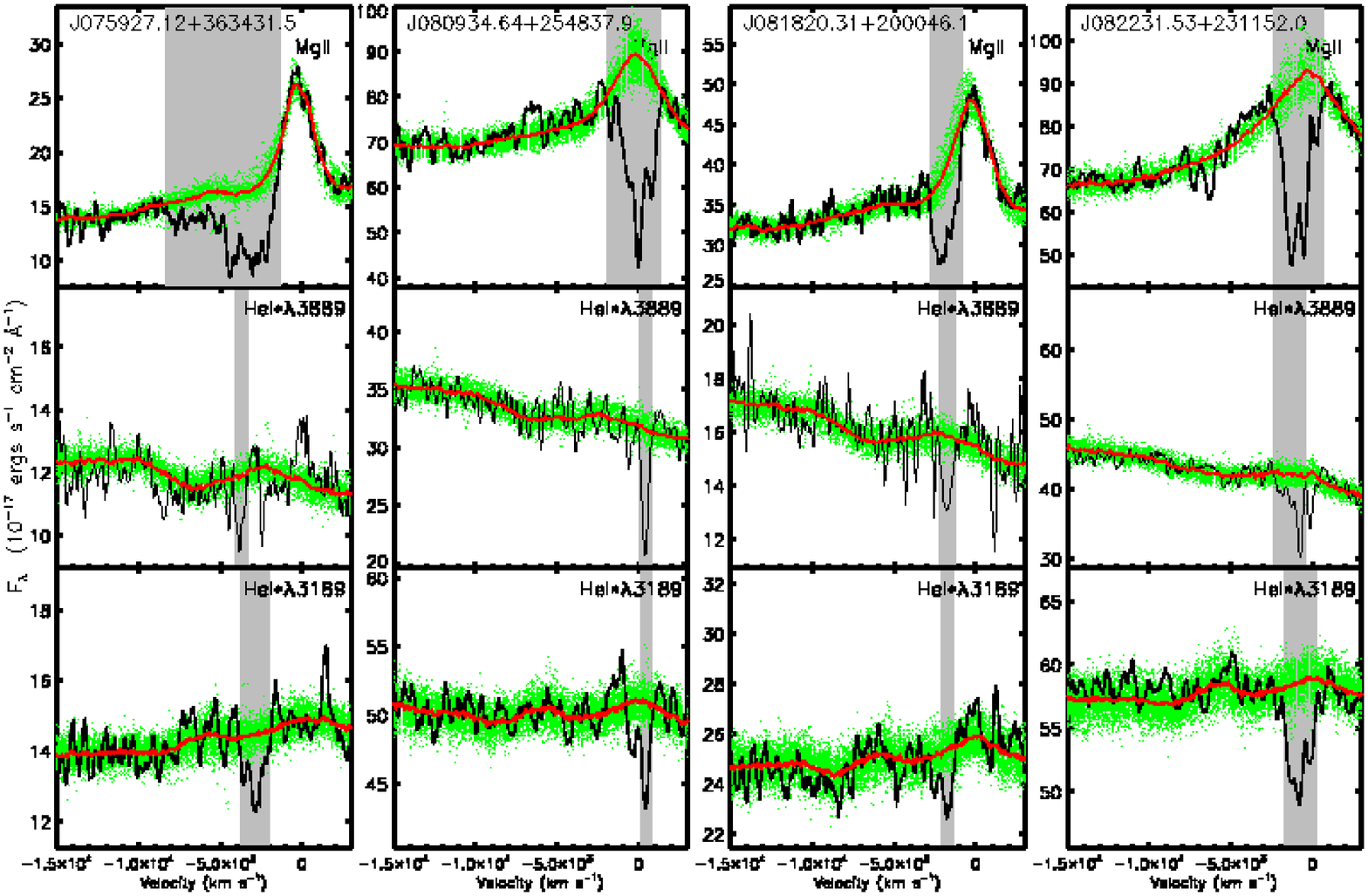}
   \includegraphics[width=5.8in]{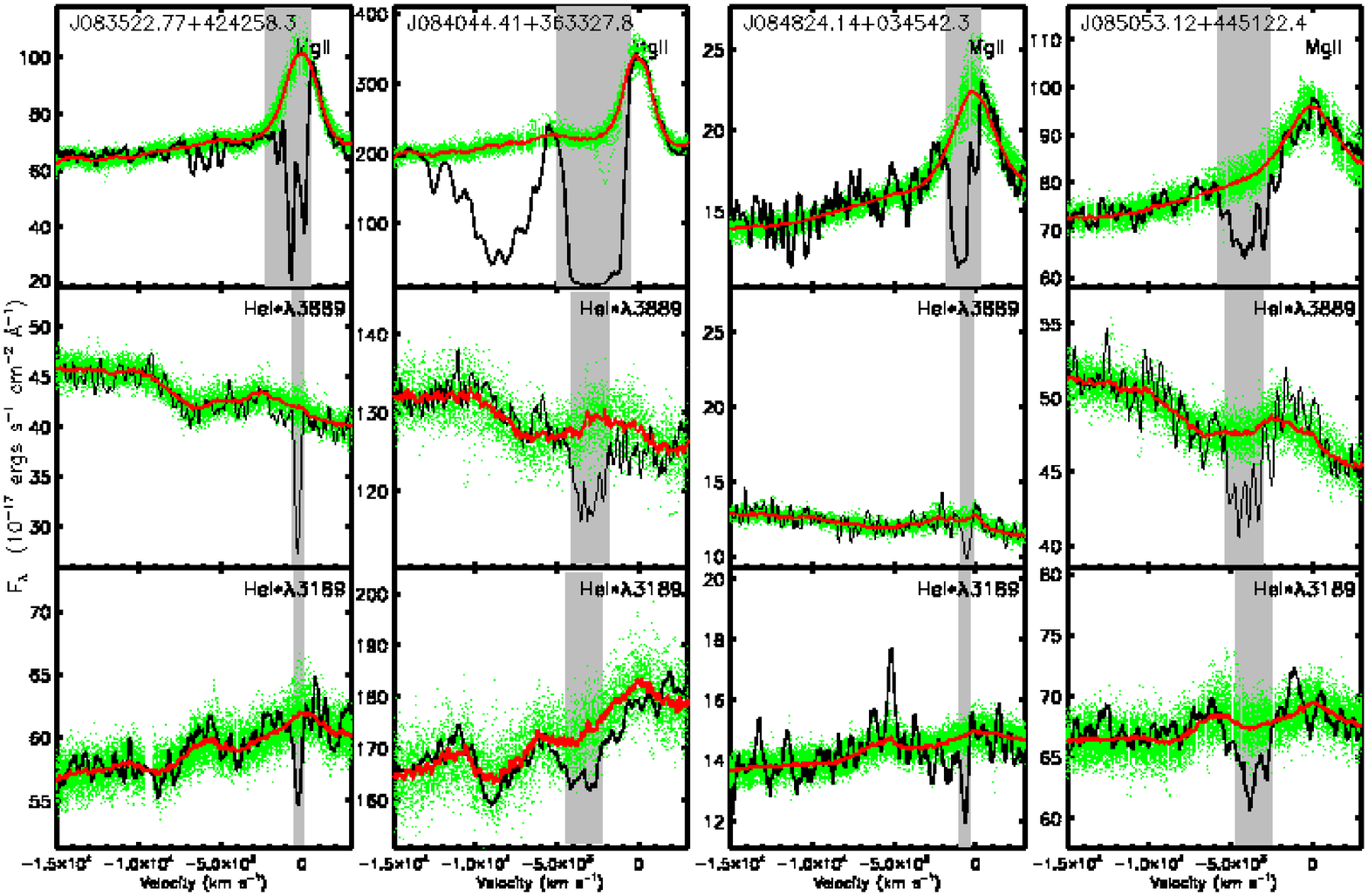}
 \end{figure}
\begin{figure}[htbp]
   \centering
   \includegraphics[width=5.8in]{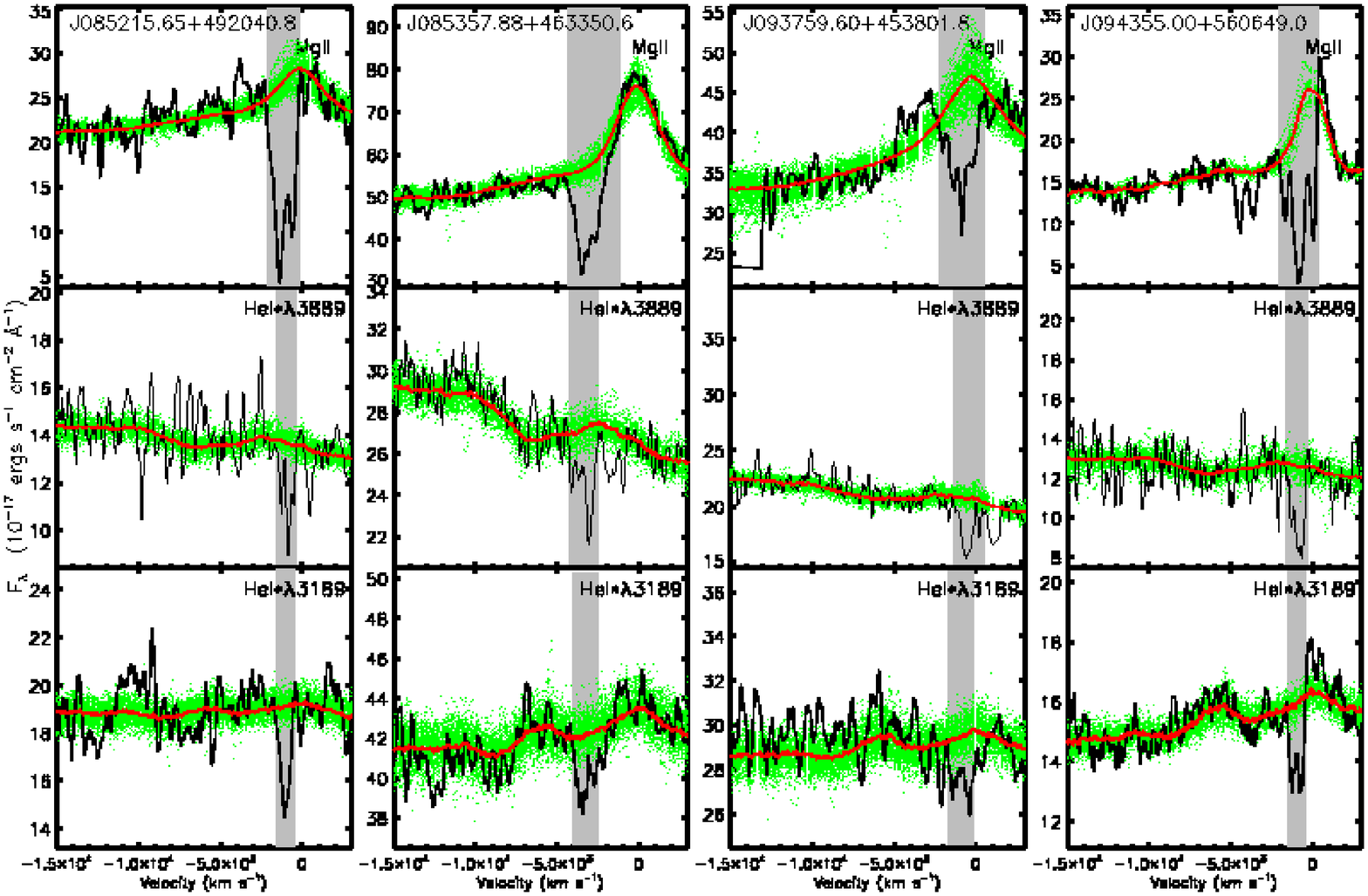}
   \includegraphics[width=5.8in]{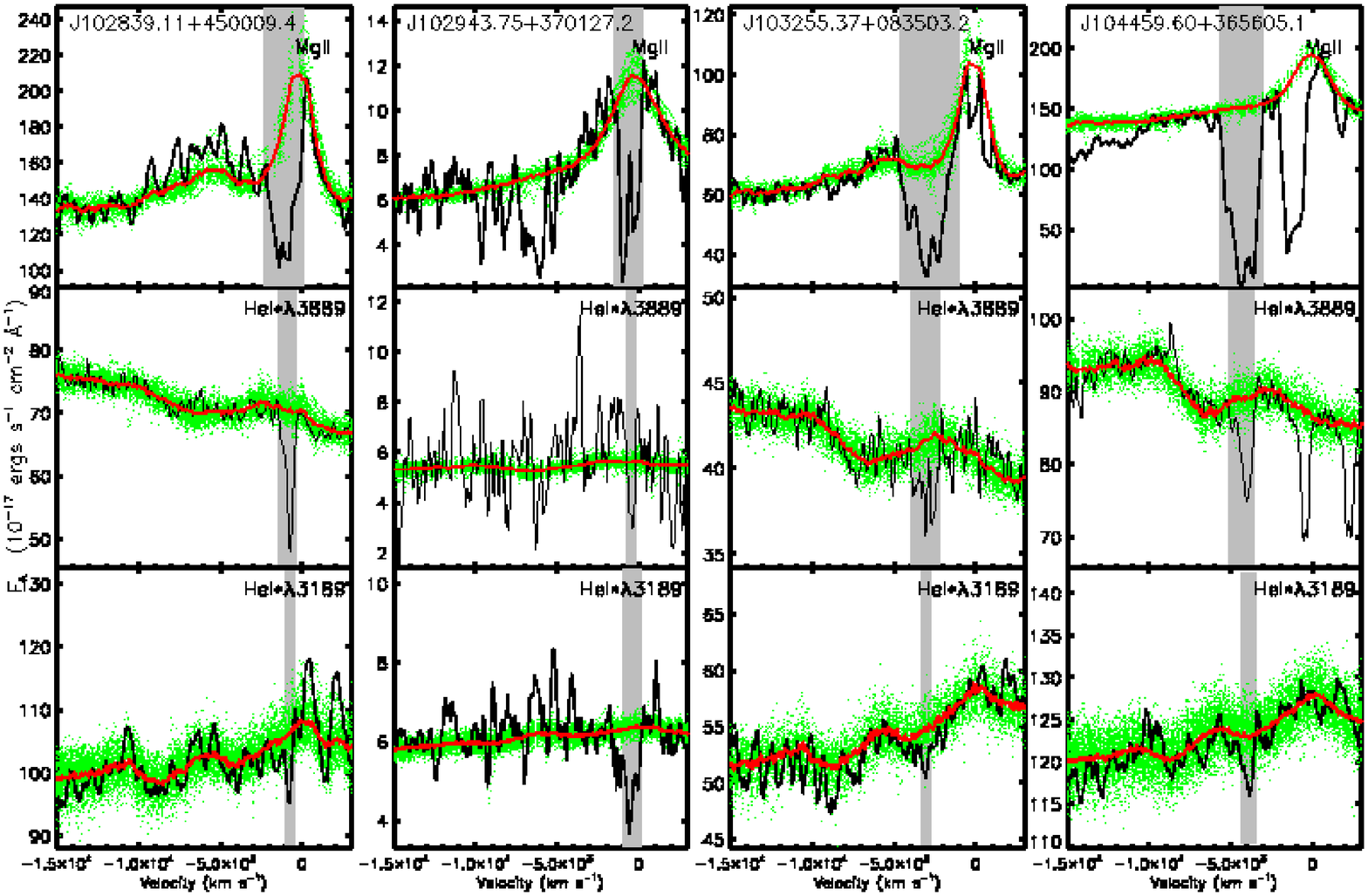}
 \end{figure}
\begin{figure}[htbp]
   \centering
   \includegraphics[width=5.8in]{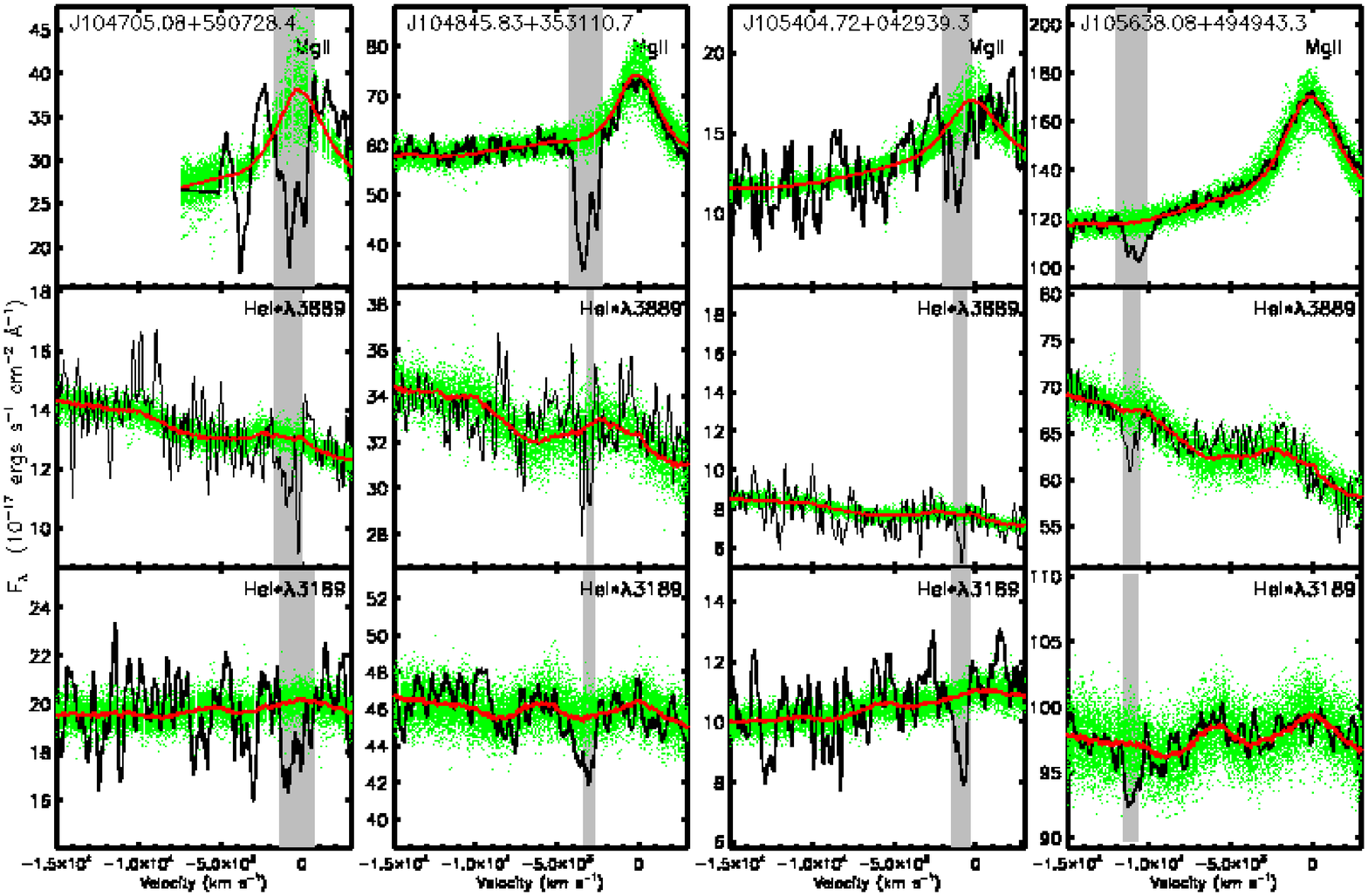}
   \includegraphics[width=5.8in]{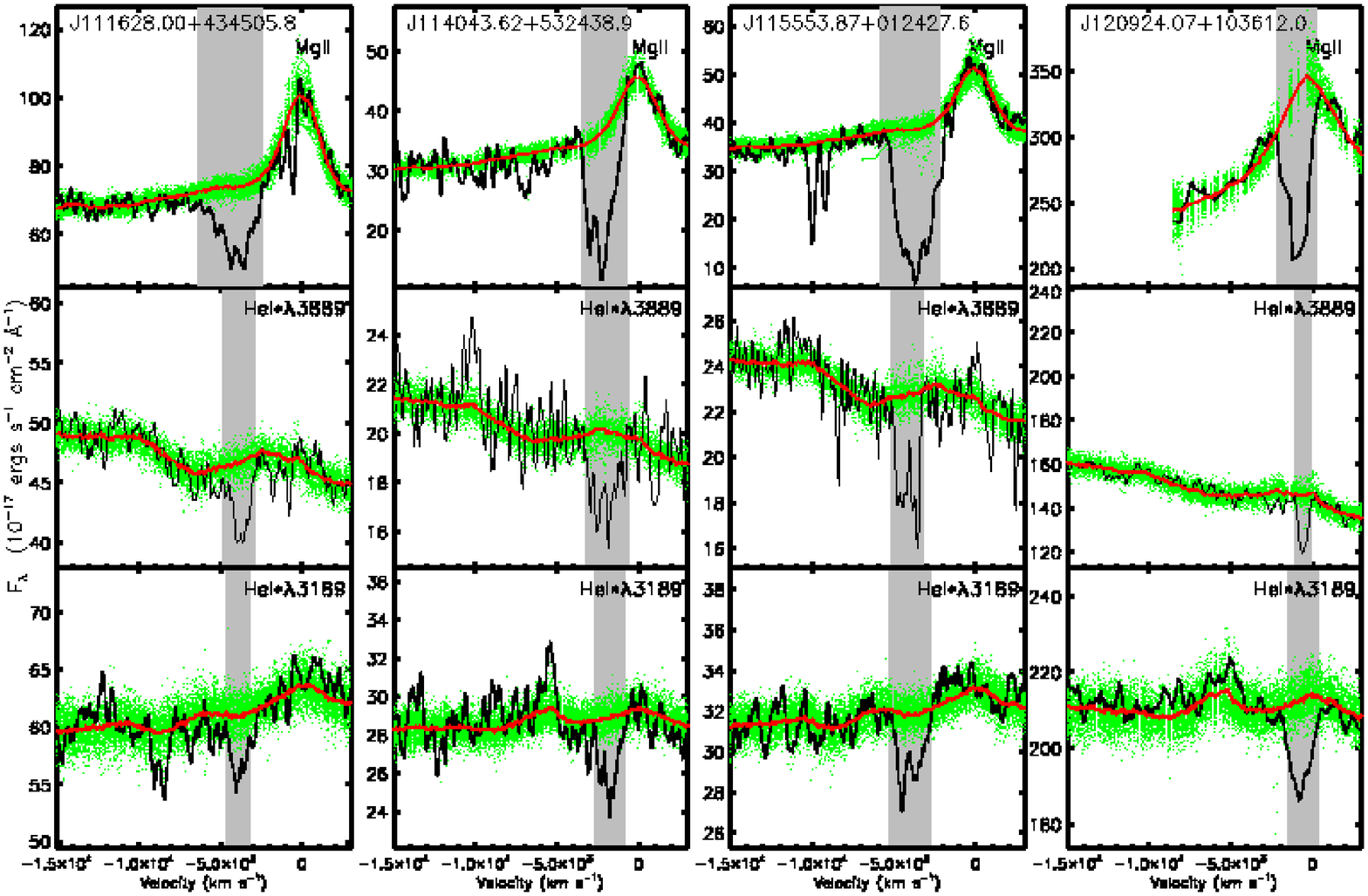}
 \end{figure}
\begin{figure}[htbp]
   \centering
   \includegraphics[width=5.8in]{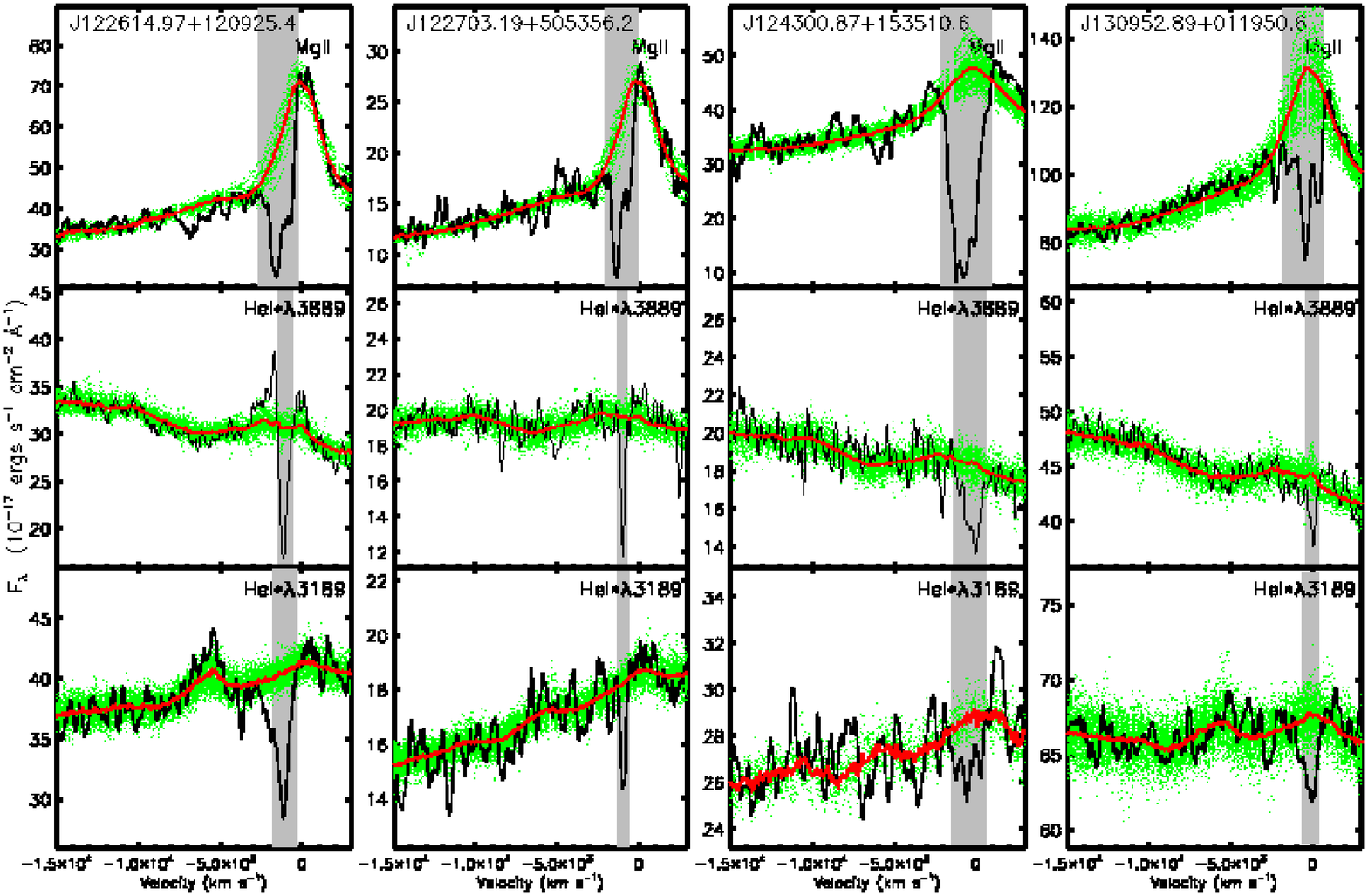}
   \includegraphics[width=5.8in]{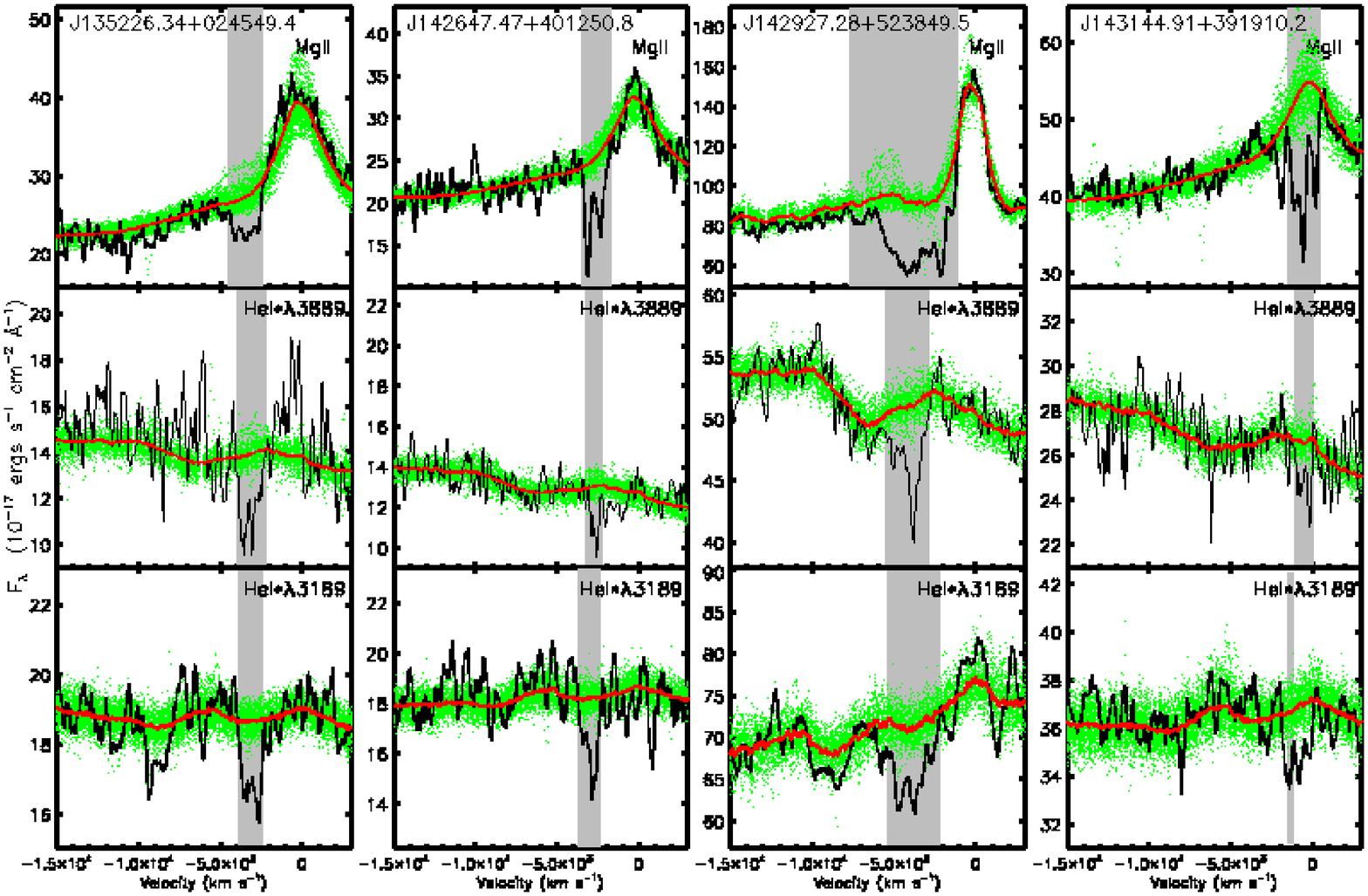}
 \end{figure}
\begin{figure}[htbp]
   \centering
   \includegraphics[width=5.8in]{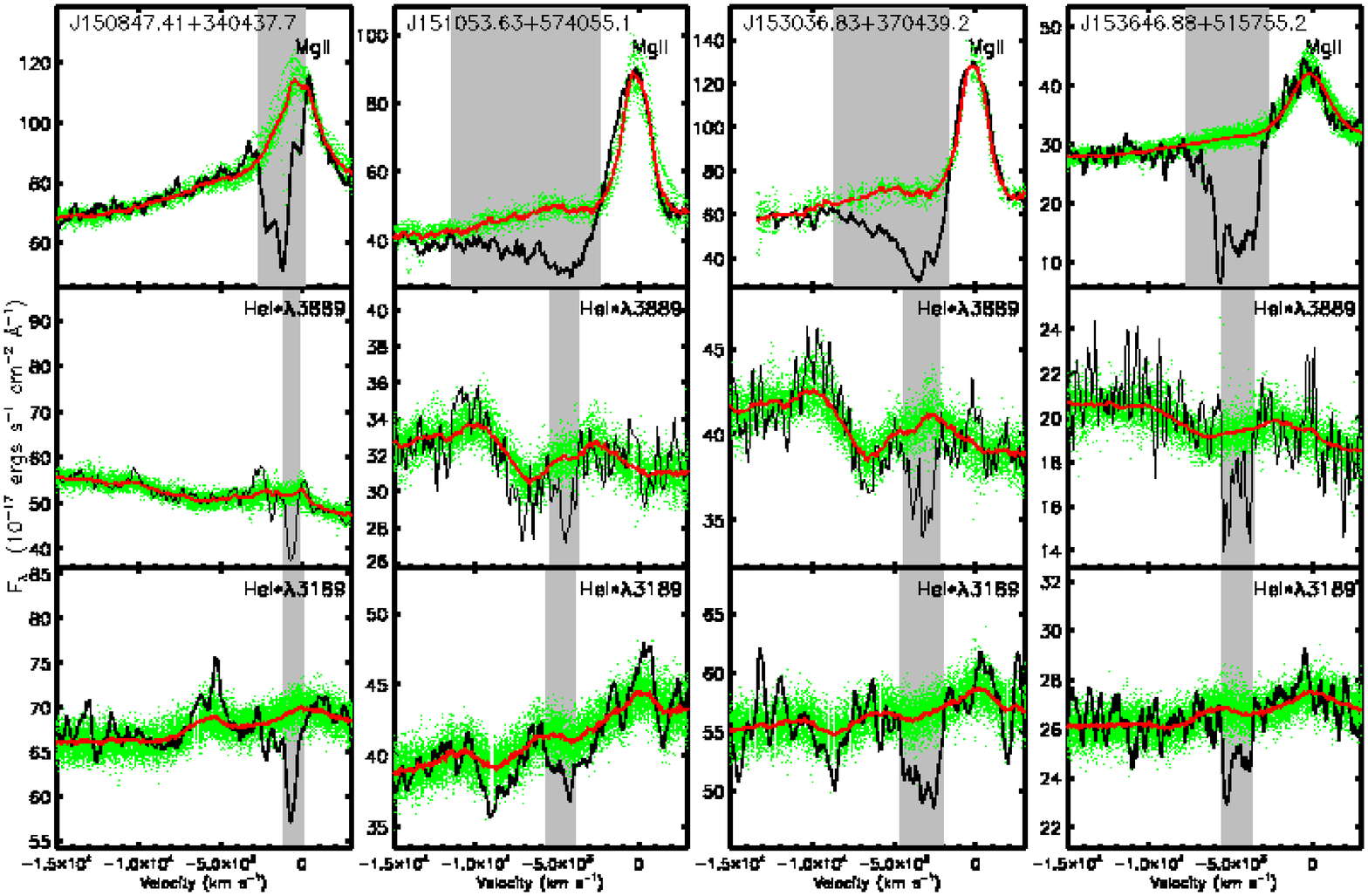}
   \includegraphics[width=5.8in]{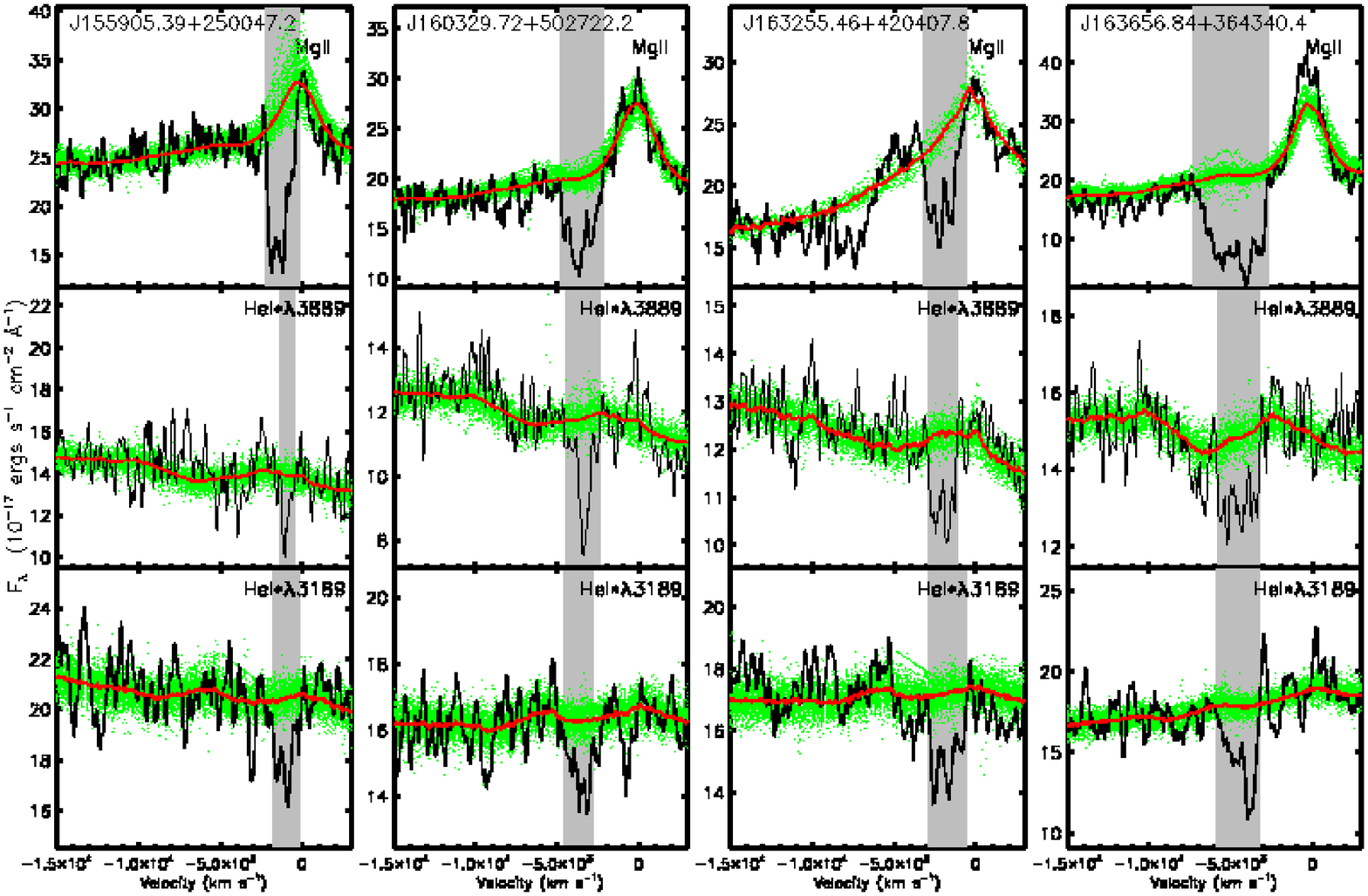}
 \end{figure}
\begin{figure}[htbp]
   \centering
   \includegraphics[width=5.8in]{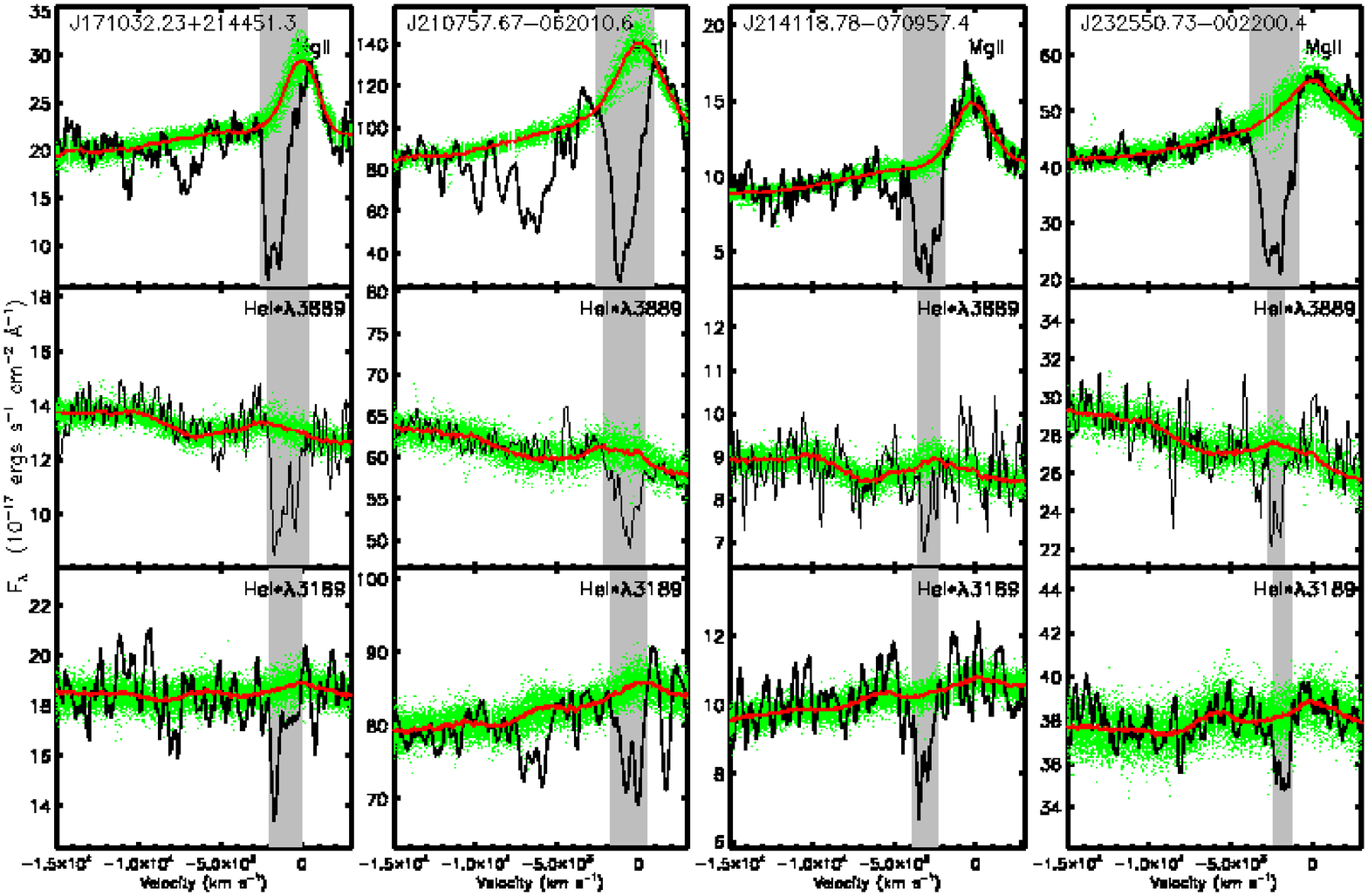}
 \end{figure}

 \clearpage

 \subsubsection{Sources without \hei$\lambda3189$ absorption line}
\begin{figure}[htbp]
   \centering
   \includegraphics[width=5.8in]{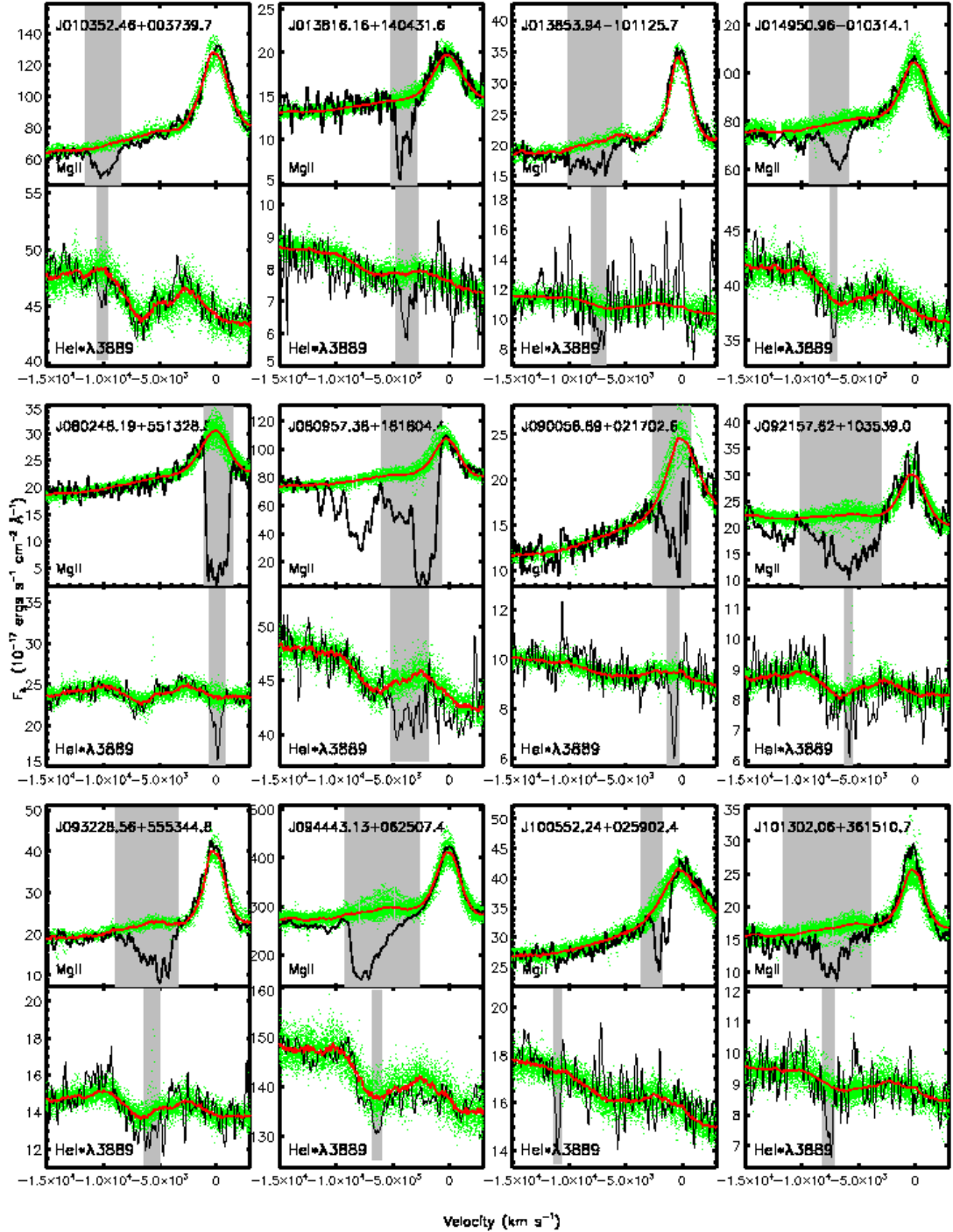}
   \caption{\footnotesize Demonstration of the spectral fitting of the 49 He\,I*$\lambda3889$ BAL quasars 
   without a He\,I*$\lambda3189$ absorption line detected. The symbols are the same as Figure~\ref{fig:fitresult}. }
 \end{figure}

\begin{figure}[htbp]
   \centering
   \includegraphics[width=5.8in]{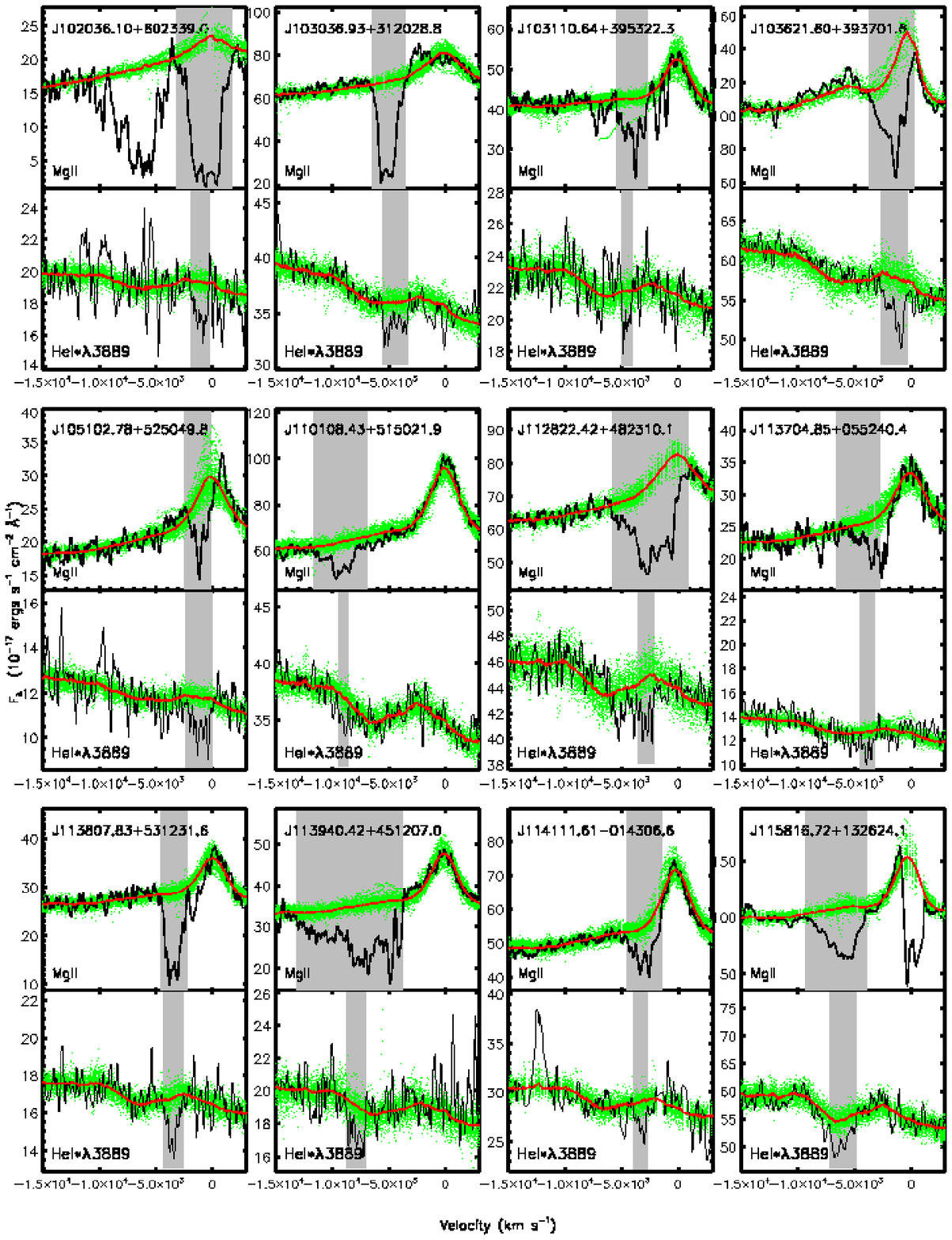}
 \end{figure}

\begin{figure}[htbp]
   \centering
   \includegraphics[width=6.2in]{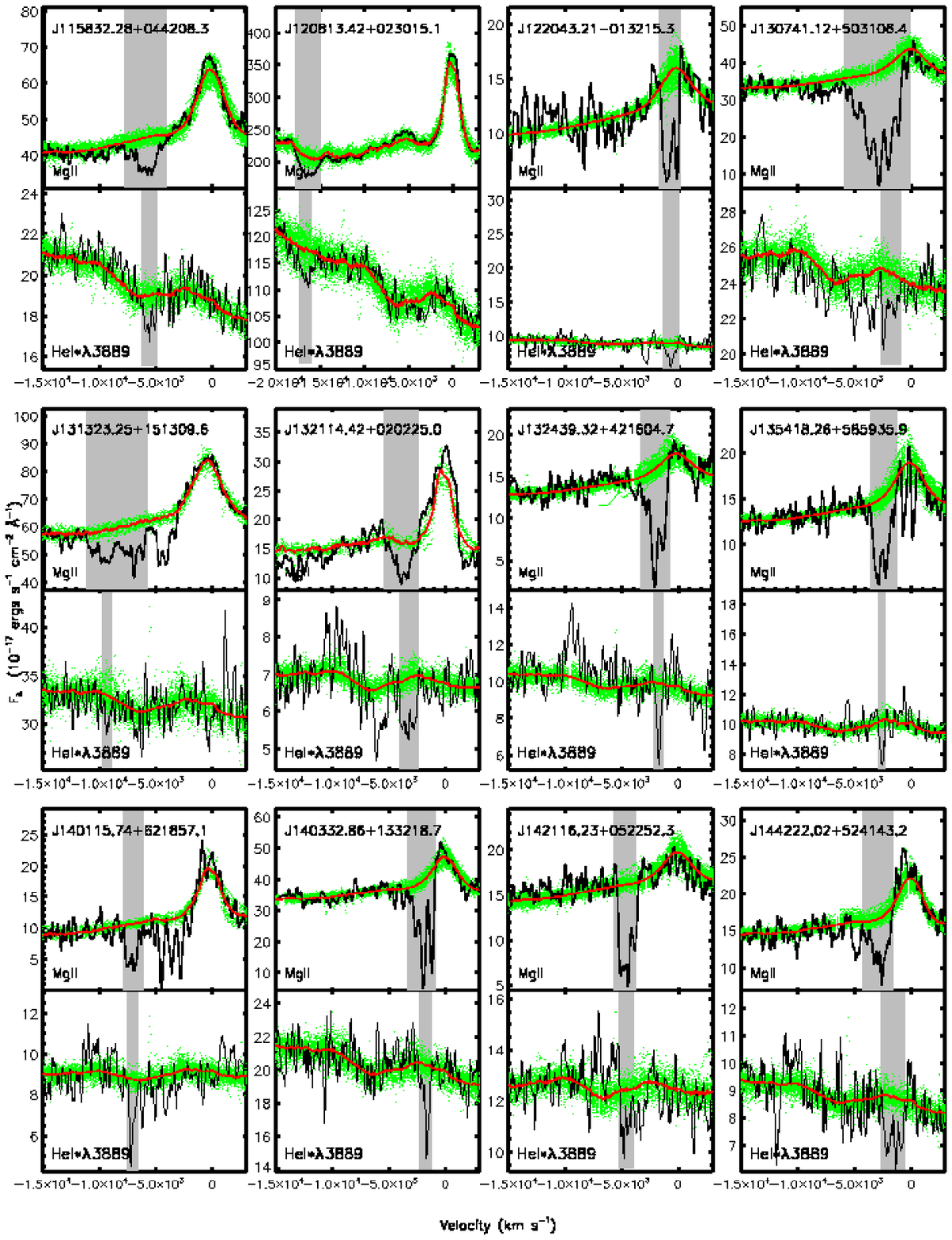}
 \end{figure}

 \begin{figure}[htbp]
   \centering
   \includegraphics[width=6.2in]{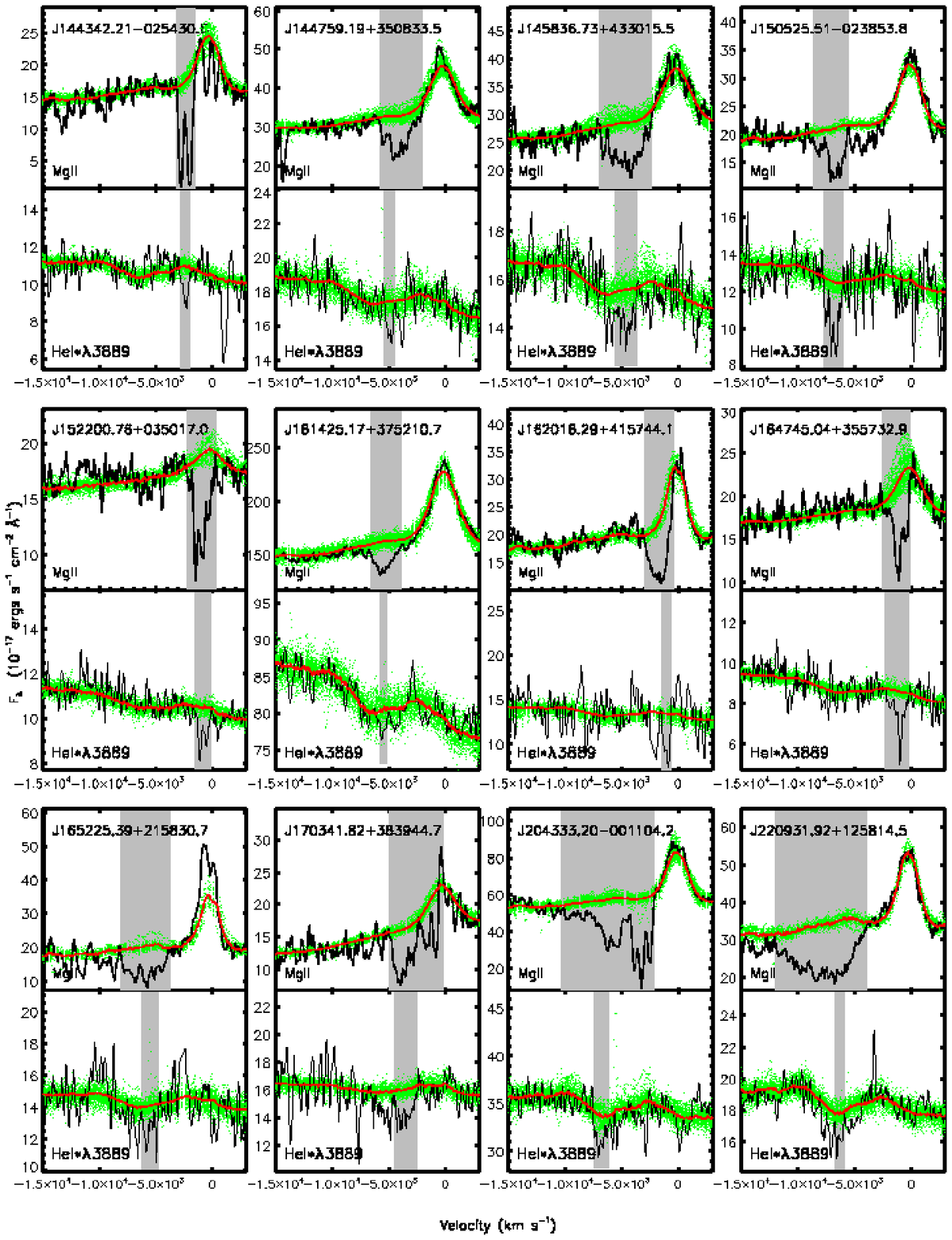}
 \end{figure}

\begin{figure}[htbp]
   \centering
   \includegraphics[width=6.2in]{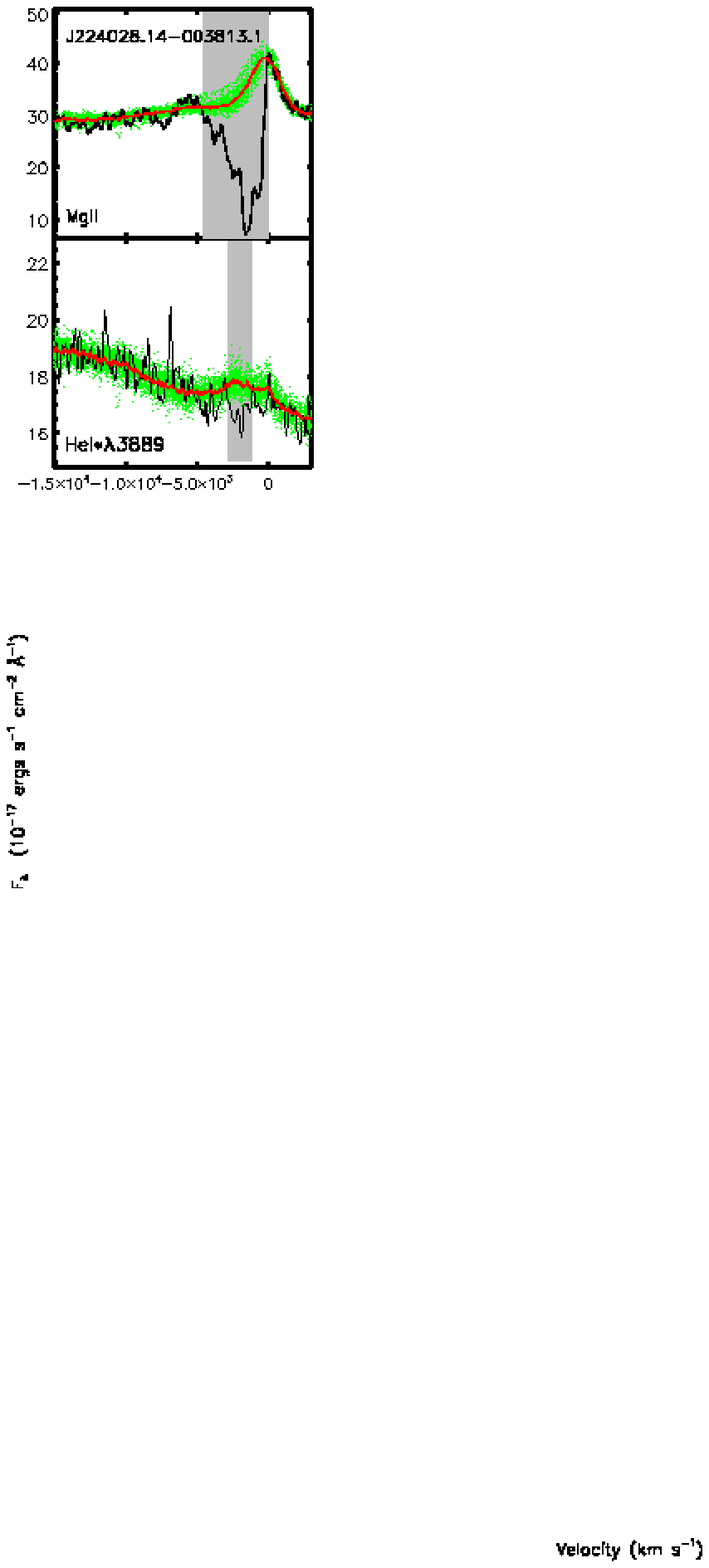}
 \end{figure}

\clearpage


\end{document}